\documentclass[fleqn,usenatbib]{mnras}

\usepackage{newtxtext,newtxmath}
\usepackage{textgreek}

\usepackage[T1]{fontenc}
\usepackage{ae,aecompl}

\usepackage{graphicx}   
\usepackage{siunitx}
\usepackage{amsmath}    
\usepackage{amssymb}    
\usepackage{subfig}

\usepackage{bm}

\usepackage[para, online, flushleft]{threeparttable}

\newcommand{\name}{SDSS 1206+4332~}

\newcommand{\Hunit}{km s$^{-1}$ Mpc$^{-1}$}
\newcommand{\HTOMlarge}{66.7$^{+6.1}_{-5.6}$}
\newcommand{\HTWE}{68.8$^{+5.4}_{-5.1}$}
\newcommand{\HFOU}{72.5$^{+2.1}_{-2.3}$}
\newcommand{\Om}{$\Omega_{\rm m}$}

\title[Cosmographic analysis of the doubly imaged quasar \name]{H0LiCOW - IX. Cosmographic analysis of the doubly imaged quasar \name and a new measurement of the Hubble constant}

\author[S. Birrer et al.]{\parbox{\textwidth}{
S. Birrer,$^{1}$\thanks{E-mail: sibirrer@astro.ucla.edu}
T. Treu,$^{1}$ 
C. E. Rusu,$^{2,3}$
V. Bonvin,$^{4}$
C. D. Fassnacht,$^{3}$
J.~H.~H.~Chan,$^{4}$
A. Agnello,$^{5}$
A. J. Shajib,$^{1}$
G.~C.-F.~Chen,$^{3}$
M. Auger,$^{6}$ 
F. Courbin,$^{4}$
S. Hilbert,$^{7, 8}$
D. Sluse,$^{9}$
S.~H.~Suyu,$^{10, 11, 12}$
K.~C.Wong,$^{13}$
P. Marshall,$^{14}$ 
B. C. Lemaux,$^{3}$
G. Meylan$^{4}$ 
} 
\\
\\
$^{1}$Department of Physics and Astronomy, University of California, Los Angeles, CA 90095, USA\\
$^{2}$ Subaru Fellow; Subaru Telescope, National Astronomical Observatory of Japan, 650 N Aohoku Pl, Hilo, HI 96720, USA\\ 
$^{3}$Department of Physics, University of California, Davis, CA 95616, USA\\
$^{4}$Laboratoire d'astrophysique Ecole Polytechnique Fédérale de Lausanne (EPFL), Observatoire de Sauverny, CH-1290 Versoix, Switzerland\\
$^{5}$European Southern Observatory, Karl-Schwarzschild-Strasse 2, 85748 Garching bei Muenchen, Germany \\
$^{6}$Institute of Astronomy, University of Cambridge, Madingley Road, Cambridge CB3 0HA, UK\\
$^{7}$ Exzellenzcluster Universe, Boltzmannstr. 2, D-85748 Garching, Germany \\
$^{8}$Ludwig-Maximilians-Universit{\"a}t, Universit{\"a}ts-Sternwarte, Scheinerstr. 1, D-81679 M{\"u}nchen, Germany\\
$^{9}$STAR Institute, Quartier Agora - All\'ee du six Ao\^ut, 19c B-4000 Li\`ege, Belgium\\
$^{10}$Max-Planck-Institut f{\"u}r Astrophysik, Karl-Schwarzschild-Str.~1, 85748 Garching, Germany\\
$^{11}$Physik-Department, Technische Universit\"at M\"unchen, James-Franck-Stra\ss{}e~1, 85748 Garching, Germany\\
$^{12}$Institute of Astronomy and Astrophysics, Academia Sinica, 11F of ASMAB, No.1, Section 4, Roosevelt Road, Taipei 10617, Taiwan\\
$^{13}$EACOA Fellow; National Astronomical Observatory of Japan, 2-21-1 Osawa, Mitaka, Tokyo 181-8588, Japan\\
$^{14}$Kavli Institute for Particle Astrophysics and Cosmology, Stanford University, 452 Lomita Mall, Stanford, CA 94035, USA\\
}

\date{Accepted XXX. Received YYY; in original form ZZZ}

\pubyear{2018}

\begin{document}
\label{firstpage}
\pagerange{\pageref{firstpage}--\pageref{lastpage}}
\maketitle

\begin{abstract}
We present a blind time-delay strong lensing (TDSL) cosmographic analysis of the doubly imaged quasar \name. We combine the relative time delay between the quasar images, \textit{Hubble Space Telescope} imaging, the \textit{Keck} stellar velocity dispersion of the lensing galaxy, and wide-field photometric and spectroscopic data of the field to constrain two angular diameter distance relations. The combined analysis is performed by forward modelling the individual data sets through a Bayesian hierarchical framework, and it is kept blind until the very end to prevent experimenter bias. After unblinding, the inferred distances imply a Hubble constant $H_0 = $ \HTWE~\Hunit, assuming a flat $\Lambda$ cold dark matter cosmology with uniform prior on $\Omega_{\text{m}}$ in [0.05, 0.5].
The precision of our cosmographic measurement with the doubly imaged quasar \name is comparable with those of quadruply imaged quasars and opens the path to perform on selected doubles the same analysis as anticipated for quads. Our analysis is based on a completely independent lensing code than our previous three H0LiCOW systems and the new measurement is fully consistent with those. We provide the analysis scripts paired with the publicly available software to facilitate independent analysis. The consistency between blind measurements with independent codes provides an important sanity check on lens modelling systematics. By combining the likelihoods of the four systems under the same prior, we obtain $H_0=$ \HFOU~\Hunit. This measurement is independent of the distance ladder and other cosmological probes.
\end{abstract}

\begin{keywords}
method: Gravitational lensing -- cosmology -- galaxies -- Hubble constant
\end{keywords}



\section{Introduction}

The standard cosmological model, $\Lambda$ Cold Dark Matter (CDM), is extremely successful in simultaneously describing the structure and scales in the very early universe (cosmic microwave background (CMB), baryogenesis) and the corresponding scales at low redshift (galaxy clustering, weak gravitational lensing, baryonic acoustic oscillations (BAO), supernovae of type Ia (SNIa)). A vital component of $\Lambda$CDM is the cosmological constant $\Lambda$ describing the late time acceleration of the universe \citep{Riess:1998qe,Perlmutter:1999gl}.

A popular approach to cosmography is to anchor the absolute physical scales at the last scattering surface of the cosmic microwave background (CMB) photons and propagate them to lower redshifts and more recent cosmic times using a cosmological model. Within this approach, the latest constraints from the CMB alone imply that the Hubble constant is $H_0 = 67.27 \pm 0.60$ \Hunit \citep[TT,TE,EE+lowE, 1-$\sigma$ limit,][]{Planck:2018_cosmo_param}, assuming a flat $\Lambda$CDM model. This is a sub-per-cent precision indirect measurement of the physical scales at recent times \citep[see, also,][]{Hinshaw:2013ya}. The CMB constraints can be combined with intermediate redshift probes, such as BAO \citep{Aubourg:2015} (requiring a prior on the sound horizon at drag epoch from the CMB) and cosmic shear \citep{DESH0:2018}. This approach is known as the inverse distance ladder.

Alternatively, using the local distance ladder method, \citet[][]{Riess:2016yp, Riess:2018a, Riess:2018_gaia} measure $H_0 = 73.48 \pm 1.66$ \Hunit (2.3 per cent precision on $H_0$) and \cite{Freedman:2012} measure $H_0 = 74.3 \pm 2.1$ \Hunit \citep[see also][]{Cao:2017, Jang:2018, Dhawan:2018}. There is currently a $\sim$3-$\sigma$ level tension between the direct and inverse distance ladder determination of the Hubble constant. If confirmed at higher level of significance, this tension would imply that new physics beyond flat $\Lambda$CDM is required \citep[see e.g. review by][]{Suyu:2018}. Independent methods with comparable precision are particularly valuable as a check against unknown systematics that may affect either or both the direct and inverse distance ladder method \citep[][]{GravSirene2017}.

A completely independent approach to measuring $H_0$ is Time-Delay Strong Lensing (TDSL). First proposed by \cite{Refsdal:1964pi}, the method has been applied by measuring the difference in arrival time of photons from multiply imaged active galactic nuclei\footnote{Recent discoveries \citep{Kelly:2015,Treu:2016Refsdal, Goldstein:2017, Goobar:2017, More:2017, Grillo:2018} are paving the way for TDSL cosmography using multiply imaged supernovae as originally suggested by \cite{Refsdal:1964pi}.} \citep[][see also review by \cite{Treu:2016} for a historic perspective and additional references]{Schechter:1997, Treu_Koopmans:2002, Suyu:2010rc, Fadely:2010, Suyu:2014aq, Birrer:2016zy, Bonvin:2017}. TDSL provides a direct measurement of the physical scales (or ratios) in a particular lens configuration along a particular line of sight (LOS) and yields a measurement of the Hubble constant fully independent of the local distance ladder and the CMB.

The keys to a precise and accurate determination of distances using TDSL are several. First, a precise time-delay measurement is needed, which typically requires multi-year monitoring campaigns with per-cent-level photometry \citep{Fassnacht:2002, COSMOGRAIL_1:2005, Kochanek:2006, Tewes:2013xr, Liao:2015, tak2017bayesian, Bonvin:2016} or high cadence monitoring with millimag photometry \citep{Courbin:2018, Bonvin:2018a}. Second, high signal-to-noise ratio and high-resolution imaging of the host galaxy of the lensed active galactic nuclei are needed to constrain the differences in gravitational potential across the images \citep{Suyu:2010rc}. Third, a spectroscopic measurement of the stellar velocity dispersion \citep{Treu_Koopmans:2002} is needed to help break the mass-sheet degeneracy \citep{Falco:1985no} and its generalizations \citep{Schneider:2014fo, Unruh:2017}. Fourth, one needs to measure and model the effect of mass inhomogeneities along the LOS and in the immediate neighborhood of the main deflector \citep{Keeton:2004fp, Fassnacht:2011zs, Collett:2013bh, Greene:2013wb, McCully2014, Wong:2018}.

Building on over a decade of efforts to develop techniques and gather data with sufficient constraining power, the H0LiCOW collaboration\footnote{\url{www.h0licow.org}} \citep[][]{Suyu:2017_h0licow} has published the analysis of three quadruply imaged active galactic nuclei \citep{Suyu:2010rc, Suyu:2014aq, Wong:2017, Rusu2017, Sluse:2017, Tihhonova:2018} with time delays measured by the COSMOGRAIL collaboration \citep{COSMOGRAIL_1:2005, Bonvin:2018a} and by \cite{Fassnacht:2002}. The combined constraints from the three lenses are presented by \citet{Bonvin:2017} and result in a measurement of the Hubble constant $H_0 =71.9^{+2.4}_{-3.0}$ km s$^{-1}$ Mpc$^{-1}$ and $\Omega_{\Lambda}=0.62^{+0.24}_{-0.35}$ in flat $\Lambda$CDM with uniform priors on $\Omega_{\Lambda}$ in $[0,1]$ and $H_0$ in $[0,150]$. Importantly, after the first pilot system, the H0LICOW analysis was performed blindly to the cosmological parameters, so as to avoid conscious or unconscious experimenter bias.

The precision of TDSL is currently limited by the small sample sizes of known lenses with all the appropriate ancillary data. To get to $\approx$ 1 per cent precision on the Hubble constant that is required to make the most of current and future dark energy experiments \citep{Weinberg:2013a} using TDSL, a sample of about 40 lenses needs to be analyzed with comparable measurement precision per system as the \citet{Bonvin:2017} sample \citep{Treu:2016, Jee:2016, Shajib:2018}. Whereas the number of quadruply lensed quasars discovered has vastly increased recently and is approaching the desired number \citep{Schechter:2017, Lin:2017DES, Jacobs:2017, Ostrovski:2017, Agnello:2018, Lemon:2018, Treu:2018}, quads only represent $\sim1/6$ of all lensed quasars in the sky \citep{Oguri:2010a, Collett:2015a}.

The inclusion of doubly lensed quasars in the TDSL analysis, which are 5 times more abundant as quads in the sky and generally easier to monitor for time delays, would substantially enlarge the final sample size thus boosting the statistical precision. Furthermore, a more diverse lens sample would allow for additional assessment of relative systematics among different subsets of the TDSL sample. A statistical approach to examine the dependence of time delays on the complexity of lens potentials based on a sample of 16 lensed quasars has been performed by \cite{Oguri:2007} and resulted in a Hubble parameter of $H_0 = 68 \pm 6$ (stat.) $ \pm 8$ (syst.) km s$^{-1}$Mpc$^{-1}$, by \cite{Read:2007} with 10 lensed quasars to result in $H_0 = 64^{+8}_{-9}$ km s$^{-1}$Mpc$^{-1}$ superceded by \cite{Coles:2008} with 11 lensed quasars yielding in $H_0 = 71^{+6}_{-8}$ km s$^{-1}$Mpc$^{-1}$. However, these old results should be taken with a grain of salt, since many of the time delays that went into that analysis are now superseded by better and improved determinations from multi-year monitoring campaigns.

As statistical precision improves, the combined systematic uncertainties must be controlled to the same level of accuracy. The agreement between the three existing measurements of $H_0$ from the H0LiCOW collaboration is encouraging. However, more work is needed to determine the systematic floor of the current approach and identify ways to reduce it.

In this work, we address two of the issues discussed above, statistical uncertainties and systematic limitations, by performing a blind cosmographic analysis of the doubly lensed quasar \name, using a lens modelling framework and code that are completely independent of those used for the first three lenses. The system is of a special kind: although the quasar is only doubly imaged, parts of the host galaxy cross the inner lensing caustic and get quadruply lensed in a fold configuration forming an extended ring. This configuration allows for a very similar analysis as recently applied for quadruply lensed quasars \citep{Suyu:2010rc, Suyu:2014aq, Birrer:2016zy, Wong:2017}. We expect that many similar examples with relatively high surface brightness parts of quasar host galaxy crossing the inner caustic can be found as hundreds of doubles are discovered, and thus our analysis can serve as a pathfinder for much larger samples.

We self-consistently incorporate new high resolution \textit{HST} imaging data
with existing kinematics data of \cite{Agnello:2016a}, quasar light curves monitoring data of  \cite{Eulaers:2013} (hereafter, E13), and a LOS analysis in
a Bayesian hierarchical model. We provide the full likelihood of the
cosmographic analysis that enables a self-consistent combined analysis
with other strong lenses and other cosmographic probes. We also
provide a new determination of the Hubble constant, independent of the
local and inverse distance ladder method. Finally, since our new blind measurement is consistent with the previous H0LiCOW collaboration measurements, we combine the likelihood from the four lenses to provide an updated TDSL measurement of the Hubble constant with $\sim3$ per cent precision in a flat $\Lambda$CDM cosmology.

The paper is structured as follows: In Section~\ref{sec:outline}, we
describe the basics of time-delay cosmography and outline the steps of
our analysis. Section~\ref{sec:data} describes the lens system \name
and the data used in our analysis. We describe the model choices and
different options we assess in our analysis in Section~\ref{sec:model_choices}. We then go through the forward modelling of
the different data sets in Section~\ref{sec:forward_modelling}. Section~\ref{sec:los_analysis} describes
the LOS analysis. We describe the combined Bayesian hierarchical
analysis in Section~\ref{sec:combined_analysis}. We present our
results in Section~\ref{sec:results} and summarize our work in Section~\ref{sec:conclusion}.

Crucially, the analysis presented in this work through Section~\ref{sec:outline} - \ref{sec:combined_analysis} was laid out and executed blindly with respect to the cosmographic result and in particular the value of the Hubble constant. The blinding is built in the software, by subtracting the average of every posterior distribution function before revealing it to the investigator. The scripts and pipelines are then frozen before the cosmological inference is unblinded. We displayed the cosmographic likelihood and the inference of the cosmological parameters only after all co-authors involved in the time-delay analysis have agreed that the analysis was satisfactory. The submission of this manuscript followed shortly after the unblinding with only minor changes in the text for clarity and updated figures.

The analysis and the lens modelling are performed with the publicly available software \textsc{lenstronomy}\footnote{\url{https://lenstronomy.readthedocs.io}} \citep{Birrer_lenstronomy, Birrer2015_basis_set} version \textsc{0.3.3} and the reduced data products and the lens modelling scripts are made publicly available after acceptance of the manuscript.

\section{Outline of the analysis} 
\label{sec:outline}
We combine time-delay measurements between the two images of the quasar, $\Delta t_{\rm AB}$, \textit{Hubble Space Telescope} (\textit{HST}) imaging data, $\boldsymbol{d_{\textit{HST}}}$, stellar kinematics of the deflector galaxy, $\sigma^\text{P}$, and wide field imaging and spectroscopy of the environment of the lens, $\boldsymbol{d_{\text{env}}}$, to measure angular diameter distances and hence the Hubble constant. We specifically denote $\boldsymbol{d_{\textit{HST}}}$ as the data vector of individual pixel values of the imaging data and $\boldsymbol{d_{\text{env}}}$ the collection of objects with their photometric and spectoscopical measurements.

This section outlines our analysis. We describe the observables and how they relate to the underlining cosmological model (Section \ref{sec:observables}), highlight the cosmographic constraining power of the combined data sets (Section \ref{sec:ang_diameter_dist}), layout the formal notation of the combined Bayesian analysis of this work (Section \ref{sec:cosmographic_analysis}), and highlight our strategy in regards to lensing degeneracies and other potential systematics (Section \ref{subsec:degeneracies}). The details of the modelling choices are presented in Sections~\ref{sec:model_choices} and~\ref{sec:forward_modelling}.

\subsection{Observables} \label{sec:observables}
The excess time delay \citep[see e.g.][]{Schneider:1992vp} of an image at $\bm{\theta}$ with corresponding source position $\bm{\beta}$ relative to an unperturbed path is
\begin{equation}
 t(\bm{\theta}, \bm{\beta}) = \frac{(1 + z_{\text{d}})}{c} \frac{D_{\text{d}}D_{\text{s}}}{D_{\text{ds}}} \left[ \frac{(\bm{\theta} - \bm{\beta})^2}{2} - \psi(\bm{\theta}) \right],
\end{equation}
where $z_{\rm d}$ is the redshift of the deflector, $c$ the speed of light, $\psi$ the lensing potential and $D_{\rm d}$, $D_{\rm s}$ and $D_{\rm ds}$ the angular diameter distances from the observer to the deflector, from the observer to the source and from the deflector to the source, respectively.

The relative time delay between two images A and B is
\begin{equation}\label{eqn:time_delay}
    \Delta t_{\rm AB} = \frac{D_{\Delta t}}{c} \left[\phi(\bm{\theta}_{\rm A}, \bm{\beta}) - \phi(\bm{\theta}_{\rm B}, \bm{\beta}) \right],
\end{equation}
where
\begin{equation}
	\phi(\bm{\theta}, \bm{\beta}) = \left[ \frac{(\bm{\theta} - \bm{\beta})^2}{2} - \psi(\bm{\theta}) \right]
\end{equation}
is the Fermat potential and
\begin{equation} \label{eqn:time_delay_distance}
 D_{\Delta t} \equiv (1+z_{\text{d}})\frac{D_{\text{d}}D_{\text{s}}}{D_{\text{ds}}}
\end{equation}
is the so-called time-delay distance.

The lensing potential, $\psi$, and the true source position, $\bm{\beta}$, required for the prediction of the time delay, can be inferred by modelling the appearance of multiply imaged structure in high resolution imaging data, $\boldsymbol{d_{\textit{HST}}}$. Comparison with the data allows us to constrain the parameters of the lens model, $\boldsymbol{\xi}_{\text{lens}}$, and the parameters of the surface brightness distribution of the deflector and lensed source model, $\boldsymbol{\xi}_{\text{light}}$, and their covariances.

The details of the mass distribution along the LOS can significantly impact observables and thus need to be taken into account \citep[see e.g.][]{McCully2017, Rusu2017, Sluse:2017, Birrer_2017los, Tihhonova:2018}. Large scale structure primarily introduces second order distortions in the form of shear and convergence.
Perturbers very close to the LOS of the main lens can induce higher order perturbations (flexion and beyond) that need to be modelled explicitly to accurately account for their effect on the observables. In our analysis, we model the nearest massive galaxies explicitly while the larger scale structure is accounted by a convergence and an external shear term \citep[see][for a similar approach]{Wong:2017}.

The LOS convergence effectively alters the specific angular diameter distances relevant to the lensing system, $D'$, relative to the homogeneous background metric, $D^{\text{bkg}}$. We take into account the external convergence factor, $\kappa_{\text{ext}}$, perturbing the time-delay distance, $D_{\Delta t}$, \citep{Suyu:2010rc}:
\begin{equation} \label{eqn:kappa_ext}
 D'_{\Delta t} \equiv \left(1 - \kappa_{\text{ext}} \right) D_{\Delta t}^{\text{bkg}},
\end{equation}
where $D'_{\Delta t}$ indicates the time-delay distance along the specific LOS corresponding to the explicit lens model and $D_{\Delta t}^{\text{bkg}}$ corresponds to the homogeneous unperturbed background metric. The factor $(1-\kappa_{\rm ext})$ is estimated by comparing the relative weighted number counts and redshifts of galaxies along the LOS of the strong lens relative to LOSs of similar statistical properties in simulations, following the work of \cite{Rusu2017}.

The LOS projected stellar velocity dispersion of the deflector galaxy, $\sigma^\text{P}$, adds valuable information to the cosmographic inference. $\sigma^\text{P}$ depends on the three-dimensional gravitational potential, the three-dimensional stellar (light) profile and the anisotropy distribution of the stellar orbits, $\beta_{\text{ani}}$. The gravitational potential and the stellar light profile can be expressed in terms of a de-projection of the lens surface mass density and surface brightness models, whose parameters, $\boldsymbol\xi_{\text{lens}}$ and $\boldsymbol\xi_{\text{light}}$, are constrained by the imaging data in combination with the cosmographic relevant angular diameter distances as
\begin{equation} \label{eqn:sigma_P}
  (\sigma^\text{P})^2 = \frac{D_{\text{s}}}{D_{\text{ds}}} c^2 J(\boldsymbol{\xi}_{\text{lens}}, \boldsymbol{\xi}_{\text{light}}, \beta_{\text{ani}}),
\end{equation}
where $J$ captures all the model components computed from angles measured on the sky and the stellar orbital anisotropy distribution. We describe the detailed modelling that goes into Equation~\ref{eqn:sigma_P}~(and thus $J$) in Section~\ref{subsec:kinematics_modelling}.

Gravitational microlensing can also produce changes in the actual
time delays measured between quasar images of order the light-crossing
time-scale of the quasar emission region \citep{Tie:2018a}. We take
into account the possible effects of this so-called microlensing time delay using the
description presented by~\cite{Bonvin:2018a} and fold it into our
analysis using the foward modelling approach of
\cite{Chen:2018_micro_lensing}. The effect is much smaller than other uncertainties for \name, as described in Section~\ref{subsec:time_delay_micro_lensing}.

\subsection{Cosmographic likelihood} \label{sec:ang_diameter_dist}

The likelihood for the cosmological relevant parameters, $\boldsymbol{\pi}$, is fully contained in the angular diameter distances inferred from the data for the particular redshift configuration of the lens, $\{D_{\text{d}}, D_{\text{s}}, D_{\text{ds}}\} \equiv D_{\text{d,s,ds}}$. We can therefore write the probability of a cosmological model, $\boldsymbol{\pi}$, given the data, $\boldsymbol{d_{\text{J1206}}}$, as
\begin{multline} \label{eqn:cosmographic_likelihood}
    P(\boldsymbol{\pi}| \boldsymbol{d_{\text{J1206}}}) \propto
    P(\boldsymbol{d_{\text{J1206}}}| \boldsymbol{\pi})P(\boldsymbol{\pi}) = 
    P(\boldsymbol{d_{\text{J1206}}}| D_{\text{d,s,ds}}(\boldsymbol{\pi}))P(\boldsymbol{\pi}),
\end{multline}
where we made it explicit that the evaluation of the likelihood of a specific cosmology, $\boldsymbol{\pi}$, folds in the likelihood of the data, $\boldsymbol{d_{\text{J1206}}}$, only through the explicit predictions of the angular diameter distances, $D_{\text{d,s,ds}}(\boldsymbol{\pi})$.
In this paper, we present a cosmological model independent likelihood $P(\boldsymbol{d_{\text{J1206}}}| D_{\text{d,s,ds}})$ that can be combined with other cosmological probes as well as posterior distributions for specific cosmological models and priors, $P(\boldsymbol{\pi})$.

The data allows us to constrain two angular diameter distance ratios. First, inverting Equation~\ref{eqn:time_delay} leads to
\begin{equation} \label{eqn:ang_dist_delay}
    (1+z_{\text{d}})\frac{D_{\text{d}}D_{\text{s}}}{D_{\text{ds}}} = \frac{c\Delta t_{\rm AB}}{\Delta\phi_{\rm AB}(\boldsymbol{\xi}_{\text{lens}})}.
\end{equation}
Second, Equation~\ref{eqn:sigma_P} leads to
\begin{equation} \label{eqn:ang_dist_kin}
  \frac{D_{\text{s}}}{D_{\text{ds}}} = \frac{(\sigma^\text{P})^2}{c^2 J(\boldsymbol{\xi}_{\text{lens}}, \boldsymbol{\xi}_{\text{light}}, \beta_{\text{ani}})}.
\end{equation}
Equation~\ref{eqn:ang_dist_delay}, containing the time-delay distance $D_{\Delta t}$ (see Equation \ref{eqn:time_delay_distance}) is the most relevant term in the TDSL analysis and is inversely proportional to the Hubble constant.

The constraints on the angular distances of Equation~\ref{eqn:ang_dist_delay} and~\ref{eqn:ang_dist_kin} share the parameters in the lens model, $\boldsymbol{\xi}_{\text{lens}}$, and as such are correlated and their covariance needs to be taken into account. Following \citet{Birrer:2016zy} we map the full covariance between the different data sets and the angular diameter distances involved.

For illustration purpose, we can also combine Equations~\ref{eqn:ang_dist_kin} and~\ref{eqn:ang_dist_delay} algebraically to solve for $D_{\text{d}}$
\begin{equation} \label{eqn:D_d}
  D_{\text{d}} = \frac{1}{(1 + z_{\text{d}})}\frac{c\Delta t_{AB}}{\Delta\phi_{AB}(\boldsymbol{\xi}_{\text{lens}})}  
  \frac{c^2 J(\boldsymbol{\xi}_{\text{lens}}, \boldsymbol{\xi}_{\text{light}}, \beta_{\text{ani}})}{(\sigma^\text{P})^2}.
\end{equation}
To account for the effect of the LOS convergence in the cosmographic likelihood, the angular diameters have to be transformed according to Equation (\ref{eqn:kappa_ext}) to be compared with a cosmological model. The total cosmographic information will always be contained in a two-dimensional plane of angular diameter distance ratios \citep{Birrer:2016zy}.

\subsection{Combined Bayesian Analysis}\label{sec:cosmographic_analysis}

The cosmographic likelihood (Equation \ref{eqn:cosmographic_likelihood}) is the product of the likelihoods of the independent data sets:
\begin{multline} \label{eqn:prob_D}
    P(\boldsymbol{d_{\text{J1206}}}| D_{\text{d,s,ds}}) = P(\Delta t_{AB}|D_{\text{d,s,ds}}) P(\sigma^\text{P}|D_{\text{d,s,ds}}) \\
    \times P(\boldsymbol{d_{\textit{HST}}}|D_{\text{d,s,ds}}) P(\boldsymbol{d_{\text{env}}}|D_{\text{d,s,ds}}).
\end{multline}
The cosmographic parameters primarily fold in the likelihoods of the time delay and the stellar kinematics. The single plane lensing kernel does not require any knowledge of the absolute scales involved and is independent of the angular diameter distances\footnote{In case of multi-plane lensing, additional relative distance scaling relations to specific redshifts have to be included in the modelling, and thus a minor cosmological dependence arises.}. The LOS analysis is marginally dependent on the specific cosmology through the lensing kernel and the amplitude of the mass power spectrum. This second-order effect has a sub-per-cent level impact on the inferred distance ratios and we ignore this dependence in our analysis.

The different likelihoods in Equation~\ref{eqn:prob_D} include `nuisance' parameters. These are the lens model parameters, $\boldsymbol{\xi}_{\text{lens}}$, and light model parameters $\boldsymbol{\xi}_{\text{light}}$ inferred from $\boldsymbol{d_{\textit{HST}}}$, as well as the external convergence $\kappa_{\text{ext}}$ inferred from $\boldsymbol{d_{\text{env}}}$ and the kinematic anisotropy $\beta_{\text{ani}}$, where a prior must be chosen. Additionally we consider a microlensing time delay effect with parameters $\boldsymbol{\xi}_{\text{micro}}$. The marginalization over the `nuisance' parameters, taking into account the specific dependence of the involved parameters, can be expressed as follows:
\begin{multline} \label{eqn:joint_prob}
    P(\boldsymbol{d_{\text{J1206}}}| D_{\text{d,s,ds}}) = 
    \int P(\boldsymbol{d_{\textit{HST}}}| \boldsymbol{\xi_{\text{lens}}}, \boldsymbol{\xi_{\text{light}}})P(\boldsymbol{\xi_{\text{lens}}}, \boldsymbol{\xi_{\text{light}}})\\
     \times P(\boldsymbol{d_{\text{env}}}|\kappa_{\text{ext}})P(\kappa_{\text{ext}})
    P(\Delta t_{AB}| D_{\text{d,s,ds}}, \boldsymbol{\xi_{\text{lens}}}, \boldsymbol{\xi}_{\text{micro}}, \kappa_{\text{ext}}) \\
    \times P(\sigma^\text{P}| D_{\text{d,s,ds}}, \boldsymbol{\xi_{\text{lens}}}, \boldsymbol{\xi_{\text{light}}}, \kappa_{\text{ext}}, \beta_{\text{ani}}) \mathrm{d}\boldsymbol{\xi}_{\text{lens,light,micro}} \mathrm{d}\kappa_{\text{ext}} \mathrm{d}\beta_{\text{ani}}. 
\end{multline}
Given the hierarchy of the parameters, the sampling of the full likelihood can be partially separated (see Section~\ref{sec:combined_analysis} for details).

\subsection{Lensing degeneracies and the assessment of systematics} \label{subsec:degeneracies}
Degeneracies are inherent in strong lens modelling \citep[e.g.,][]{Saha:2000zr, Saha:2006jb}. In particular, the mass-sheet degeneracy \citep[MSD,][]{Falco:1985no} is relevant to consider in a cosmographic analysis. As shown by \cite{Falco:1985no}, a remapping of a reference mass distribution $\kappa$ by
\begin{equation} \label{eqn:mass_sheet_transform}
    \kappa_{\lambda} (\bm{\theta}) = \lambda \kappa(\bm{\theta}) + (1 - \lambda)
\end{equation}
combined with an isotropic scaling of the source plane coordinates $\bm{\beta} \rightarrow \lambda \bm{\beta}$ will result in the same dimensionless observables (image positions, image shapes and magnification ratios) regardless of the value of $\lambda$ but with changed time-delays. This type of mapping is called mass-sheet-transform (MST).

The additional mass term in the MST (Equation~\ref{eqn:mass_sheet_transform}) can be internal to the lens galaxy (affecting the lens kinematics) or due to LOS structure (not affecting the lens kinematics) \citep[see e.g.,][]{Saha:2000zr, Wucknitz:2002he}. The external part of the MST can equivalently be expressed in terms of an external convergence, $\kappa_{\text{ext}}$, of Equation~\ref{eqn:kappa_ext}. Information about the external part of the MST must come from constraints other than the direct modelling of the lensing galaxy, such as from galaxy counts and redshifts of the LOS galaxy population \citep{Rusu2017, Birrer_2017los} or weak gravitational lensing \citep{Tihhonova:2018}. The internal part of the MST is more subtle to capture as pointed out by \cite{Schneider:2013pm} and discussed by \cite{Xu:2016} for simulated galaxies. A particular assumption of the radial form of the lens model breaks the internal part of the MST and may lead to significant biases in the cosmographic inference. A more general transform, the source position transformation \citep[SPT][]{Schneider:2014fo}, is further discussed by \cite{Unruh:2017} and \cite{Wertz:2018_spt}.

\cite{Suyu:2014aq} did a re-analysis of the lens system RXJ1131-1231 with two different mass models (a power-law mass profile and a composite model explicitly modelling the stellar and dark matter profiles) spanning a reasonable range in flexibility and concluded that adding kinematic information of the deflector galaxy is sufficient to obtain a robust cosmographic inference \citep[see also][for a recent study on the effect of power-laws in determining $H_0$]{Sonnenfeld:2018}. \cite{Birrer:2016zy} addressed the concerns of \cite{Schneider:2014fo} by mapping the internal part of the MSD in the analysis and applied priors on the reconstructed source size $\beta$.

In this work, we explore a wide range of different model choices in both lens and light models to to mitigate the impact of systematics (including choices affected by the MST) and covariances that go beyond those present within specific model choices. We note that the Fermat potential, and thus the inferred time delay distance $D_{\Delta t}$ (Equation \ref{eqn:ang_dist_delay}), is subject to the MST. The kinematic constraints of the deflector enter in the analysis through Equation (\ref{eqn:sigma_P}) and is affected differently by the MST with a angular distance ratio independent of the absolute scales involved (and thus $H_0$). The MST, paired with kinematic measurements, imposes a specific correlation in $D_{\Delta \rm{t}}$ vs $D_{\rm d}$ which limits the impact of the MST, \citep[see e.g.][in this regard]{Birrer:2016zy}.

\section{The lens \name and the data} \label{sec:data}
The gravitational lens \name was discovered by \cite{Oguri:2005}. Based on adaptive optics (AO) imaging obtained with NIRC2 at the W. M. Keck Observatory, \citet{Agnello:2016a} discovered that the lens is a doubly lensed quasar with extended source emission crossing the inner caustic forming a nearly-complete Einstein ring-like configuration that previous analyses had confused for a companion galaxy. They concluded that the combination of a large time-delay and a favorable lensing configuration make this system promising for cosmography, but deeper data with a known point spread function (PSF) and dedicated modelling were needed. \name is the brightest of only three currently known natural coronagraph of the quasar emission region, the others being MG2016+112 \citep[][]{More:2009} and SDSS~J1405+0959 \citep[][]{Rusu:2014}.

The quasar image separation is $3\farcs03$, and its high variability allowed a precise measurement of a relative time-delay of 111.3 $\pm$ 3 days \citep{Eulaers:2013}. The redshift of the lens was initially reported as $z_{\rm d}= 0.748$ and the quasar source redshift as $z_{\rm s} = 1.789$ \citep{Oguri:2005}. \cite{Agnello:2016a} used Keck-DEIMOS \citep{Faber:2003} spectroscopy to correct the redshift of the lens to $z_{\rm d}= 0.745$ and measured the projected integrated stellar velocity dispersion of the lensing galaxy to be $\sigma = 290 \pm 30$ km s$^{-1}$. 

\begin{figure*}
  \centering
  \includegraphics[angle=0, width=120mm]{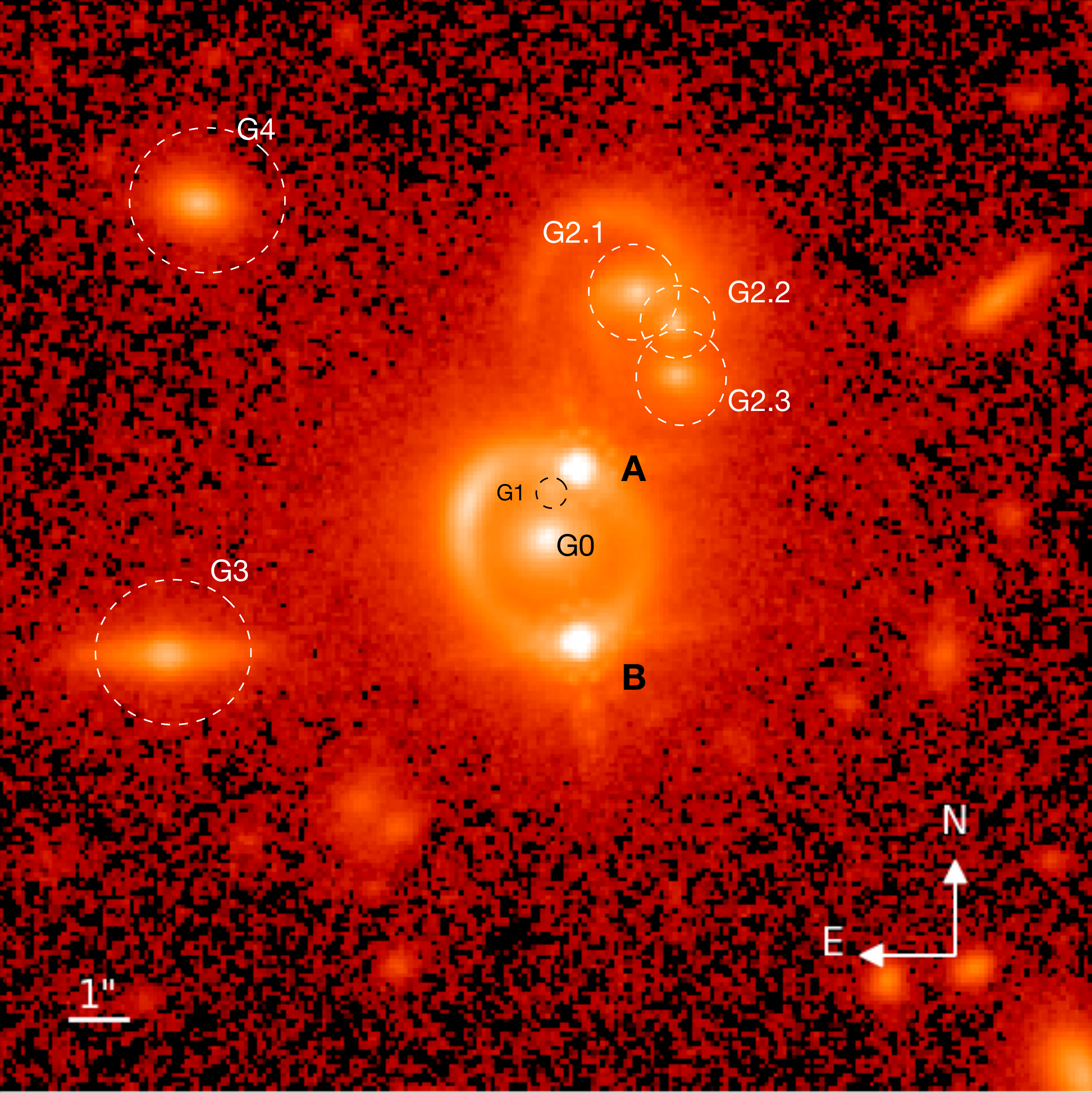}
  \caption{Drizzled \textit{HST}-WFC3 image through filter F160W of the lens \name. The doubly lensed quasar is embedded in a source galaxy, parts of which are quadruply lensed in a fold configuration. We label the different galaxies that we explicitly model. Prominently visible is a galaxy triplet in direction N-W, G2, and two other less massive nearby galaxies in direction E and N-E, G3 and G4.}
\label{fig:hst_data}
\end{figure*}

In this work, we use new high resolution deep \textit{HST} WFC3 images through the F160W filter  (PID:14254, PI: T. Treu) to trace the extended Einstein ring at high signal-to-noise ratio and derive precise astrometry of the quasar positions, with a stable PSF. The total exposure time is 8456 seconds. The single exposures (pixel size of $0\farcs13$) were drizzled and combined on a pixel scale of $0\farcs08$. The \textit{HST} image is presented in Figure~\ref{fig:hst_data}.

The detailed modelling of the extended source structure observed in the deep \textit{HST} image allows us to precisely estimate the relative lensing potential between the positions of the quasar images (see Section~\ref{subsec:image_modelling}).

To obtain information on the environment, $\boldsymbol{d_{\text{env}}}$, and thus determine $\kappa_\mathrm{ext}$, we have conducted the following photometric and spectroscopic observing runs: Gemini/GMOS-N \citep{hook04} imaging in the $g,r,i$-bands, Gemini/NIRI \citep{Hodapp:2003} imaging in $Ks$-band (Proposal ID GN-2017A-Q-39, PI: C. E. Rusu), CFHT/WIRCAM \citep{Puget:2004} imaging in $Ks$-band (Proposal ID 17at99, PI: K. Wong), WIYN/ODI \citep{Jacoby:2002} imaging in $u$-band (Proposal ID 2017A-0108, PI: C. E. Rusu), and Keck/DEIMOS optical spectroscopy (Proposal ID 2017A-0120, P.I. C. D. Fassnacht). The WIYN/ODI run was lost due to telescope technical problems, and the CFHT/WIRCAM data, too shallow compared to the Gemini/NIRI $Ks$-band data, are not used in our analysis. We note that there is also archival Spitzer/IRAC \citep{Fazio:2004} data available (Proposal ID 80025, PI: L. v. Zee), but we do not make use of it, as it only partially overlaps with our field.

The Gemini/GMOS-N run resulted in exposures of $1\times170$s in $g$-band, $6\times120$s in $r$-band, $15\times120$s in $i$-band on 2017 April 5, and $5\times170$s additional exposures in $g$-band on 2017 April 3. These were taken at airmass $\sim1.2$, and the seeing was $\sim0.45\arcsec-0.60\arcsec$ in $g$-band, $\sim0.45\arcsec$ in $ri$-bands. The Gemini/NIRI data consist of $84\times30$ s usable exposures obtained on 2017 February 15, at airmass $\sim1.1$, with seeing $\sim0.35\arcsec$. As the NIRI field of view (FOV) is only $119.9\arcsec\times119.9\arcsec$ in size and as we are interested in the galaxies within $120\arcsec$ of the lensing system (see Section \ref{sec:lostechnique}), we observed four regions (quadrants), non-overlapping except for small patches due to dithering, and with the lensing system at one edge of each of them. All Gemini data were observed in photometric conditions, reduced using recommended techniques with the Gemini \textsc{IRAF\footnote{\textsc{IRAF} \citep{Tody:1986} is distributed by the National Optical Astronomy Observatory, which is operated by the Association of Universities for Research in Astronomy (AURA) under cooperative agreement with the National Science Foundation.}} 1.14 package\footnote{\url{http://www.gemini.edu/node/11823}}, and photometrically calibrated using standard stars. Additional details on the data reduction and analysis are provided in Appendix \ref{sec:datawght}.

The \name field was observed with the Deep Imaging Multi-Object Spectrograph
(DEIMOS) on the Keck II telescope on 2017 March 29 UT.  
The instrument was configured with the 600ZD grating and a central
wavelength of 7150 \AA, yielding a nominal dispersion of 0.65 \AA\
pix$^{-1}$ and a wavelength range of roughly 4500--9800 \AA, depending
on the slit position. The DEIMOS field of view allowed us to survey
galaxies within 14\farcm5 of the lens system, with a higher spatial
concentration close to the lens system. We used four total
slitmasks, targeting 263 objects in total. We obtained three
exposures through each slitmask, with integration times of 1200~s or
1800~s per exposure.  The total exposure time used for the first three
slitmasks was 4800~s, while the fourth slitmask was observed for
4800~s.

The data were reduced with a modified version of the {\tt spec2d} pipeline that was used for the DEEP2 \citep{Newman:2013}, as described by Lemaux et al. (in prep). We visually inspected each of the 263 output spectra and the resulting redshifts were given a quality score, $Q$, where galaxies with $Q=3$ and $Q=4$ are considered to be usable for science \citep{Newman:2013}. In all, 226 galaxies had $Q \geq 3$ and an additional 3 objects were unambiguously identified as stars, so 87 per cent of the slits produced a usable redshift.  
We supplemented the DEIMOS spectra with the redshift of the lensing galaxy from \cite{Agnello:2016a} and 64 additional spectra from the Sloan Digital Sky Survey \citep{Abolfathi:2018} within the field of view of the DEIMOS masks. We present the redshift distribution in Appendix \ref{sec:groups}.

\section{Model choices} \label{sec:model_choices}
In this section, we present our modelling choices in detail. We go through the parameterization of the main deflector galaxy (\ref{subsec:main_deflector}), the source galaxy (\ref{subsec:source_light}), a sub-clump identified in the data (\ref{subsec:subclump}), the description of the nearby perturbing galaxies (\ref{subsec:nearby_perturbers}), the LOS structure (\ref{subsec:los_modelling}) and the modelling of the deflector stellar kinematics (\ref{subsec:kinematics_modelling}). The functional form and parameterization of all the model ingredients follow the definitions of \textsc{lenstronomy}.

The different modelling choices do not all require to fit the data sets equally well. The aim is to provide the inference a sufficient range in exploring solutions of various complexities. Later on in Section \ref{subsec:marginalize_options}, before the unblinding, we apply a statistical measure to weight the different models that go into our final posteriors.

All the choices were blind to the cosmographic likelihood. We displayed the cosmographic likelihood and the inference of the cosmological parameters only after all co-authors involved in the analysis have agreed that the analysis and model choices were satisfactory and the analysis was frozen.

\subsection{Main deflector galaxy, G0}\label{subsec:main_deflector}
The main deflector, G0 in Figure \ref{fig:hst_data}, is a massive elliptical galaxy. We consider two options in this analysis:
\begin{enumerate}
  \item Option \texttt{SPEMD\_SERSIC}: The mass distribution is modelled as a singular power-law elliptical mass distribution (\texttt{SPEMD}) and the light distribution as two superposed elliptical Sersic profiles with shared centroids and free relative position angles and ellipticities.
  \item Option \texttt{COMPOSITE}: We split the luminous and dark component of the lens model into two composite parts \citep[see, e.g.,][]{Dutton:2011}. The luminous component (light and lens) are modelled with two superposed elliptical \texttt{CHAMELEON} models \citep[following][]{Suyu:2014aq} with shared centroids and free relative position angles and ellipticities. The normalization between light and convergence [effectively a mass-to-light (M/L) ratio] is held free. The dark matter mass is modelled as an elliptical \texttt{NFW} profile where the ellipticity is introduced in the lensing potential and the centroid is free with respect to the light centre.
\end{enumerate}

\subsection{Quasar host galaxy} \label{subsec:source_light}
The quasar and its host galaxy are modelled with two different components, a point source representing the quasar emission region and an extended component representing the host galaxy light profile.
The quasar point source is modelled directly in the image plane. We follow \cite{Birrer2015_basis_set} and assign sufficient freedom in the lens model to solve for a unique solution that maps the image plane positions back to a source plane position. The amplitudes of the images are left free to allow for the significant contribution of stellar microlensing and quasar variability.
The extended quasar host is modelled with a range in complexity and freedom assigned to the light distribution. We explore the following 4 options:
\begin{enumerate}
  \item Option \texttt{DOUBLE\_SERSIC}: Two elliptical Sersic profiles with joint centroid at the quasar position.
  \item Option \texttt{DOUBLE\_SERSIC}$+2 n_{\text{max}}$: Additionally to \texttt{DOUBLE\_SERSIC} we add shapelet functions \citep{Refregier:2003eg, Birrer2015_basis_set} with maximal polynomial order $n_{\text{max}}=2$ centered at the quasar and with free scale parameter, $\beta$.
  \item Option \texttt{DOUBLE\_SERSIC}$+5 n_{\text{max}}$: Addition of maximal polynomial order $n_{\text{max}}=5$ on top of \texttt{DOUBLE\_SERSIC}.
  \item Option \texttt{DOUBLE\_SERSIC}$+8 n_{\text{max}}$: Addition of maximal polynomial order $n_{\text{max}}=8$ on top of \texttt{DOUBLE\_SERSIC}.
\end{enumerate}
This approach is similar to the one chosen by \cite{Shajib:2018b} in their automated approach to model a set of quadruply lensed quasar images.
The galaxy host parameterization is explicitly scale invariant. \cite{Birrer:2016zy} demonstrated that enforcing a fixed source reconstruction scale can artificially break the SPT (and as such the MST) that can underestimate the uncertainties in the inferred value of the Hubble constant.

\subsection{Sub-clump near image A: G1} \label{subsec:subclump}
Initial models with only the main deflector (\ref{subsec:main_deflector}) left significant residuals in the models, in particular near image $A$ (at position G1 in Figure \ref{fig:hst_data}). Subtraction of the modelled light components revealed an additional light component in the image plane. This extra component is also visible in the AO assisted image presented by \citet{Agnello:2016a}.
Including a circular Sersic light model and a singular isothermal sphere ($\texttt{SIS}$) model with joint centroids, we find significant improvements of the goodness of fit values and reasonable values for the model components. We can not confirm the redshift of the clump. Throughout this work, we forward model a single-plane lens model, effectively setting the redshift of the additional light component to the redshift of the main deflector galaxy.

In case the perturber had a different redshift, the leading order effect is a change in the lensing efficiency which our parameterization incorporates. We also explore the first order non-linear coupling of a foreground shear field (see Section \ref{subsec:los_modelling}) and conclude that this term has no significant effect on the cosmographic analysis (Section \ref{subsec:marginalize_options}).

\subsection{Nearby perturbing galaxies: G2-G4} \label{subsec:nearby_perturbers}

The galaxy triplet located about $4\farcs4$ from the main deflector center can impact significantly the lens model and has to be modelled explicitly.

The galaxy triplet was covered by one of the slits in the DEIMOS observations described in Section~\ref{sec:data}. The resulting spectrum shows clear [O\,\textsc{ii}] emission as well as weaker H$\beta$ and [O\,\textsc{iii}] emission and several absorption features, giving a secure redshift of $z_{\rm G2} = 0.7472$. This redshift is consistent with that of the main deflector, G0, and places at least one of the triplet galaxies in a galaxy group that contains the primary lensing galaxy (see Section~\ref{sec:los_analysis}). The tidal arm of the northern component is circumstantial evidence of interaction, supporting the physical association hypothesis.

Two other galaxies may or may not have a significant impact on the
cosmographic analysis, G3 and G4 (see Figure \ref{fig:hst_data}) in direction East and North-East.  These galaxies were also targets of the DEIMOS observations and
we obtained a spectrum for each of them.  The data reduction pipeline
uses galaxy templates to assign the most likely redshifts for each
spectrum.  Our visual examination of the spectra resulted in quality scores
of $Q=1$ for G3 and $Q=2$ for G4.  However, their tentative redshifts,
$z_{\rm G3} \sim 0.748$ and $z_{\rm G4} \sim 0.751$
also place them in the galaxy group that contains G0 and G2. We optionally also include them explicitly in our analysis. We model the nearby galaxies as singular isothermal spheres \texttt{SIS} with fixed centroid at the light center.

In order to minimize degeneracies between the large number of parameters, we set priors on the parameters that describe the individual contribution of the 3 (respectively 5 when including G3 and G4) nearby perturbers. We thus introduce a relative M/L ratio prior of the perturbers by measuring the flux of the perturbers and parameterize their Einstein radii with the scaling law 
\begin{equation}
 \theta_E \propto \sigma^2 \propto L_*^{1/2},
\end{equation}
where the first proportionality is coming from the isothermal profile and its associated velocity dispersion, $\sigma$, and the second relation is the Faber-Jackson \citep{Faber:1976} relation, $L_* \propto \sigma^{\gamma}$, relating the luminosity, $L_*$, with the velocity dispersion, $\sigma$, through a power-law with exponent $\gamma=4$.

We assume a typical 0.1 dex scatter in this relation and a free overall M/L scaling parameter with a uniform prior in the units of Einstein radius. The prior in the scatter in this relation is implemented by drawing a realization from this distribution for each sampling of the full parameter space and then fixing the relative profiles through an individual sampling.

The M/L scaling imposed may not be very accurate for describing the galaxies G2-G4. The scaling relation however, needs only be satisfied within the dynamic range of the galaxies (about 1.5 dex in measured flux) and the imposed scatter on the scaling relation effectively produces a wide dynamic range in scaling parameters $\gamma$.

To summarize, we chose two options for the nearby perturbers:
\begin{enumerate}
  \item Option \texttt{TRIPLET}: The nearby galaxy triplet is modelled with three individual \texttt{SIS} profiles based on a fixed M/L ratio among them and an overall free scaling parameter.
  \item Option \texttt{TRIPLET+2}: In addition to option \texttt{TRIPLET}, the two perturbers in the East and North-East are also modelled explicitly with the same M/L prior.
\end{enumerate}

\subsection{LOS structure} \label{subsec:los_modelling}
The collective effect of additional LOS halos and large scale structure introduce linear lensing distortion. The reduced shear terms can be explicitly modelled and lead to measurable imprints in the imaging data of extended sources.
The lens equation implemented in \textsc{lenstronomy}, following \cite{Birrer_2017los}, is

\begin{equation} \label{eqn:shear_modeling}
	\boldsymbol{\beta} = -\boldsymbol{\alpha}_{\text{d}}\left(\Gamma_{\rm d} \boldsymbol{\theta}\right)
	+ \Gamma_{\rm s} \boldsymbol{\theta},
\end{equation}
with $\boldsymbol{\alpha}_{\text{d}}$ as the scaled deflection of the main deflector and  
\begin{equation} \label{eqn:tidal_distortion}
	\Gamma = \left[{\begin{array}{c c }1 -\gamma _{\rm ext,1}&-\gamma _{\rm ext,2}\\-\gamma _{\rm ext,2}&1 +\gamma _{\rm ext,1}\end{array}}\right]
\end{equation}
as the shear distortion, applicable for both, $\Gamma_{\rm d}$ and $\Gamma_{\rm s}$ with different parameters, $\gamma _{\rm ext,1}$ and $\gamma _{\rm ext,2}$. The subscript $_{\rm s}$ denotes the distortion induced along the entire LOS from the observer to the source and the subscript $_{\rm d}$ is the distortion induced from the observer to the main deflector with different parameters for the distortions \citep{Birrer_2017los, Birrer2018_cosmic_shear}. With this definiton, the external shear strength is
\begin{equation}
\gamma_{\rm ext} = \sqrt{\gamma _{\rm ext,1}^2 + \gamma _{\rm ext,2}^2}
\end{equation}
and the external shear angle is
\begin{equation}
\phi_{\rm ext} = \text{tan}^{-1}\left[\gamma _{\rm ext,2}, \gamma _{\rm ext,1} \right] / 2.
\end{equation}

In this work, we consider two different descriptions of the LOS distortions:

\begin{enumerate}
  \item Option \texttt{SIMPLE\_SHEAR}: We only consider the shear distortion to the source plane, $\Gamma_{\rm s}$, and set $\Gamma_{\rm d}$ to unity. This is the standard external shear implementation in the literature.
  \item Option \texttt{FOREGROUND\_SHEAR}: In addition to \texttt{SIMPLE\_SHEAR}, we include non-linear shear terms affecting the main deflector plane, $\Gamma_{\rm d}$. This option is effectively a multi-plane lens model. The rays in the background ray-tracing get first deflected by the foreground shear field before they enter the main deflection plane.
\end{enumerate}

The effect of the LOS convergence is described by a single number, $\kappa_{\text{ext}}$, \citep{Suyu:2010rc} acting on the time-delay distance, $D_{\Delta t}$ (Equation \ref{eqn:kappa_ext}).

\subsection{Stellar kinematics of the deflector galaxy}\label{subsec:kinematics_modelling}
To model the stellar velocity dispersion, we consider spherical models with the only distinction in radial, $\sigma_{\rm r}^2$, and tangential, $\sigma_{\rm t}^2$, dispersion. The spherical Jeans equation of the 3-dimensional luminosity distribution $\rho_*$ in a gravitational potential $\Phi$ is then
\begin{equation} \label{eqn:jeans}
  \frac{\partial ( \rho_*\sigma_{\rm r}^2)}{\partial r} + \frac{2 \beta_{\text{ani}}(r) \rho_*\sigma_{\rm r}^2}{r} = - \rho_*\frac{\partial \Phi}{\partial r},
\end{equation}
with the stellar anisotropy parameterized as
\begin{equation}
  \beta_{\text{ani}}(r) \equiv 1 -  \frac{\sigma_{\rm t}^2}{\sigma_{\rm r}^2}.
\end{equation}
The same approach was chosen by, e.g., \cite{Suyu:2010rc}. The modelled luminosity-weighted projected velocity dispersion $\sigma_{\rm s}$ is given by
\begin{equation} \label{eqn:I_R_sigma2}
  I(R) \sigma_{\rm s}^2 = 2\int_R^{\infty} \left(1-\beta_{\text{ani}}(r)\frac{R^2}{r^2}\right) \frac{\rho_* \sigma_{\rm r}^2 r \mathrm{d}r}{\sqrt{r^2-R^2}}
\end{equation}
where $R$ is the projected radius and $I(R)$ is the projected light distribution. In this work, $I(R)$ is a function of the parameters $\boldsymbol{\xi_{\rm light}}$.

Massive elliptical galaxies are assumed to have isotropic stellar motions in the center of the galaxy ($\beta_{\text{ani}}=0$) and radial motions in the outskirts ($\beta_{\text{ani}}=1$). A simplified description of the transition can be made with an anisotropy radius parameterization, $r_{\text{ani}}$, defining $\beta_{\text{ani}}$ as a function of radius $r$ \citep{Osipkov:1979, Merritt:1985}
\begin{equation} \label{eqn:r_ani}
  \beta_{\text{ani}}(r) = \frac{r^2}{r_{\text{ani}}^2+r^2}.
\end{equation}
Equation~\ref{eqn:I_R_sigma2} can be restated as \citep[see][Appendix]{MamonLokas2005}
\begin{equation} \label{eqn:I_R_sigma2_mamon}
  I(R) \sigma_{\rm s}^2 = 2G\int_R^{\infty} K\left(\frac{r}{R}, \frac{r_{\text{ani}}}{R}\right) \rho_*(r) M(r) \frac{\mathrm{d}r}{r},
\end{equation}
where $M(r)$ the 3-dimensional enclosed mass distribution and $K$ is a function specific to the anisotropy model provided by \cite{MamonLokas2005} in equation A16. Not all lens and light profiles we use in the modelling (in 2 dimensions) have analytical de-projections in 3 dimensions available. To perform the de-projections, we use a multi-Gaussian decomposition \citep{Cappellari2002} of the modelled projected light and lens model and perform the de-projection on the Gaussian functions.

\section{Forward modelling the data sets} \label{sec:forward_modelling}
In this section, we describe the forward modelling of the data sets based on the choices outlined in Section~\ref{sec:model_choices}. We provide details of the imaging modelling (Section \ref{subsec:image_modelling}), the projected stellar kinematics (Section \ref{subsec:spectra}) and the time delay (Section \ref{subsec:time_delay_micro_lensing}), and provide the priors and likelihood associated with the different data sets. Every decision made in this section was taken before the unblinding of the cosmographic results. The analysis of the LOS contribution will be presented separately in Section~\ref{sec:los_analysis}.

\subsection{Imaging modelling} \label{subsec:image_modelling}

The imaging data are modelled with the \texttt{ImSim} module of \texttt{lenstronomy} \citep{Birrer_lenstronomy, Birrer2015_basis_set}. For a proposed set of lens model parameters, $\boldsymbol{\xi_{\text{lens}}}$, and light model parameters, $\boldsymbol{\xi_{\text{light}}}$, we render the linear response functions on the data (i.e. amplitudes of light profiles, point sources and shapelet coefficients) on the image plane and optimize the linear parameters with a linear minimization based on the imaging likelihood \citep[see e.g.][]{Birrer2015_basis_set}. To accurately compute the response in the observed image plane for each component, we perform the ray-tracing through a higher resolution grid relative to the pixel sizes, by a factor of $3\times3$ per pixel.

The PSF convolution is performed on the higher resolution ray-tracing grid to accurately account for sub-pixel scale features in the brightness distribution and its response through the convolution kernel. This requires a higher resolution sub-pixel sampled PSF. When the size of the kernel and the image is inflated by a factor of $3\times3$, the convolution dominates the computational cost of the image modelling. To mitigate the computational cost, we only apply the sub-sampled kernel on inner most $9\times9$ pixels (in the \textit{HST} units) and the convolution by the tails and larger extent of the PSF is performed on the regular image pixel grid. This saves significant computational cost without loss of accuracy. We perform an iterative PSF estimate. For details we refer to Appendix~\ref{app:psf_iteration}.

The imaging likelihood, $P(\boldsymbol{d_{\textit{HST}}}| \boldsymbol{\xi_{\text{lens}}}, \boldsymbol{\xi_{\text{light}}})$, is computed based on a Gaussian background noise level estimated from an empty patch of the \textit{HST} data and a Poissonian component based on the excess flux paired with the CCD gain of the instrument. Possible error covariance due to the drizzling procedure in co-adding multiple single exposures are neglected, which leads to a slight under-estimation of our errors in regions of high flux gradients in the data.

To test the sensitivity of our analysis to the specific region where we evaluate the imaging likelihood, we chose two different circular regions: 3\farcs0 and 3\farcs2 radius centered at the main deflector. We also exclude pixels at a region where the impact of the nearby galaxy triplet is expected. We refer in this work to the assignment of pixels to be included/excluded in the likelihood as masking and pixels included in the likelihood are in the mask.

\begin{figure*}
  \centering
  \includegraphics[angle=0, width=190mm]{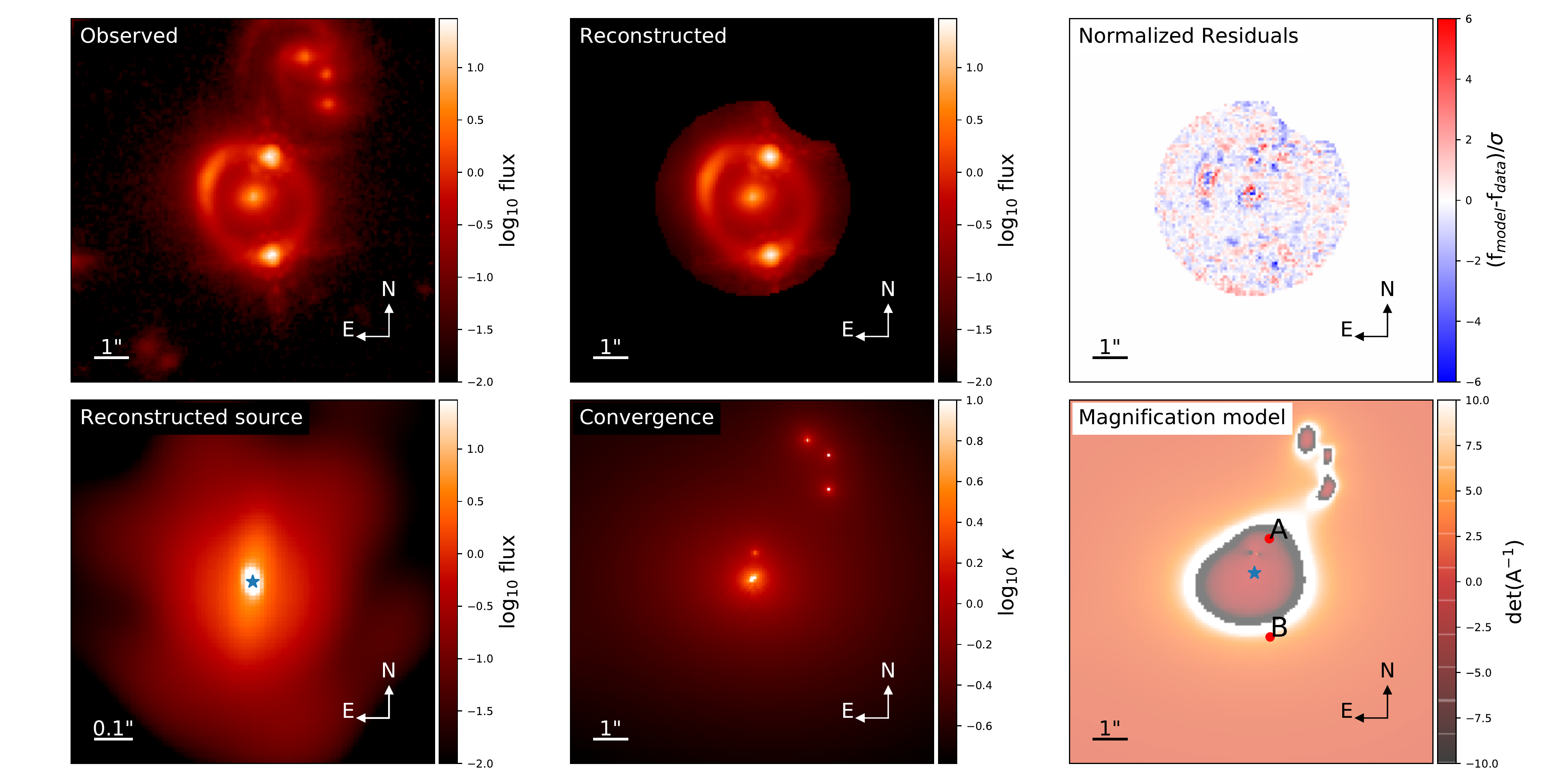}
  \caption{Example of a lens model drawn from the posterior sample and its ability to reconstruct the \textit{HST} image. \textbf{Upper left:} Reduced \textit{HST} image data. \textbf{Upper middle:} Reconstructed image within the chosen mask region. \textbf{Upper right:} Normalized residuals of the model compared to the data based on the noise map. \textbf{Lower left:} Reconstructed source from the model with a double Sersic profile and $n_{\text{max}}=8$ shapelet coefficients. \textbf{Lower middle:} Convergence of the lens model. \textbf{Lower right:} Magnification of the lens model and indicated image positions A and B as well as the intrinsic source position of the quasar (marked as a star).}
\label{fig:lens_model}
\end{figure*}

\begin{figure*}
  \centering
  \includegraphics[angle=0, width=190mm]{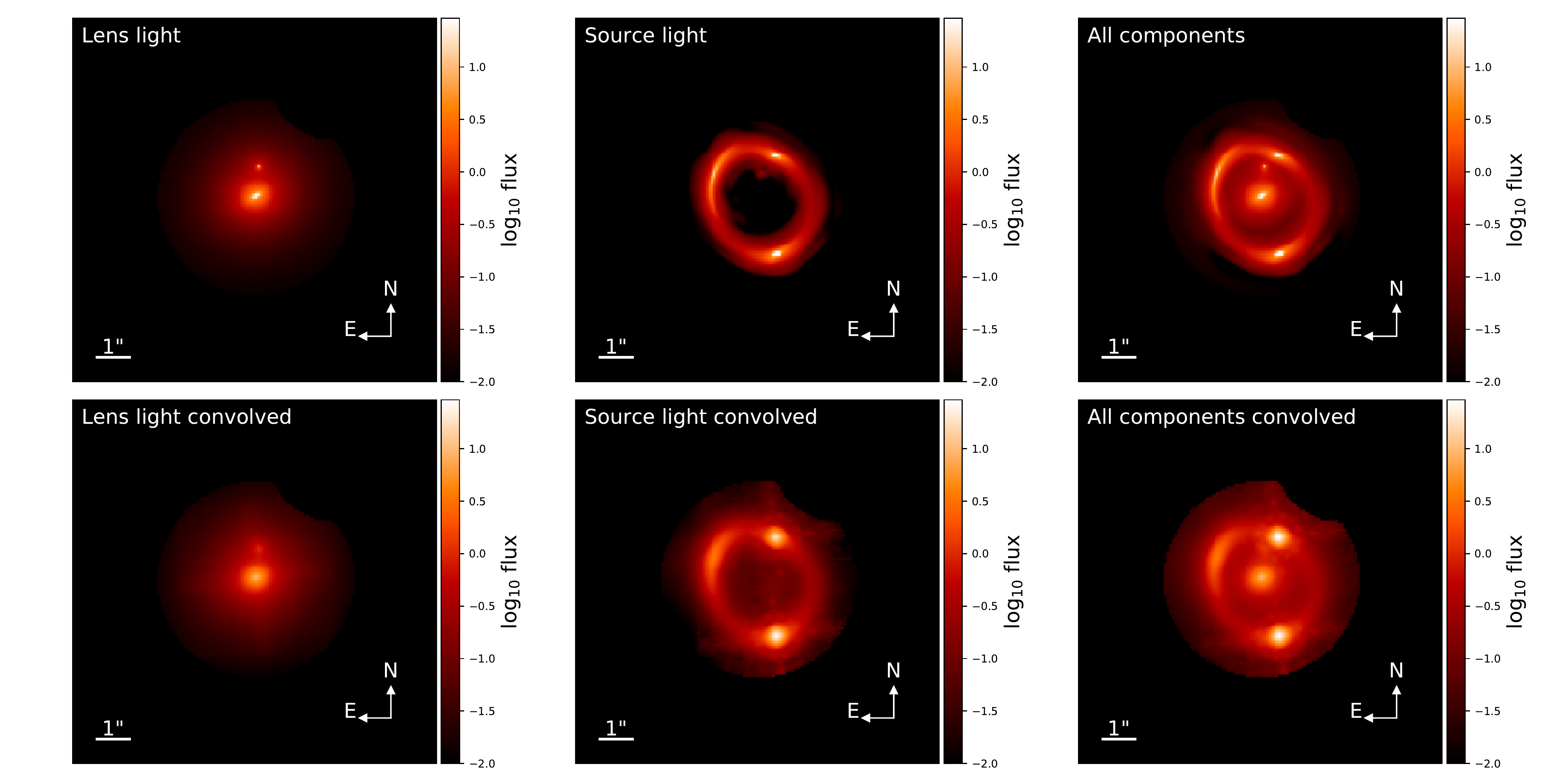}
  \caption{The same model as presented in Figure~\ref{fig:lens_model} decomposed in its individual components. \textbf{Upper panels: } Model components without the instrumental convolution applied. \textbf{Lower panels:} Model components with the PSF convolution applied. \textbf{Left:} Lens light component as modelled by a double Sersic profile for G0 and a spherical Sersic profile for G1. \textbf{Middle:} Lensed extended source light, modelled with a double Sersic profile and $n_{\text{max}}=8$ shapelet coefficients. \textbf{Right:} Lensed source and lens light components combined. The lower panel also includes the components of the point sources.}
\label{fig:lens_model_decomposed}
\end{figure*}

Figure~\ref{fig:lens_model} presents for illustration a typical result of the \textit{HST} image modelling drawn from the posterior distribution. Figure~\ref{fig:lens_model_decomposed} presents the same model decomposed into its components.

\subsection{Spectra of the deflector galaxy} \label{subsec:spectra}
To compare the LOS stellar velocity dispersion of a model with measurements, the details of the observational conditions have to be taken into account. In particular, we model the slit aperture $\mathcal{A}$ and the PSF convolution of the seeing, $* \mathcal{P}$. The luminosity-weighted LOS velocity dispersion within an aperture, $\mathcal{A}$, is then \citep[see also Equation (20) in][]{Suyu:2010rc}
\begin{equation} \label{eqn:sigma_convolved}
  (\sigma^\text{P})^2 = \frac{\int_{\mathcal{A}}\left[I(R) \sigma_{\rm s}^2 * \mathcal{P} \right]\mathrm{d}A }{\int_{\mathcal{A}}\left[I(R) * \mathcal{P} \right]\mathrm{d}A}
\end{equation}
where $I(R) \sigma_{\rm s}^2$ is taken from Equation~\ref{eqn:I_R_sigma2}. We model the integrated velocity dispersion given by \citet{Agnello:2016a} with a Gaussian PSF of full width at half maximum (FWHM) $1\farcs0$ and a slit aperture of $3\farcs8\times1\farcs0$ centered on the deflector galaxy where only the central $1\farcs0\times1\farcs0$ area is selected to measure the spectral dispersion. The convolution and integrals of the expression above are performed with spectral rendering \citep{Birrer:2016zy} implemented in the \texttt{Galkin} module of the \textsc{lenstronomy} package.

To compute the likelihood $P(\sigma^\text{P}| D_{\text{d,s,ds}}, \boldsymbol{\xi_{\text{lens}}}, \boldsymbol{\xi_{\text{light}}}, \kappa_{\text{ext}}, \beta_{\text{ani}})$, we assume Gaussian errors on the uncertainties presented by \cite{Agnello:2016a}.

\subsection{Time-delay measurement and microlensing effects} \label{subsec:time_delay_micro_lensing}

E13 presented light curves of the two lensed images from seven years of monitoring. Averaging over four different curve-shifting techniques, they obtained a time delay of 111.3 $\pm$ 3 days. Here, we re-analyse the light curves from E13 using the \textsc{PyCS} software \citep{Tewes:2013xr, Bonvin:2016} and the new analysis framework introduced by \cite{Bonvin:2018a}. The main improvement with respect to E13 resides in the inclusion of a number of consistency tests in the final time-delay estimate, namely marginalizing over various microlensing models and curve-shifting techniques parameters.

\begin{figure} 
  \centering
  \includegraphics[angle=0, width=80mm]{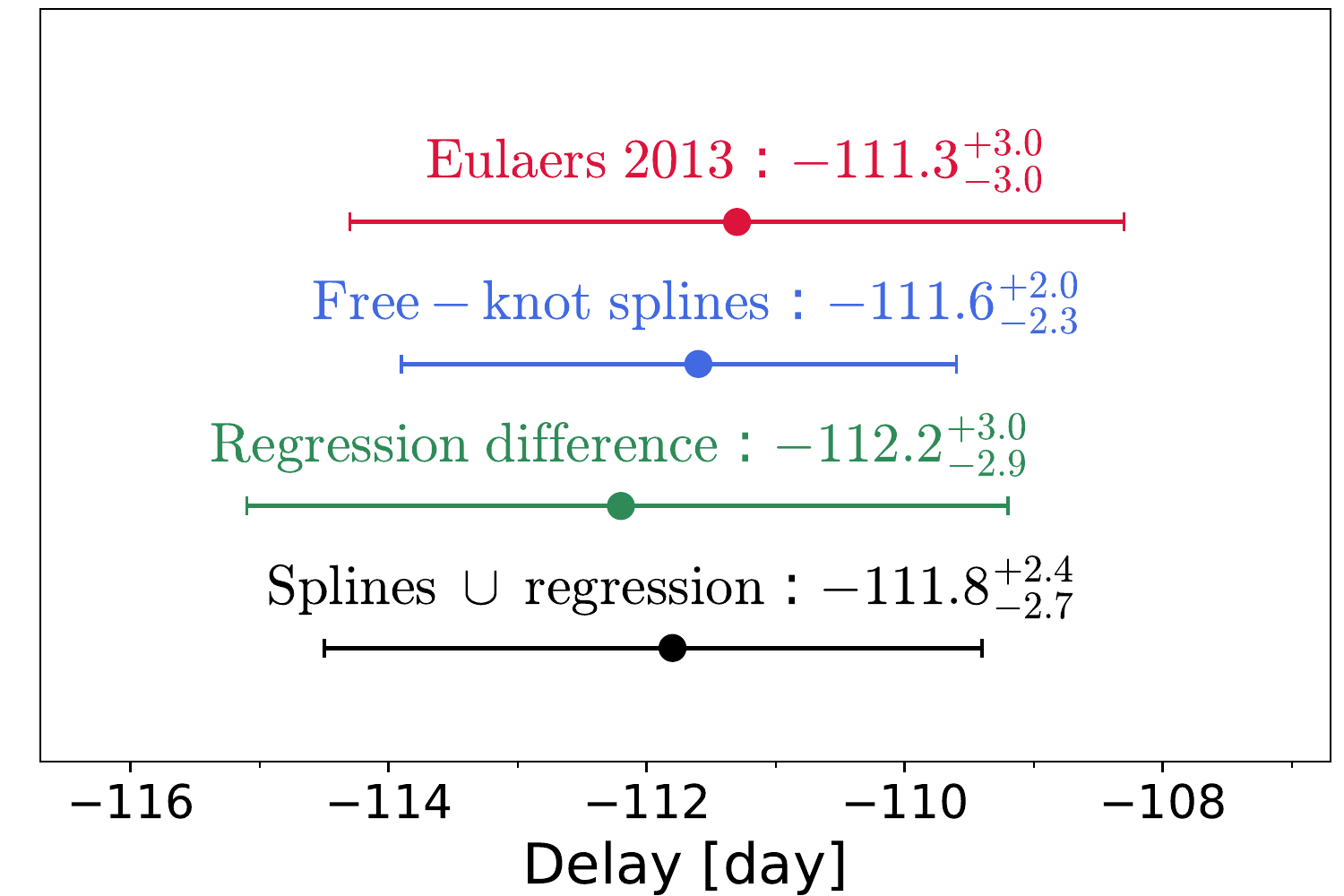}
  \caption{Measured time delay between images A and B of \name from the data set of \citet{Eulaers:2013}. Indicated are the mean and 1-$\sigma$ errors of the original analysis by \citet{Eulaers:2013} and two different updated re-analysis methods. We chose the equal weight marginalized measurement for this work.}
  \label{fig:time_delay}
\end{figure}

Two different curve-shifting techniques were combined. The free-knot splines technique explicitely models the quasar intrinsic luminosity variations from the two light curves, as well as the per-image extrinsic luminosity variations, attributed to microlensing. The regression difference technique uses Gaussian processes to model the variability of each of the two light curve, that are then shifted in time in order to minimize their variability difference. The most stringent difference between the two techniques resides in the explicit modeling of microlensing in the free-knot splines technique, in contrast with the regression difference technique that is by construction insensitive to the presence of smooth microlensing in the data. The resulting time delays obtained after marginalization over the technique parameters are presented in Fig. \ref{fig:time_delay}, along with the original estimate from E13 and a combined estimate marginalizing over the result of the free-knot spline and regression difference technique that we use in this work. The details of the marginalization process can be found in \cite{Bonvin:2018a}. Our final time-delay estimate reads $\Delta t_{\rm AB} = 111.8^{+2.4}_{-2.7}$ days, in good agreement with the original work of E13 and improved precision (see Fig. \ref{fig:time_delay}). In this work, we take the full non-Gaussian distribution of the uncertainty into account.

The measured time delay between two quasar images may deviate from the cosmographic delay (Equation \ref{eqn:time_delay}) due to microlensing on the quasar accretion disc \citep{Tie:2018a}. The microlensing time-delay effect on the two images, $t_{{\rm A}, m_k}$ and $t_{{\rm B}, m_k}$, depends on the quasar accretion disc, the local magnification tensor of the lens model and local stellar densities and the mass function thereof.

In this work, we follow \cite{Bonvin:2018a} and \cite{Chen:2018_micro_lensing} to estimate and marginalize over the expected microlensing time delay. The lensing parameters at the images, A and B, are presented in Table \ref{table:images}. The estimates are an average over all best fit parameters of the lens model choices. The stellar convergence, $\kappa_*$, is estimated from the composite models that impose a M/L scaling.

The accretion disc size and shape is estimated following \cite{Tie:2018a} as a standard, non-relativistic, thin disc model emitting as a blackbody \citep{Shakura:1973}. The accretion disk scale, $R_0$, is a function of black hole mass, $M_{\rm bh}$ and the accretion luminosity, $L$, with respect to the Eddington luminosity, $L_{\text{E}}$.

For our study, the black hole mass estimate comes from Sloan Digital Sky Survey (SDSS) spectra \citep{Shen:2011} based on Mg\,\textsc{ii} and results in a black hole mass of log$(M_{\text{bh, Mg\,\textsc{ii}}}/ M_{\odot}) = 8.93$. Based on the limitations of the Mg\,\textsc{ii} technique, we assign a $\pm0.25$ dex uncertainty to this measurement \citep{Woo:2018}. As this measurement was based on the magnified image, we apply a magnification correction
\begin{equation}
  \log M_{\rm bh} = \log M_{\text{bh, Mg\,\textsc{ii}}} - b \log(\mu),
\end{equation}
where we chose $\mu = 4$ as the fiducial magnification within the SDSS fibre and $b=0.5$ corresponds to the black hole calibration factor for Mg\,\textsc{ii} of \cite{Vestergaard_Peterson:2006}. This leads to $\log (M_{\rm bh} / M_{\odot}) = 8.62 \pm 0.25$. The Eddington ratio based on the same work \citep{Shen:2011} is estimated to be $\log (L'_{\text{bol}}/L'_{\text{Edd}}) = -1.18$. Applying the magnification corrections leads to
\begin{equation}
	\log \frac{L_{\text{bol}}}{L_{\text{Edd}}} = \log \frac{L'_{\text{bol}}}{L'_{\text{Edd}}} + (b - 1) \log \mu.
\end{equation}
This results in a intrinsic Eddington ratio of $\log (L_{\text{bol}}/L_{\text{Edd}}) = -1.48$. The parameters that went into the model are presented in Table \ref{table:micro_lensling_specs}. A smaller $M_{\rm bh}$ or a smaller Eddington ratio will lead to smaller predicted disk sizes and the microlensing component will be smaller than assumed in this work. Small changes in the lensing parameter within the scatter of Table \ref{table:images} creates no significant changes in the predicted microlensing time-delay.

\begin{figure}
  \centering
  \includegraphics[angle=0, width=80mm]{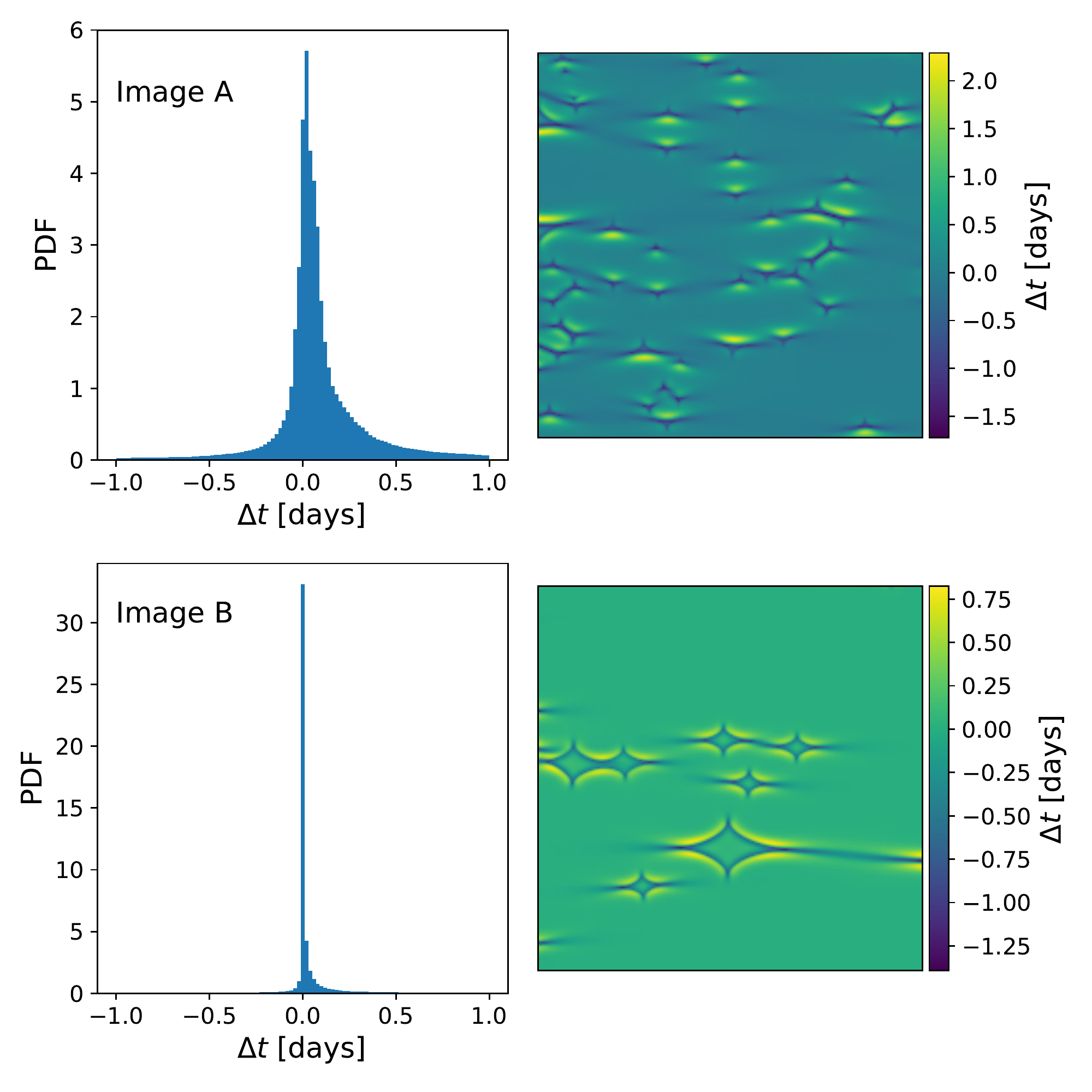}
  \caption{Microlensing time-delay maps and statistical distribution for the two quasar images A and B. The maps (right panel) are based on the magnification tensor of the lens model (Table~\ref{table:images}), an estimate of the  stellar initial mass function (IMF) and the normalization estimated from the stellar contribution to the lensing mass, and accretion disc properties summarized in Table \ref{table:micro_lensling_specs}. The distributions of the expected microlensing time-delay of the two images are shown on the left panels. The microlensing time delay is small compared to the measurement uncertainties of the relative time delay between the two images.}
\label{fig:micro_lensing_delays}
\end{figure}

With this description, we create microlensing time-delay maps around images A and B (Figure \ref{fig:micro_lensing_delays}). We take the microlensing time delay and its uncertainty into account by simply sampling the delay distributions from the maps (Figure \ref{fig:micro_lensing_delays}) and subtract the expected delay from the measured time-delay
\begin{equation}
 \Delta t_{{\rm AB}, \text{corrected}} = \Delta t_{{\rm AB}, \text{measured}} - \left(t_{{\rm A}, m_k} - t_{{\rm B}, m_k}\right).
\end{equation}
The predicted microlensing time-delay effect is significantly smaller than a day on both images for \name due to the small accretion disk size estimated and is thus sub-dominant with respect to the measurement uncertainty on the cosmological delay.

\begin{table}
\caption{Lensing quantities at the quasar image positions as used to predict the microlensing time delay. All values and uncertainties in this table are based on the combined distributions of all the model options considered in this work.}
\begin{center}
\begin{threeparttable}
\begin{tabular}{l r r}
    
    \hline
    Image A \\
    $\kappa$ & = 0.65  & $\pm$ 0.03 \\
    $\gamma$ & = 0.66  & $\pm$ 0.05 \\
    $\mu$ & = 3.24 & $\pm$ 0.53 \\
    $\kappa_*$ & = 0.095 & $\pm$ 0.023 \\

    \hline
    Image B \\
    $\kappa$ & = 0.43  & $\pm$ 0.04 \\
    $\gamma$ & = 0.35  & $\pm$ 0.03 \\
    $\mu$ & = 5.22 & $\pm$ 0.73 \\
    $\kappa_*$ & = 0.020 & $\pm$ 0.005 \\
      \hline
\end{tabular}
\begin{tablenotes}
\end{tablenotes}
\end{threeparttable}
\end{center}
\label{table:images}
\end{table}

\begin{table}
\caption{Quasar accretion model parameters used to compute the microlensing time delays.}
\begin{center}
\begin{threeparttable}
\begin{tabular}{l r r}

    \hline
    $\left<M_*\right>$ $[M_{\odot}]$  & = 0.3  \\
    $ \log [M_{\rm bh} / M_{\odot}]$ & = 8.62 \\
    $ \log [L_{\text{bol}}/L_{\text{Edd}}]$ & = $-1.48$ \\
    $ \eta$ & = 0.1 \\
    $R_0$ [cm] & = $5.52\times 10^{14}$ \\
    $\lambda$ [micron (obs)] & = 0.664  \\
      \hline
\end{tabular}
\begin{tablenotes}
\end{tablenotes}
\end{threeparttable}
\end{center}
\label{table:micro_lensling_specs}
\end{table}

\section{LOS analysis and the external convergence} \label{sec:los_analysis}
In this section, we describe the inference of the LOS convergence posterior distribution given the wide field photometric and spectroscopic data, $p(\kappa_{\mathrm{ext}}|\mathbf{\boldsymbol{d_{\text{env}}}})$. Our analysis follows the technique presented by \citet{Rusu2017}. We briefly summarize it here, point out our modifications and formulate the inference problem from the likelihood and prior $P(\boldsymbol{d_{\text{env}}}|\kappa_{\text{ext}})P(\kappa_{\text{ext}})$ as an application of Approximate Bayesian Computing (ABC).

We present the resulting posterior $P(\kappa_{\text{ext}}| \boldsymbol{d_{\text{env}}})$ and present robustness tests thereof. Furthermore, we discuss the integration and separability assumptions of the specific modelling of nearby perturbers and the statistical LOS analysis.

\subsection{LOS: Description of the technique} \label{sec:lostechnique}

The likelihood $P(\boldsymbol{d_{\text{env}}}|\kappa_{\text{ext}})$ is not directly accessible from the environmental data, $\boldsymbol{d_{\text{env}}}$, describing projected positions, luminosity and redshift estimates of several hundreds of galaxies in the field of \name. Instead of finding an expression of this likelihood, we circumvent the problem by putting the weight on simulations through the ABC framework. We chose a summary statistic that compresses the data in terms of weighted number counts and compare this information with mock data generated by numerical simulations where the underlining convergence, $\kappa_{\text{ext}}$, is accessible.
To construct the mock data, we use the Millennium Simulation \citep[][hereafter MS]{Springel:2005mu}, which consists of simulated dark matter halos in a cosmologically representative volume. The MS has been augmented with catalogs of galaxies with synthetic photometry, painted on top of the dark matter halos \citep{DeLucia:2007}, and with convergence and shear maps corresponding to a grid of source redshift planes \citep{Hilbert:2009}. 

To calibrate the mock data rendered from the MS, we use a control field which contains data of similar quality as $\boldsymbol{d_{\text{env}}}$ but of sufficiently large size to overcome cosmic variance. We use the Canada-France-Hawaii Telescope Legacy Survey \citep[CFHTLS;][]{Gwyn:2012}, in the form of object catalogs provided by CFHTLenS \citep{Heymans:2012}, as the control field.

We produce mock data (and in particular the summary statistic) at each location (LOS) of the MS over a grid of apertures, but this time relative to the whole MS. This procedure guarantees a prior, $P(\kappa_{\text{ext}})$, that reflects the global distribution of the MS.

Our summary statistic is a weighted galaxy number density (with weights specified by $q$) within a choice of aperture and limiting observed magnitude, $i$, stated as $\mathcal{A}_i$, Following \citet{Rusu2017}, the relative density we use as our summary statistic is

\begin{equation}
\overline{\zeta_q}^{\mathcal{A}_i} \equiv \mathrm{median}\frac{N_\mathrm{gal, lens}^{\mathcal{A}_i}\cdot\mathrm{median}(q_\mathrm{gal, lens}^{\mathcal{A}_i})} {N_\mathrm{gal, CFHTLenS}^{\mathcal{A}_i}\cdot\mathrm{median}(q_\mathrm{gal, CFHTLenS}^{\mathcal{A}_i})},
\end{equation}
where for each CFHTLenS subfield $\mathcal{A}_i$ of aperture and depth equal to that around the lensing system, the median of the weighted galaxy property $q^{\mathcal{A}_i}_\mathrm{gal, CFHTLenS}$ inside the aperture is multiplied by the number of galaxies inside the aperture. The same expression derived on the lens system is stated as $q^{\mathcal{A}_i}_\mathrm{gal, lens}$. Given that our aperture is defined by its radius, and we quantify the environment in terms of the statistics of galaxies with redshift smaller than the one of the quasar source $z_{\rm s}$, we adopt empirical weights defined in terms of this minimal set of quantities: $q=1,1/r,\left(z_{\rm s} \cdot z - z^2\right)/r$. Here, 1 refers to the case when no weight is used, $r$ is the projected distance of a given galaxy to the lens, and $z$ is the galaxy redshift (for most galaxies estimated with photo-z).

\citet{Rusu2017} found that the derived external convergence is almost insensitive to the choice of limiting aperture, limiting magnitude, and weight. We therefore limit our analysis to a subsample of the choices tested in that work, namely the $45\arcsec$- and $120\arcsec$-radius apertures, and $i\leq23$, $i\leq24$, where the deeper and wider limits come from the analysis of \citet{Collett:2013bh} and the narrower limit comes from \cite{Fassnacht:2011zs}.

\begin{figure*}
\begin{center}
\includegraphics[angle=0,scale=.95]{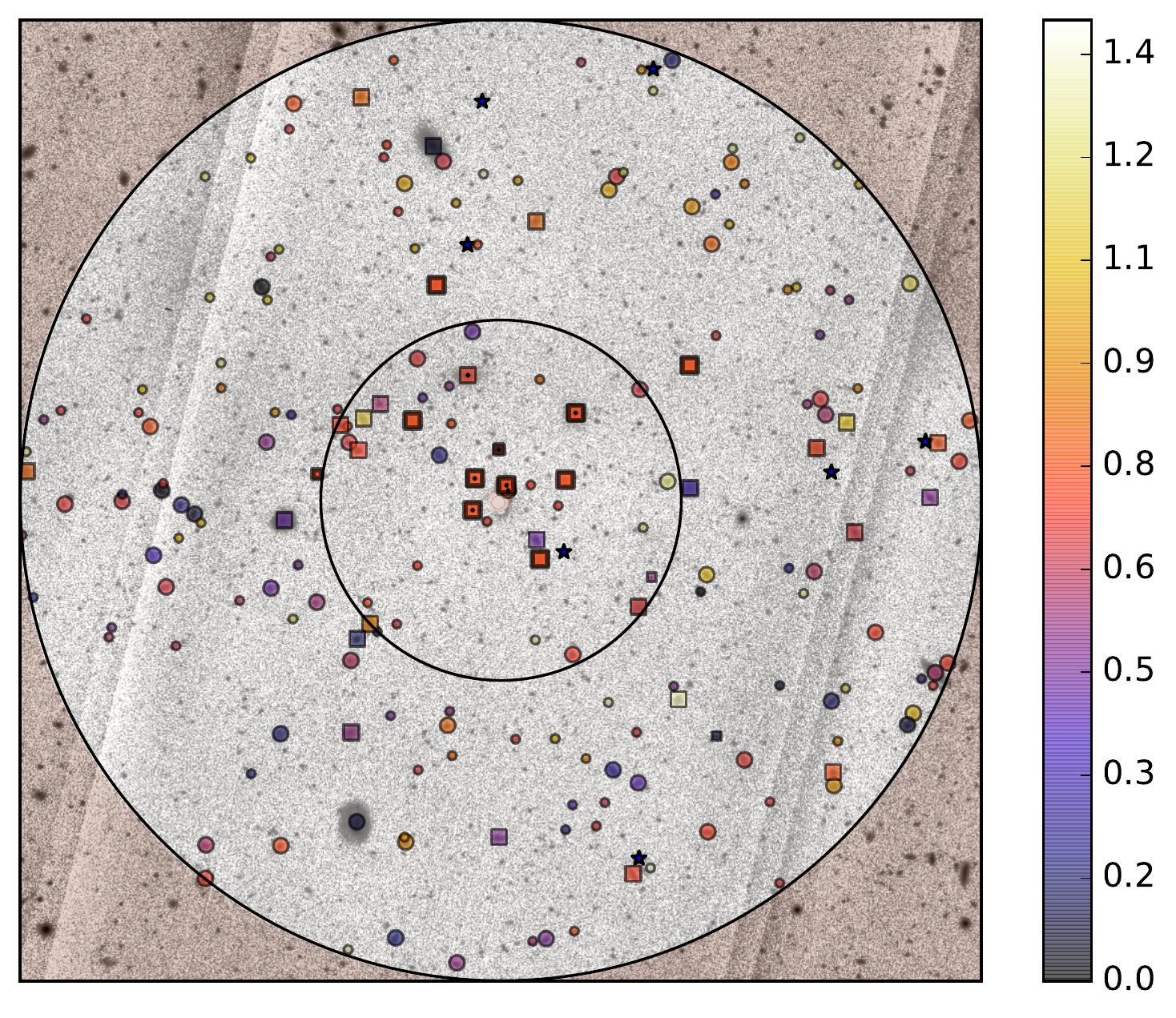}
\caption{$i-$band image of \name, showing the environment of the lensing system. The lens is masked with a 5\arcsec radius. The two circles mark the $45\arcsec$ and $120\arcsec$ apertures, respectively. North is up and East is to the left. Galaxies with spectroscopic redshifts are marked with squares (combined sample of DEIMOS + SDSS redshifts), and those without are marked with circles. Stars are marked with empty star symbols. Galaxies identified as part of the group containing the lensing galaxy (see Appendix \ref{sec:groups}) are enclosed inside a black contour. Galaxies with the largest flexion shifts as computed using the methodology of \citet{Sluse:2017} (up to $\sim 1$ order of magnitude smaller than that of the nearby triplet) are marked with a black dot at the center. Large symbols mark objects with $i<23$ mag, and small symbols mark objects with $23 < i \leq 24$ mag. The colors corresponds to the photometric, or when available, the spectroscopic redshift values. Only objects with $z<z_l$ are marked.}
\label{fig:fov}
\end{center}
\end{figure*}

In Appendix \ref{sec:datawght}, we give further details on the estimation of the weighted count ratios from the data, taking systematics into account, and in Appendix \ref{sec:groups} we explore the existence of galaxy groups around the lensing system. We show the distribution of galaxies around the lensing system in Figure \ref{fig:fov}, and the results of our estimate of the weighted galaxy number count ratios in Table \ref{tab:wghtcounts}, including our uncertainties propagated from the observations. Our results show that, depending on the chosen weights, the LOS to the lens is mostly overdense (i.e., $\overline{\zeta_\mathrm{q}} >1$) inside the $45\arcsec$ aperture, and mostly underdense inside the $120\arcsec$ aperture. As the weights incorporating the galaxy redshifts invariably lead to larger densities, and the members of the group hosting the lensing galaxy, which fall inside the FOV (see Appendix \ref{sec:groups} and Figure \ref{fig:fov}), are mostly confined within the $45\arcsec$ aperture, we conclude that the lens resides within an underdense large-scale environment, with an overdensity at the center, due to a local group. 

\begin{table}
 \centering
  \begin{minipage}{85mm}
  \caption{Weighted galaxy count ratios $\overline{\zeta_q}^{\mathcal{A}_i}$}
  \begin{tabular}{@{}lcccccccc@{}}
  \hline 
$q$ & $45\arcsec, i < 24$ & $120\arcsec, i < 24$ & $45\arcsec, i < 23$ & $120\arcsec, i < 23$  \\
 \hline
$1$               	       & $1.11^{+0.15}_{-0.08}$ & $0.86^{+0.05}_{-0.04}$ & $1.22^{+0.07}_{-0.04}$ & $0.81^{+0.02}_{-0.01}$  \\ 
$z$                          & $1.22^{+0.14}_{-0.10}$ & $0.92^{+0.04}_{-0.06}$ & $1.39^{+0.05}_{-0.06}$ & $0.88^{+0.03}_{-0.02}$  \\ 
$1/r$                        & $0.92^{+0.13}_{-0.07}$ & $0.87^{+0.05}_{-0.04}$ & $0.95^{+0.03}_{-0.04}$ & $0.80^{+0.01}_{-0.02}$  \\ 
$z/r$                        & $1.01^{+0.13}_{-0.07}$ & $0.94^{+0.06}_{-0.06}$ & $1.08^{+0.04}_{-0.03}$ & $0.85^{+0.03}_{-0.01}$  \\ 
\hline
\end{tabular}
\\ 

Medians of weighted galaxy counts, inside various aperture radii and limiting magnitudes.
\label{tab:wghtcounts}
\end{minipage}
\end{table}

Finally, we use the relative weighted density between the data, $\overline{\zeta_\mathrm{q,}}_\text{data}$, and the simulations, $\overline{\zeta_\mathrm{q,}}_\text{sim}$, as the metric distance to apply the ABC selection criteria $\left|\overline{\zeta_\mathrm{q,}}_\text{data} -\overline{\zeta_\mathrm{q,}}_\text{sim} \right|<\epsilon$, with $\epsilon$ being sufficiently small.

The propagated measurement uncertainties to the errors reflected in the different weighted counts of the summary statistics are given in Table \ref{tab:wghtcounts} and further details are provided in Appendix \ref{sec:datawght}. We can use those estimates of the uncertainties (Gaussian approximation) as informative weights on the ABC selection criteria directly, avoiding the explicit sampling of the measurement uncertainty in the ABC process.

The distribution of underlying convergence values, $\kappa_{\text{ext}}$, chosen for the redshift plane closest to $z_{\rm s}$, of the samples passing this criteria is the estimate of the posterior $p(\kappa_{\mathrm{ext}}|\mathbf{\boldsymbol{d}_{\text{env}}})$ given the observed data $\mathbf{\boldsymbol{d}_{\text{env}}}$.

ABC allows us to apply conjoint sets of summary statistics. In our specific application, we can apply different conjointly used weights (summary statistics), $\zeta_{q_1}, ..., \zeta_{q_n}$, in the sense that the lines of sight selected from the MS must be similar to the LOS of the lensing system in terms of each of the relative densities corresponding to $q_i$ passing the threshold in $\epsilon_i$. We refer the reader to \citet{Rusu2017} for details of the numerical implementation. The use of multiple conjoined weights can make use of additional information present in the data, and therefore may narrow down the width of the resulting $p(\kappa_{\mathrm{ext}}|\mathbf{\boldsymbol{d}_{\text{env}}})$. Here, we add to this approach by not only considering conjoined weights, but also conjoined aperture sizes. Following \citet{Rusu2017} and \citet{Greene:2013wb}, where $q=1$ is always employed, we therefore compute $p( \kappa_\mathrm{ext}|\zeta_{q, 1}^{\mathcal{A}_i}, ..., \zeta_{q, n}^{\mathcal{A}_i};\mathbf{\boldsymbol{d}_{\text{env}}})$ with at most four conjoined constraints. This limit is due to the finite number of LOS available inside the MS, and due to computational speed.

\subsection{LOS: Results from the summary statistics} \label{sec:res}

\begin{figure}
\begin{center}
\includegraphics[angle=0,width=80mm]{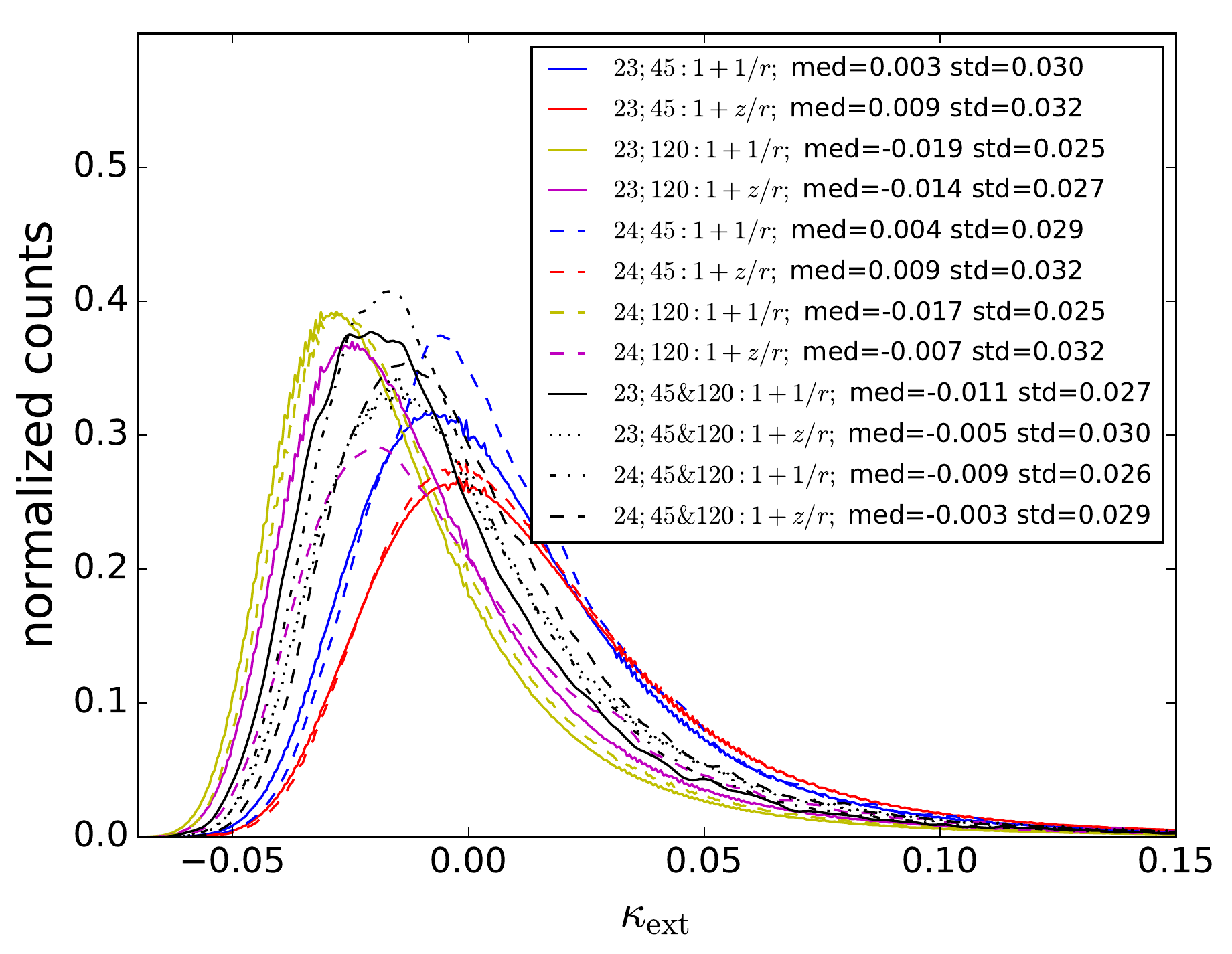}\\
\caption{Convergence distributions for the limits of $i<23$ mag, $i<24$ mag, aperture radii of $45\arcsec$ and $120\arcsec$, and conjoined number counts weighted by $1$, $1/r$ and $z/r$. The median and standard deviation of each distribution is quoted.}\label{fig:kappa}
\end{center}
\end{figure}

Figure~\ref{fig:kappa} shows the results of our $p(\kappa_{\mathrm{ext}}|\mathbf{\boldsymbol{d}_{\text{env}}})$ estimation based on different summary statistics and the ABC procedure. The distributions, corresponding to different conjoined weights as well as limiting aperture radius and magnitude, have a standard deviation of $\sim 0.025$-$0.032$, and medians distributed around zero, which vary by $\lesssim1$ standard deviation. In agreement with the expectations based on the measured relative densities, the inferred convergence is larger when measured inside the $45\arcsec$ aperture, and smaller otherwise. The distributions vary little with limiting magnitude, given the same constraints and apertures. We therefore conclude that we have reached the necessary depth to perform this analysis. The distributions are further brought into agreement if we use constraints based on both aperture sizes, and the $p(\kappa_{\mathrm{ext}}|\zeta_1^{45\arcsec},\zeta_{z/r}^{45\arcsec},\zeta_1^{120\arcsec},\zeta_{z/r}^{120\arcsec};\mathbf{\boldsymbol{d}_{\text{env}}})$ distribution is also tighter than those computed from either aperture, reflecting the fact that the knowledge of the LOS being locally overdense but globally underdense provides useful information. We chose the above-mentioned distribution with $\kappa_{\mathrm{ext}} = -0.003\pm0.029$ as our fiducial distribution to use for the cosmographic inference, as this distribution is the most informative in terms of the deeper magnitude limit, the use of both aperture radii, and the use of a weight incorporating redshift information, therefore information about the presence of the group. This distribution is also a good approximation for the mean of the distributions we explored.

\subsection{LOS: Robustness checks}
The summary statistics employed does not explicitly select LOS in the MS conditioned on having a lens present in its centre. We expect the LOS in the MS to be representative excluding the lens plane. Lenses are common in group environments and even very nearby correlation is observed \citep{Huterer:2005, Oguri:2005a,Treu:2009}. The special environment that we are faced with when inferring the statistics about lenses may be biased with respect to the convergence distribution selected of the MS.

In particular, we need to quantify the local environment of the lens with respect to the LOS captured by the summary statistics applied on the MS. We expect the summary statistics to behave as follows:

\begin{enumerate}
	\item Large scale over-densities specific to the lens around 45\arcsec and on larger scales are well captured.
	\item Correlated structure nearby the lens is not well captured.
	\item The shot noise of the specific alignment of nearby structure (in projection irrespective of the redshift) is representative of the mean galaxy number density within the inner mask region of the summary statistics.
\end{enumerate}

We perform two analyses to test these assumptions and if necessary to apply additional corrections to the LOS estimate: (i) we look at how well the summary statistics can capture the nearby group and (ii) we generate mock realisations with a rendering process that quantitatively captures the galaxy density within the weight region and the statistics in the convergence distribution in the MS and investigate with respect to these models, whether there is excess structure present around the lens.

Certain specific and impactful mass distributions present in our universe might not sufficiently be well captured by the summary statistics in the ABC framework, or the mass distribution may be so specific that the sample statistics within ABC do not allow us to explore its effect. In those cases, a specific model based on the information available is required.

\subsubsection{Nearby group}
There is spectroscopic evidence that the lensing galaxy is at the center of a small group (see details about the observations and derived group properties in Appendix \ref{sec:groups}).
To test the reliability of the summary statistic in this case, we compare the original summary statistics approach with a composite one, consisting of (i) a modified summary statistics excluding all galaxies spectroscopically confirmed to be part of the group, including the lensing galaxy and (ii) an explicit rendering of the group properties and their uncertainties provided by the spectroscopic campaign (see Appendix \ref{sec:groups}).

The new convergence distribution estimated from the summary statistics excluding the group members has a median shift of $\Delta\kappa_\mathrm{ext, no group}\sim0.006-0.014$ towards negative values. The rendering of the group halo results in a median convergence value of $\kappa_\mathrm{group}=0.01$. The combined distributions of $\kappa_\mathrm{group}$ and $\kappa_\mathrm{ext, no group}$ are fully consistent with the summary statistics including all objects without making the distinction of a group being present.

This demonstrates the ability of the chosen summary statistics and the sufficient sample statistics in the MS to capture the impact of the group environment statistically sufficiently well in determining the external convergence value to our required accuracy.

What speaks in favor of the ABC approach in this case is that priors are easier to quantify and effectively reflect the distribution available in a large simulation box. Alternatively if very detailed information were available and a precise location and mass structure could be inferred, a direct model may be more precise and possibly reduced further the uncertainty on the LOS estimate. In this work we go with the ABC approach, lacking the additional information. We note however, that the uncertainty on our LOS estimate is already a subdominant contribution to our overall error budget.

\subsubsection{Local environment} \label{subsec:los_double_counting}
To test the ability of our summary statistics in describing the very local environment, particularly the perturbing effect of the nearest galaxies (in projection) of the lens, we perform the following test: we exclude the galaxies G3 and G4 from the catalogue entering the summary statistics and compare how the selected LOS of the MS change with respect to the baseline statistical distribution. The test shows no significant change in the selected LOS from the MS which confirms our intuition that below the scales of the aperture, the LOS selected from the MS are random pointings within the environment specified at a larger scale.

The local environment of lenses is often overdense, since they are massive early-type galaxies \citep{Treu:2009}. Our LOS summary statistics may not capture this effect sufficiently (since it assumes a random pointing consistent with the weighted number counts regardless of the presence of a strong lens), and we may have to explicitly include the lensing effect of such close structures present in the lens plane. It is important to quantify the local effect with respect to the LOS selected in the MS from the summary statistic. We thus randomly sample the galaxy positions within the field (in projection) and quantify the chance alignment rates as a function of radius. We conclude that the galaxy triplet G2 is a clear outlier in this statistic and its effect is not represented in the LOS selected in the MS. In contrast, the nearby galaxies G3 and G4 are statistically well represented as a chance alignment expected to be captured by the LOS selected by the MS.

Based on these arguments, we chose to explicitly model the convergence of G2 on top of the pre-quantified LOS effect but decide to avoid the convergence effect of G3 and G4.

In practice, when modelling G3 and G4, we subtract the convergence term induced on the lens by the models associated with G3 and G4 and compute an effective external convergence, $\kappa_{\text{ext, eff}}$ as
\begin{equation} \label{eqn:los_correction}
	\kappa_{\text{ext, eff}} = \kappa_{\text{ext}} - \kappa_{\rm G3+G4}
\end{equation}
Simple mock renderings of LOS structure assures that this procedure can guarantee a sub-per-cent accuracy on the LOS effect. Our approach potentially over-predicts the scatter in the LOS but not on the cost of a systematic shift.

\section{Combined analysis} \label{sec:combined_analysis}
In this section, we specify how we sample the cosmographic likelihood of the combined analysis (Equation \ref{eqn:joint_prob}). We describe how we can subdivide the sampling of the parameter space within the hierarchical model (Section \ref{subsec:sampling_likelihood}). We then present all the different model options that we consider in this work from Section~\ref{sec:model_choices} (Section~\ref{subsec:option_summary}) and how we marginalize over the different model choices introduced in Section~\ref{sec:model_choices} (Section~\ref{subsec:marginalize_options}). All the decisions listed in this section were made before the unblinding of the cosmographic results.

\subsection{Sampling the likelihood} \label{subsec:sampling_likelihood}
The full joint likelihood over all data sets and marginalized over all nuisance parameters (Equation \ref{eqn:joint_prob}) can be separated into several independent tasks. The partial separability makes certain covariances explicit and improves the convergence and sampling speed significantly.

We first compute the posterior values of $\boldsymbol{\xi_{\text{lens}}}$ and $\boldsymbol{\xi_{\text{light}}}$ from the imaging likelihood only, $P(\boldsymbol{d_{\textit{HST}}}| \boldsymbol{\xi_{\text{lens}}}, \boldsymbol{\xi_{\text{light}}})P(\boldsymbol{\xi_{\text{lens}}}, \boldsymbol{\xi_{\text{light}}})$. This part of the sampling contains between 39 to 44 non-linear parameters and an additional 6-51 linear parameters (flux amplitudes), depending on the model chosen. To sample the high-dimensional parameter space efficiently, we first apply a Particle Swarm Optimizer (PSO) to find a maximum in the likelihood. Through this process, we add incrementally the complexity in the model and apply three times a PSF re-optimization. After this process is completed, we run a Markov Chain Monte Carlo (MCMC) algorithm (\textsc{emcee} \citep{emcee} implemented in \textsc{CosmoHammer} \citep{Akeret:2013nl}) to sample the posterior distribution. This step-by-step procedure is facilitated with the \texttt{Workflow} module built into \textsc{lenstronomy}.

The terms containing the likelihood of the time-delay, $\Delta t_{AB}$, and of the velocity dispersion, $\sigma^\text{P}$ can be folded in by simple sampling the data based on their uncertainties to map the $\boldsymbol{\xi_{\text{lens}}}$ and $\boldsymbol{\xi_{\text{light}}}$ into the angular diameter distance posteriors through Equation (\ref{eqn:ang_dist_delay}) and (\ref{eqn:ang_dist_kin}).

The likelihood of the LOS, $P(\boldsymbol{d_{\text{env}}}|\kappa_{\text{ext}})P(\kappa_{\text{ext}})$, is folded in through displacing the posterior samples of $D_{\Delta t}$ with the distribution of external convergences, $P(\kappa_{\text{ext}}| \boldsymbol{d_{\text{env}}})$, of Section \ref{sec:los_analysis} according to Equation (\ref{eqn:kappa_ext}). The relevant likelihood in cosmology, $P(\boldsymbol{d_{\text{J1206}}}| D_{\text{d,s,ds}})$, is directly proportional to the sample distribution obtained through the process described above.

\subsection{Summary of the different modelling options and procedures} \label{subsec:option_summary}

Here we summarize all the options that we consider:
\begin{enumerate}
  \item Two choices for the main deflector, option \texttt{SPEMD} or option \texttt{COMPOSITE} (\ref{subsec:main_deflector}).
  \item Four choices with increasing flexibility in the source surface brightness, \texttt{DOUBLE\_SERSIC}, \texttt{DOUBLE\_SERSIC}+2n$_{\text{max}}$, \texttt{DOUBLE\_SERSIC}+5n$_{\text{max}}$, \texttt{DOUBLE\_SERSIC}+8n$_{\text{max}}$ (\ref{subsec:source_light}).
  \item We chose the M/L ratio of the nearby perturbers from a scaling law and each individual model realization has an intrinsic scatter of 0.1 dex drawn randomly from a log normal distribution (\ref{subsec:nearby_perturbers}).
  \item modelling the galaxy triplet, \texttt{TRIPLET}, or also including two other nearby galaxies, \texttt{TRIPLET}+2 (\ref{subsec:nearby_perturbers}).
  \item We optionally add a non-linear shear term to the lens model, \texttt{SIMPLE\_SHEAR} or \texttt{FOREGROUND\_SHEAR} (\ref{subsec:los_modelling}).
  \item We choose two different pixel masks to evaluate the imaging likelihood, \texttt{MASK 3.0"} or \texttt{MASK 3.2"} (\ref{subsec:image_modelling}).
  \item We iteratively refine the PSF in the fitting process. This is performed independently for all the different runs (\ref{subsec:image_modelling}).
\end{enumerate}
Considering all possible combinations of options, there are 64 individual configurations. Additionally, stochasticity is expected from the rendering of the scatter in the  M/L ratio among the nearby perturber and possibly due to the iterative PSF reconstruction. Accounting for the additional stochasticity, we run each model configuration twice with a different seed in the M/L ratio and sampler. This final step yields 128 distinct sets of model configurations overall.  

For each of the configurations listed in this section, we perform the full hierarchical sampling described in Section \ref{subsec:sampling_likelihood}. This leads to 128 distinct angular diameter distance posteriors.

\subsection{Assessing trends between different model options} \label{subsec:model_options_trend}

We keep the absolute values of the angular diameter distance posteriors blind and compare for consistency among the model configurations with respect to an overall median subtraction. We do not expect fully consistent posterior samples among any two of the configurations, as the nearby perturbers in particular do not follow the exact same relative M/L scaling and their iterative PSF reconstruction is performed independently. Instead, to assess statistical consistency, we compare the subsets of model configuration sharing certain model options against other specific model options.

\begin{figure*}
  \centering
  \includegraphics[angle=0, width=150mm]{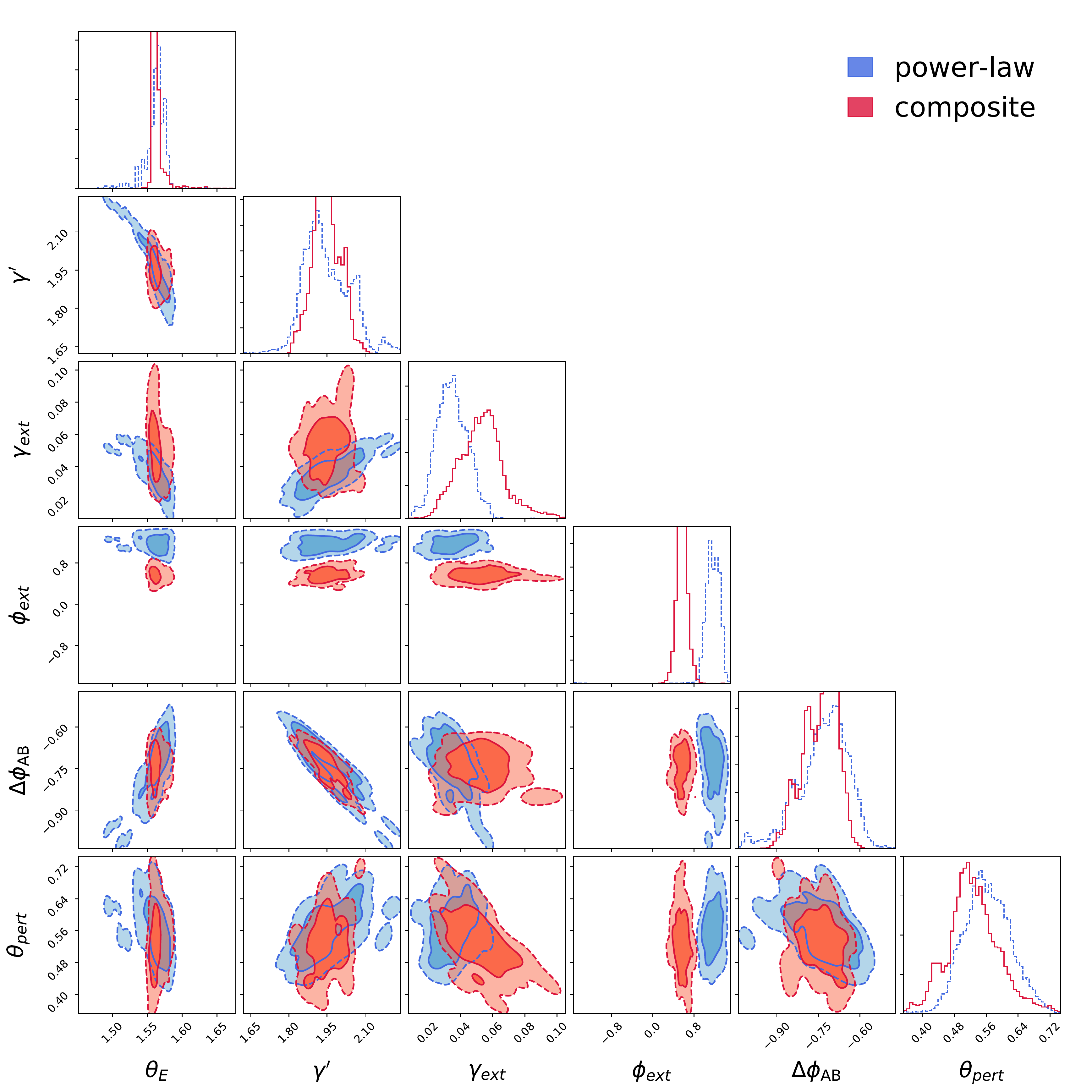}
  \caption{Lens model parameter posteriors for the set of \texttt{SPEMD} and \texttt{COMPOSITE} overlaid. The Einstein radius, $\theta_{\rm E}$, is defined as the azimutally averaged radius with a mean convergence of unity within its area. The power-law slope, $\gamma'$, is defined as the azimutally averaged logarithmic slope at the Einstein radius (applicable also for the \texttt{COMPOSITE} model). $\theta_{\text{pert}}$ is the summed Einstein radii of the galaxy triplet, G2. We derive consistent constraints with respect to the relative Fermat potential between the images A and B, $\Delta \phi_{\text{AB}}$, and other lensing quantities. The shear position angle and amplitude are discrepant by about 30$\deg$.}
  \label{fig:spemd_composite_compare}
\end{figure*}

In Figure~\ref{fig:spemd_composite_compare} we show a selection of lens model posteriors, $\boldsymbol{\xi_{\text{lens}}}$, of the full collection of samples of the \texttt{SPEMD} model vs the \texttt{COMPOSITE} model. We get statistically consistent but not fully equivalent lens model properties. The relative Fermat potential, $\Delta \phi_{\rm AB}$, is of particular relevance for our analysis due to its direct impact on the cosmographic result. $\Delta \phi_{\rm AB}$ directly depends on the logarithmic slope of the mass profile at the Einstein radius ($\gamma'$ in Figure~\ref{fig:spemd_composite_compare}). The local slope is better constraint for the \texttt{COMPOSITE} model due to anchor of the baryonic component to the light component and the assumed slope of the NFW profile relative to the \texttt{SPEMD} model.

\begin{figure*}
  \centering
  \includegraphics[angle=0, width=160mm]{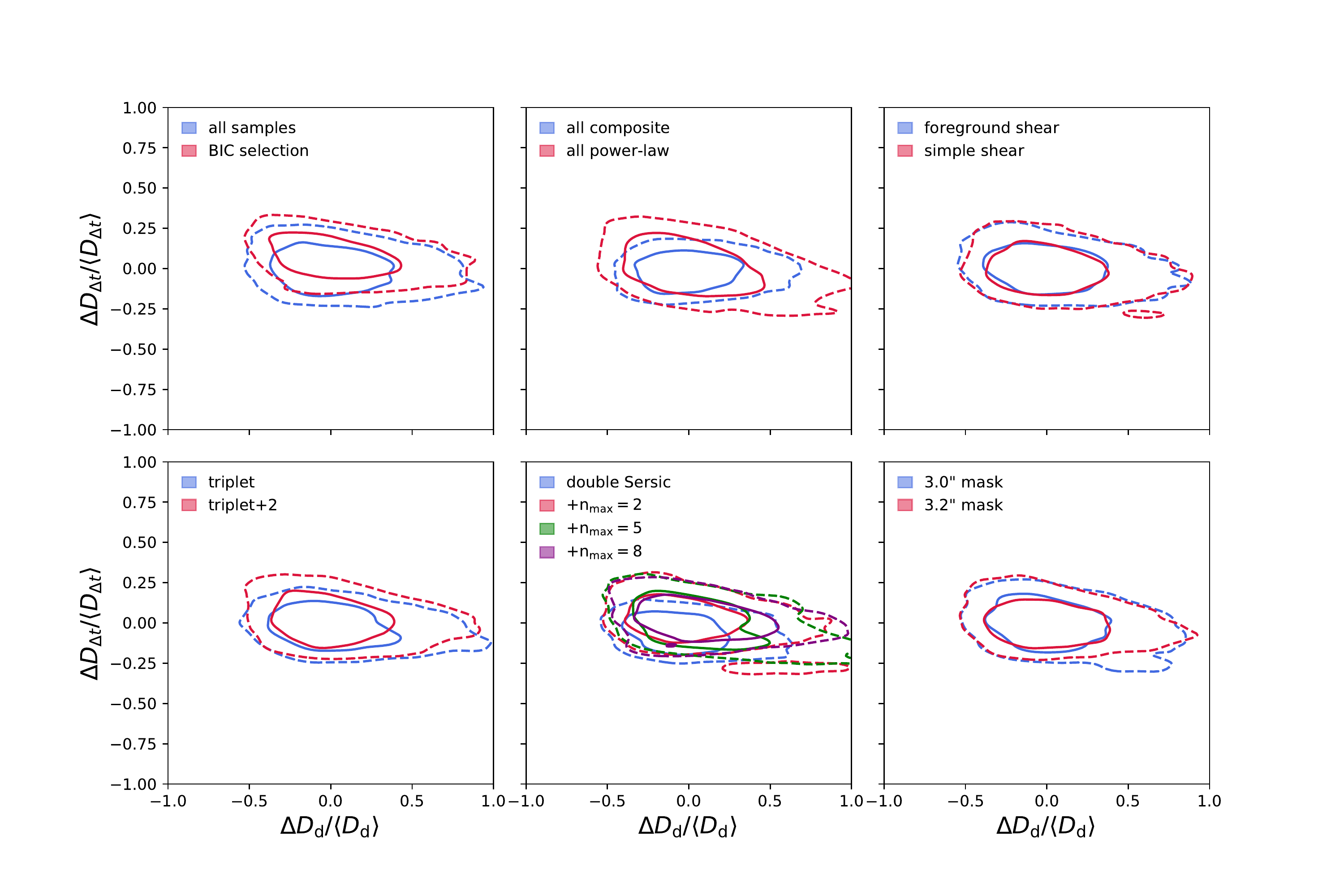}
  \caption{Median-subtracted and mean-divided relative angular diameter distance posteriors, $D_{\Delta t}$ and $D_{\rm d}$, to assess systematics within our blinded analysis. The different panels show different splits of the 128 total model configurations in different subsamples. The median value was kept blinded and kept fixed for the different panels. \textbf{Top left:} All samples (128) compared with those passed the BIC criteria. \textbf{Top middle:} \texttt{SPEMD} vs \texttt{COMPOSITE} models. \textbf{Top right:} \texttt{FOREGROUND\_SHEAR} vs \texttt{SIMPLE\_SHEAR} models. \textbf{Bottom left:} \texttt{TRIPLET} vs \texttt{TRIPLET}+2 models. \textbf{Bottom middle:} The four different source reconstruction options. \textbf{Bottom right:} The two different mask applied for the fitting.}
\label{fig:angular_diameter_blinded}
\end{figure*}

Figure~\ref{fig:angular_diameter_blinded} shows the blinded median-subtracted relative angular diameter distance constraints of the different subsets of choices against each other. Apart from the small differences between the \texttt{SPEMD} and \texttt{COMPOSITE} we observe the following trends between the subset of sample choices:

\begin{enumerate}
	\item The posteriors resulting from the two masking options are statistically identical. Our inference is robust against fine tuning of the mask.
	\item The model with the addition of G3 and G4, \texttt{TRIPLET}+2, has a significant higher $D_{\Delta t}$ value with a mean shift of $\approx 1.8$ per cent. The blinded posteriors do not yet include the correction factor applied on the excess convergence induced by G3 and G4 on the lens model.
	Applying the correction factor (Equation \ref{eqn:los_correction}) brings the samples of model \texttt{TRIPLET} and \texttt{TRIPLET+2} in statistical agreement to sub-per-cent level precision, suggesting that the flexion and higher order terms in G3 and G4 do not significantly impact the lens modelling in terms of constraints on the distances.
	\item The addition of a foreground shear term does not change the results of the cosmographic analysis. This may be simply the fact that the inferred non-linear shear terms are very small, consistent with the expectation of the LOS study.
	\item The source model complexity shows a trend of several per cent on $D_{\Delta t}$. This is in particular the case for the two lowest complexity models, \texttt{DOUBLE\_SERSIC} and \texttt{DOUBLE\_SERSIC} + $n_{\text{max}}=2$ models. The two more complex models, \texttt{DOUBLE\_SERSIC} + $n_{\text{max}}=5$ and \texttt{DOUBLE\_SERSIC} + $n_{\text{max}}=8$ are in very good agreement and no significant bias is introduced when adding additional shapelet components in the source reconstruction when going from $n_{\text{max}}=5$ to $n_{\text{max}}=8$.
\end{enumerate}

In our analysis, we explored one single parameterisation of the stellar orbital anisotropy (see Section~\ref{subsec:kinematics_modelling}) with an anisotropy radius, $r_{\text{ani}}$, and its uniform prior in the range $[0.5, 5]\times r_{\text{eff}}$. It has been noted by several works \citep{Jee:2016, Birrer:2016zy, Shajib:2018} that the specific parameterisation and prior can have a significant effect on the cosmographic result, in particular on the inference of the angular diameter distance of the deflector, $D_{\rm d}$. We checked for a dependency of the angular diameter posteriors on the anisotropy radius, $r_{\text{eff}}$ and find no significant trend, even for the largest and smallest values in $r_{\text{eff}}$. Therefore, we expect no significant effect depending on the specific parameterisation of the prior for this work. We note that this is mainly a consequence of the spectra taken at seeing conditions (1\arcsec) significantly larger than the half-light radius of the deflector, $r_{\text{eff}} = 0.36\pm 0.02$. When more precise and better spatially resolved kinematics is available, the dependence on the anisotropy prior needs to be revisited.

\subsection{Model selection criteria} \label{subsec:marginalize_options}
The models considered in this work cover a significant range in complexity and ability in reconstructing the data. A simple inspection of the $\chi^2$ values of the different models revealed an expected significant variability in the goodness of fit measures.

We use the Bayesian Information Criterion (BIC) to perform a statistical weighting of our 128 models. The BIC is defined as 
\begin{equation}
 \text{BIC} = \ln(n) k - 2 \ln(\hat{L}),
\end{equation}
where $n$ is the number of data points, $k$ the number of free parameters in a model and $\hat{L}$ the maximum likelihood of the model.

In our case, the number of model parameters, $k$, (including the linear coefficients) range between 45-90 and the number of (independent) pixels, $n$, within the 3".0 aperture mask is 4296. We are in the regime of $n \gg k$ and deliberately want a uniform weight on all models prior to the BIC criterion - the regime where BIC is applicable for Bayesian model selection. BIC also penalizes additional model complexity more than the Akaike information criterion (AIC). This allows for a better representation of models with lower complexity to have significant weight in our posteriors.

In our models, we considered two different aperture masks (Section \ref{subsec:image_modelling}). To set all models on equal footing in terms of the number of data points, $n$, and the likelihood, $\hat{L}$, we adopt the same mask to evaluate BIC for all models. This procedure is similar to \cite{Wong:2017} in their comparison of the $\chi^2$ values between different models.

The relative probability of two models, M$_1$ and M$_2$, given the data, can be expressed in terms of their relative BIC values, $\text{BIC}_1$ and $\text{BIC}_2$ as
\begin{equation}
 p({\rm M}_1)/p({\rm M}_2) \propto \exp \left(-\left(\text{BIC}_1 -  \text{BIC}_2\right) / 2 \right).
 \label{eqn:BIC_prob}
\end{equation}
We define the BIC weight in respect to the minimal BIC value in our sample, BIC$_{\rm min}$, as

\begin{equation}
f_{\rm BIC}(x) \equiv \begin{cases}
1 & x \le \text{BIC}_{\rm min} \\
\exp \left(-\frac{x -  \text{BIC}_{\rm min}}{2} \right) & x > \text{BIC}_{\rm min}
\end{cases}
\end{equation}
to make sure that the acceptance ratio is bound by 1 even when applying a convolution in $x$ (see Equation \ref{eqn:big_conv}).
We are aware that the samples we have are a finite representation drawn from an underlying more continuous distribution. These samples are representative and sufficiently densely sampled to allow an adequate posterior estimate of the angular diameter distances.

However, in terms of the BIC values, we notice a very sparse sampling that can lead to significant sample variance. To apply the BIC weighting of the samples (Equation \ref{eqn:BIC_prob}), the models to be considered have to be sufficiently well-sampled within $\Delta {\rm BIC}$ values that allow significant relative weights. This is not the case for our collection of models.

A good estimator for the underlying stochasticity comes from the relative BIC values of those samples that share exactly the same model options except the seed in the M/L scatter, the sampling of the PSO and MCMC, and, as such, the PSF iteration. These ``twins'' are expected to have the same BIC value and their difference is directly linked to the stochasticity described above. Over all the 64 sample pairs, we get an intrinsic scatter in the BIC of $\hat\sigma_{\text{BIC}} \approx 140$. This intrinsic scatter requires a significant number of samples sharing the same underlying models, $n_{\rm twins}$, to allow for a stable BIC selection according to Equation \ref{eqn:BIC_prob}.

Instead, for computational reasons, we chose to approximate the underlying smooth BIC distribution with a kernel density estimator, $h$, acting on the sparsly sampled BIC values, $x$. This leads to a new, more conservative acceptance ratio, $f^*_{\rm BIC}$, given by
\begin{equation} \label{eqn:big_conv}
	f^*_{\rm BIC}(x) \approx  h * f_{\rm BIC}(x).
\end{equation}
The estimator, $f^*$, must recover the true underlying distribution in the regime of infinitely fine sampled points $(n_{\rm twins}\rightarrow \infty)$ and when the distribution itself is a delta function without a variance in the BIC values of the twins $(\sigma_{\text{BIC}} \rightarrow 0)$.

An estimator that satisfies those properties in terms of number of twins $(n_{\rm twins})$ and taking into account our estimate of the variance in our distribution, $\hat\sigma^2_{\text{BIC}}$, is the Gaussian kernel, $h_{\sigma}$, with width, $\sigma_h$, given by
\begin{equation}
	\sigma_h^{2} = \frac{\hat\sigma^2_{\text{BIC}}}{n_{\rm twins}}.
\end{equation}
This is a conservative approach to the BIC selection and is explicitly designed to overcome the sample variance, $\sigma_{\text{BIC}}$, in our finite set of model realisations, $n_{\rm twins}$.

A single BIC selection for all samples leads to a significant different weight of \texttt{SPEMD\_SERSIC} and \texttt{COMPOSITE} models in our posteriors. To avoid being biased due to the specific model parameterization (See Section~\ref{subsec:degeneracies}), we deliberately chose two different BIC selection for the two categories of main deflector models (by splitting the samples in the two respective model branches) in order to keep the equal weights of \texttt{SPEMD\_SERSIC} and \texttt{COMPOSITE} models in our remaining posterior. We note that the two options \texttt{SPEMD\_SERSIC} and \texttt{COMPOSITE} consist also of two different lens light profiles (\texttt{DOUBLE\_SERSIC} vs \texttt{CHAMELEON}). Parts of the relative BIC differences may come from the fact that the lens light distribution is better represented by one model over the other.

We made two checks of the robustness of our approach on the cosmographic posterior sample:
\begin{enumerate}
	\item We changed the width of the kernel, $\hat\sigma^2_{\text{BIC}}/n$, by a factor of two (smaller and larger) and registered no significant change in the posterior distribution. The uncertainty in the precision on the sampling variance does not impact our (still blinded) result.
	\item Alternatively, we applied a strict binary cut of the sample at number $n_{\text{cut}}$ in sorted increasing BIC and check for the behavior of the added posteriors as a function of $n_{\text{cut}}$. We observe a slight stochastic behavior from $n_{\text{cut}}=1$ to $n_{\text{cut}}=3$. Then the posteriors remain stable up to $n_{\text{cut}}=10$ (in each subset \texttt{DOUBLE\_SERSIC} and \texttt{CHAMELEON} adding up to 20 samples selected). For any choice in $n_{\text{cut}}$ in the range [3, 10], the posteriors on the angular diameter distances are consistent with the weighted posteriors according to the kernel $\hat\sigma^2_{\text{BIC}}/\sqrt{2}$. The effective sample number selected by the kernel is $\approx 5$.
\end{enumerate}

\begin{figure*}
  \centering
  \includegraphics[angle=0, width=150mm]{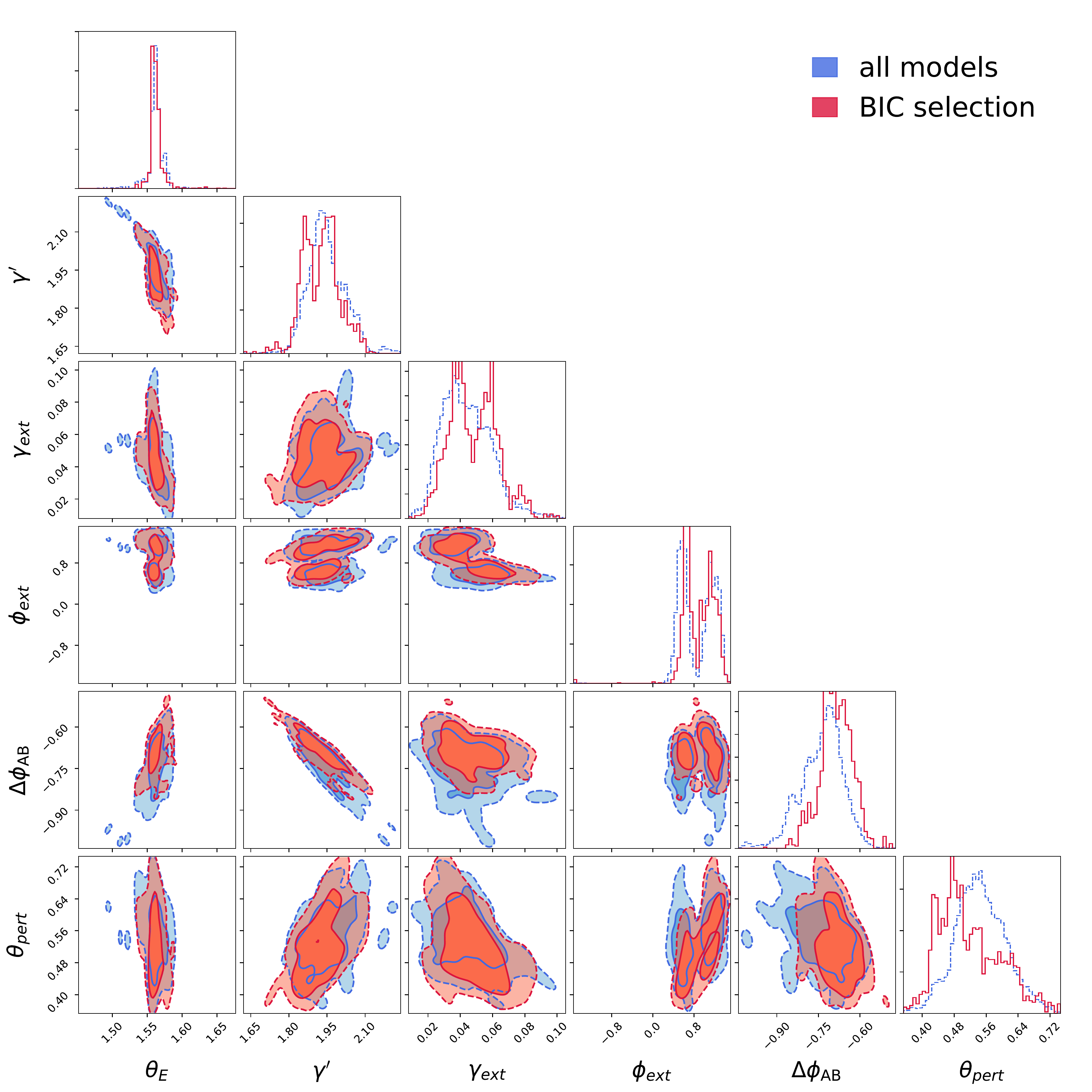}
  \caption{Lens model parameter posteriors (same parameters as for Figure \ref{fig:spemd_composite_compare}) of the BIC selected sample (red contours) and the full 128 samples (blue contours). The application of even a conservative Bayesian Information Criteria leads to an improvement on the precision on the relative Fermat potential, $\phi_{\text{AB}}$. The \texttt{SPEMD} and \texttt{COMPOSITE} models passing the BIC cut have consistent posterior values.}
  \label{fig:lens_model_bic}
\end{figure*}

\begin{figure*}
  \centering
  \includegraphics[angle=0, width=120mm]{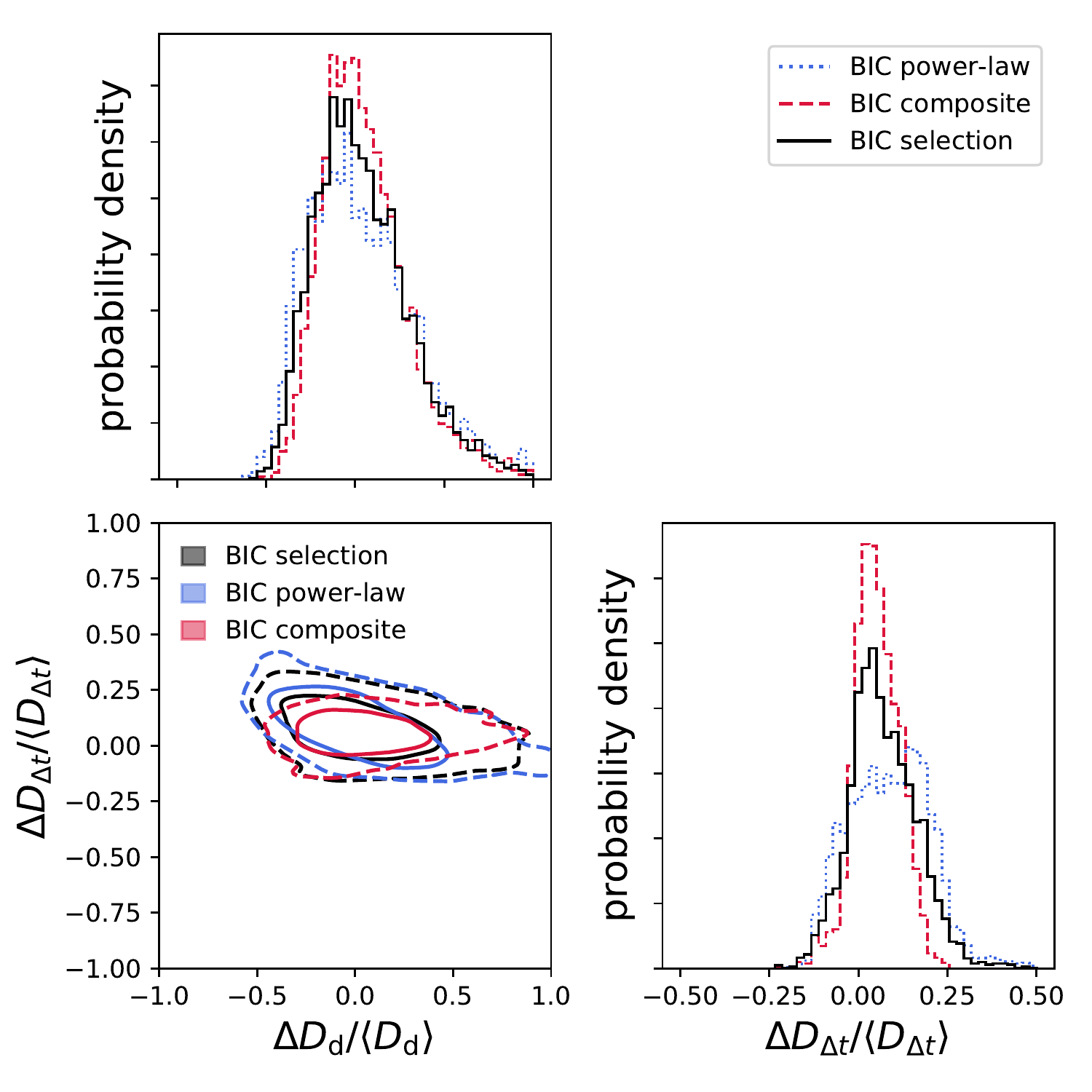}
  \caption{The comparison of the \texttt{SPEMD} and \texttt{COMPOSITE} models passing the BIC selection in terms of the angular diameter distance posteriors.}
\label{fig:diameter_distance_blind_bic_pl_comp_split}
\end{figure*}

Figure~\ref{fig:lens_model_bic} reflects the BIC selection criteria for the lens model parameters. We notice an improved precision in the parameters and in particular on the Fermat potential by applying the BIC selection. Figure~\ref{fig:diameter_distance_blind_bic_pl_comp_split} compares the BIC weighted samples split in \texttt{SPEMD\_SERSIC} and \texttt{COMPOSITE} models. In Appendix \ref{app:bic_models} we present the BIC sorted list of all models including their posterior weights entering to our final BIC-weighted result.

\section{Results} \label{sec:results}

Throughout the previous sections, the analysis is kept blind with respect to the absolute scale of the angular diameter posteriors. This section presents the unblinded results of the angular diameter distances (\ref{subsec:angular_diameter_results}) which defines the cosmographic measurement of this work and thus the cosmological likelihood. We emphasize that the analysis was frozen prior to unblinding and that the results are presented as they appeared after unblinding without any changes.

As a baseline illustration of the cosmographic content of our measurement, we present in Section~\ref{subsec:H0_j1206} the predicted value of the Hubble constant in a flat $\Lambda$CDM universe with free \Om\ obtained for SDSS 1206+4332, completely independent of any prior work, including the direct and inverse distance ladder methods, and prior time-delay cosmography work.

Then, in Section~\ref{subsec:H0_combined} we combine our new measurement with the cosmographic constraints previously published by our collaboration, and present an updated H0LiCOW measurement of $H_0$ precise at the 3 per cent level.

\subsection{Angular diameter distance posteriors from \name} \label{subsec:angular_diameter_results}
\begin{figure*}
  \centering
  \includegraphics[angle=0, width=150mm]{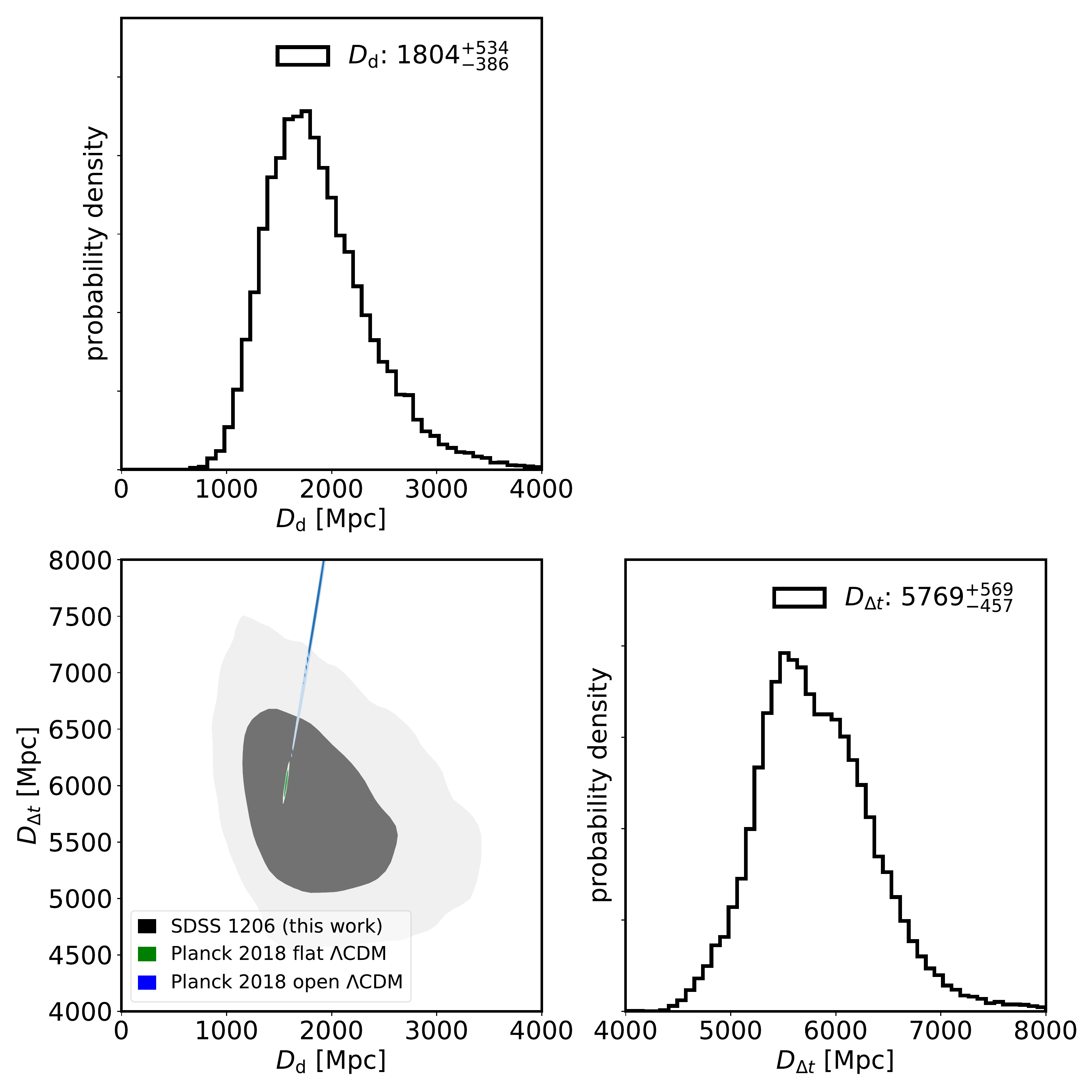}
  \caption{Unblinded angular diameter distance posteriors inferred from \name (gray contours). Over-plotted are the CMB constraints within a flat $\Lambda$CDM model (green contours) and within an open  $\Lambda$CDM model with curvature $\Omega_{\rm k}$ (blue contours) of the Planck collaboration 2018 (from TT+lowL+lowE data only), evaluated in the set of TDSL observables.}
  \label{fig:D_dt_D_d_result}
\end{figure*}

Figure~\ref{fig:D_dt_D_d_result} reveals the unblinded absolute angular distance posteriors inferred from our analysis. We measure for the time-delay distance $D_{\Delta t} = 5769_{-457}^{+569}$ Mpc (8.9 per cent marginalized uncertainty) and for the angular diameter distance to the lens $D_{\rm d} = 1804_{-386}^{+534}$ Mpc (25 per cent marginalized uncertainty). The posteriors in $D_{\Delta t}$ and $D_{\rm d}$ are correlated through the specific dependence of the lens model on the stellar kinematics of the lensing galaxy and the time delay between the two quasar images (see Equations~\ref{eqn:ang_dist_delay} and \ref{eqn:ang_dist_kin}). The angular diameter \textbf{distance} constraints are independent of the cosmological model and reflect the joint constraining power of time delay, imaging, kinematic and wide field data presented in this work.

The posteriors are fully consistent with the CMB measurements within a flat $\Lambda$CDM cosmological model. Within the strict assumptions of a flat $\Lambda$CDM model, our measurement is significantly less constraining than the CMB. However, relaxing some of the assumptions clearly illustrates the power of TDSL. For example, in Figure~\ref{fig:D_dt_D_d_result} we over-plot the Planck 2018 posteriors in a open (non-flat) $\Lambda$CDM model (from TT+lowL+lowE data only). It is clear that just a single lens adds valuable constraining power, in this case on curvature, as it measures the absolute distances at lower redshifts.

\subsection{The Hubble constant from \name} \label{subsec:H0_j1206}

The angular diameter distance posteriors presented in Section~\ref{subsec:angular_diameter_results} reflect the cosmographic likelihood, $P(\boldsymbol{d_{\text{J1206}}}| D_{\text{d,s,ds}})$ (Equation \ref{eqn:cosmographic_likelihood}). The likelihood requires the full 2-dimensional distribution and covariances in $D_{\Delta t}$ and $D_{\rm d}$. We use a kernel density estimator to access a continous evaluation of the likelihood in parameter space. The kernel is chosen to be sufficiently narrow so as not to impact the likelihood estimate and resulting posteriors.

We choose to sample a flat $\Lambda$CDM cosmology with very uninformative uniform priors on $H_0$ ( [0, 150] \Hunit) and on $\Omega_{\rm m}$ ([0, 1] or [0.05, 0.5]). The former prior on $\Omega_{\rm m}$ is chosen for consistency with our previous H0LiCOW analysis. The latter prior reflects more reasonable assumptions, since we know that the universe is not empty nor closed from a variety of robust and independent arguments.

\begin{figure*}
  \centering
  \includegraphics[angle=0, width=150mm]{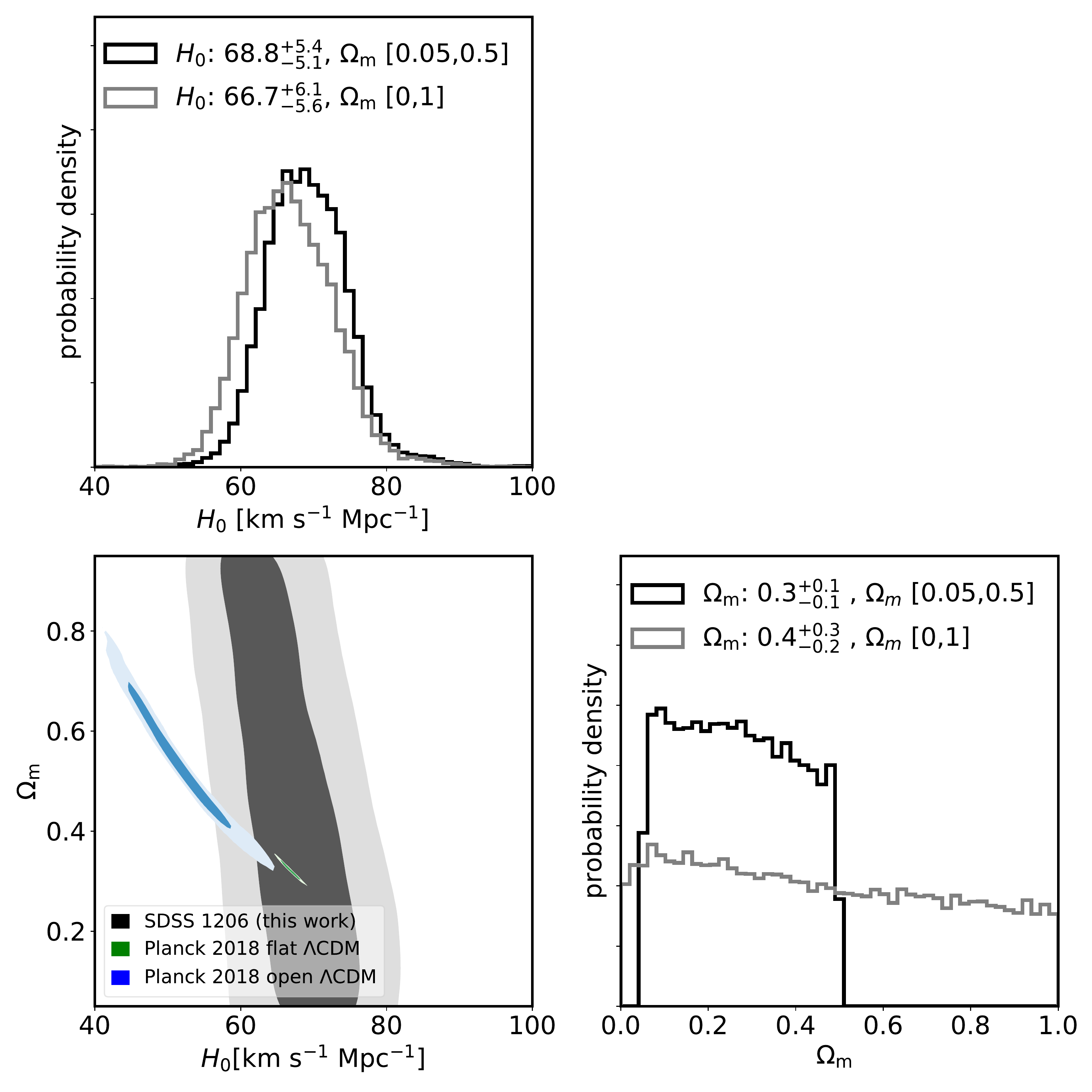}
  \caption{Inferred cosmological parameters from the sampling of the \name posteriors in a flat $\Lambda$CDM model. We sample a uniform prior on $H_0$ in [0, 150] \Hunit and on $\Omega_{\rm m}$ in [0, 1] or [0.05, 0.5] and show marginalizations assuming both priors on $\Omega_{\rm m}$. Over-plotted are the CMB constraints within a flat $\Lambda$CDM model (green contours) and within an open  $\Lambda$CDM model with curvature $\Omega_{\rm k}$ (blue contours) of the Planck collaboration 2018 (from TT+lowL+lowE data only).}
  \label{fig:H0_omega_m_result}
\end{figure*}

Figure~\ref{fig:H0_omega_m_result} shows the resulting posterior distribution functions in $H_0$ vs $\Omega_{\rm m}$. We identify a marginal tilt in the posterior degeneracy in the $H_0$ vs $\Omega_{\rm m}$ plane due to the late time cosmic acceleration parameter folding into $\Omega_{\rm m}$ when flatness is imposed. With the extremely conservative prior on $\Omega_{\rm m}$ in [0, 1], we measure from \name a Hubble constant of H$_0 = $~\HTOMlarge\  \Hunit. With a more reasonable prior on $\Omega_{\rm m}$ flat in [0.05, 0.5], we measure a Hubble constant $H_0 = $~\HTWE\ \Hunit, i.e. 7.2 per cent precision from a single lens.

It has been noted that $D_{\rm d}$ can be measured from gravitational lenses \citep{Paraficz:2009, Jee:2015, Jee:2016} and provides cosmological constraining power. Ignoring $D_{\rm d}$ and using only the marginalized $D_{\Delta t}$ as measurement (effectively ignoring the kinematics information), we achieve a precision of 9.4\% on $H_0$ with priors on $\Omega_{\rm m}$ in [0.05, 0.5]. The joint posteriors of $D_{\Delta t}$ and $D_{\rm d}$, however, add significant information to improve the precision on $H_0$ in our study. This has been explored by \cite{Birrer:2016zy} and been used to mitigate the MSD in the inference of $H_0$.

\subsection{Combined analysis with the previous H0LiCOW lenses} \label{subsec:H0_combined}
\begin{figure*}
  \centering
  \includegraphics[angle=0, width=160mm]{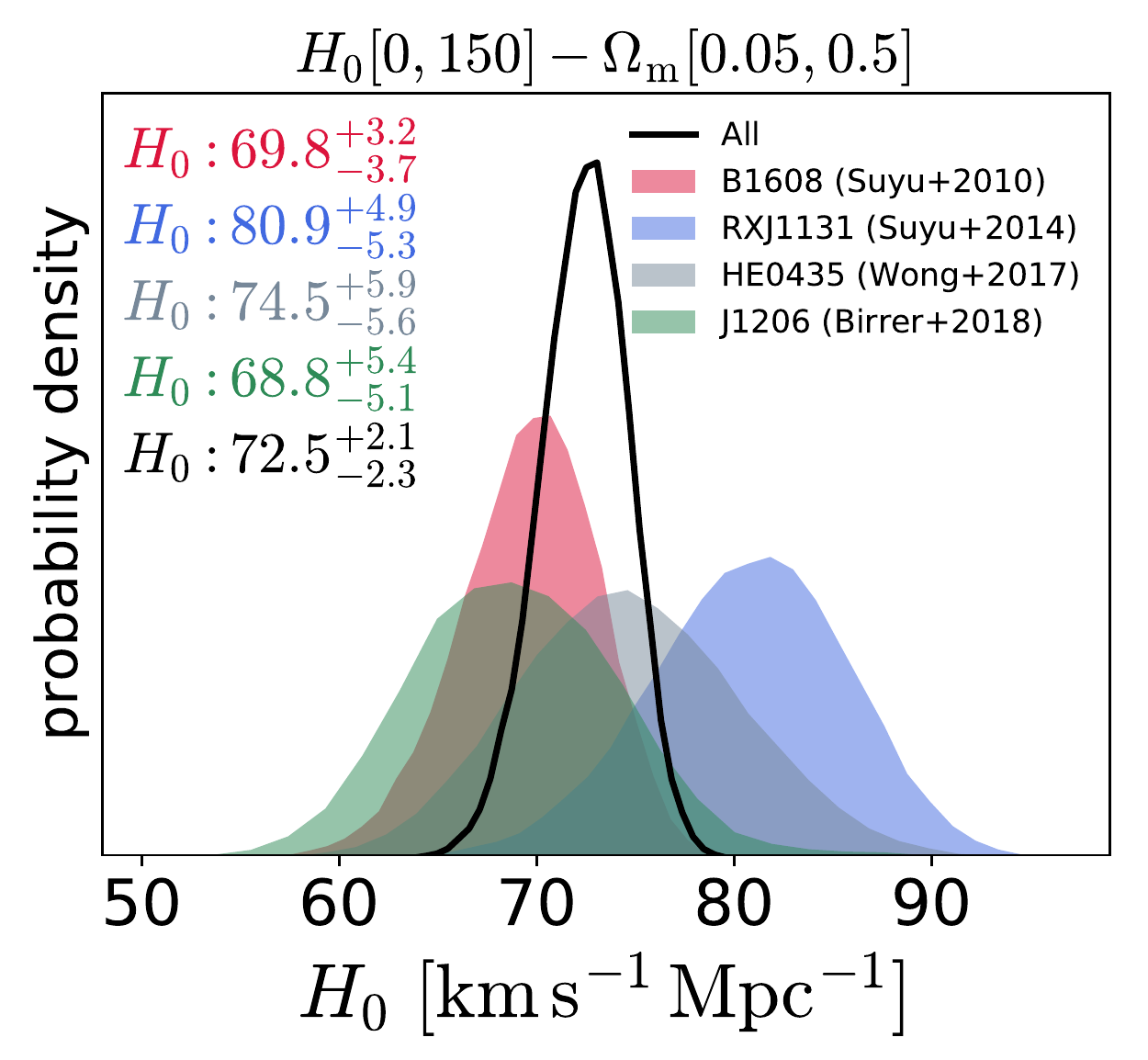}
  \caption{Combined result of the current H0LiCOW sample in terms of $H_0$ in flat $\Lambda$CDM. Shown are the 1D marginalized posterior distribution form the measurements of the four individual lenses with uniform priors on $H_0$ [0, 150] in the range \Hunit and $\Omega_{\rm m}$ in the range [0.05, 0.5]. The combined posterior is shown in black.}
\label{fig:H0_combined_narrow}
\end{figure*}

The combined analysis of the previous three H0LiCOW lenses, B1608+656 \citep{Suyu:2010rc}, RXJ1131-1231 \citep{Suyu:2013ni, Suyu:2014aq} and HE0435-1223 \citep{Wong:2017} was presented by \cite{Bonvin:2017}. In this section, we update the combined constraints on the Hubble constant adding the likelihood of \name to the combined sampling of the cosmological parameters. The result presented by \cite{Bonvin:2017} were sampled assuming a uniform prior on $\Omega_{\rm m}$ in the range [0,\,1]. In this work, we impose a more realistic and mildly more informative prior with $\Omega_{\rm m}$ uniform in the range [0.05, 0.5], which our collaboration adopts as our new baseline to quote our measurement of the Hubble constant. Figure~\ref{fig:H0_combined_narrow} presents the individual constraints of the four systems and the combined constraints. 

\begin{figure}
  \centering
  \includegraphics[angle=0, width=80mm]{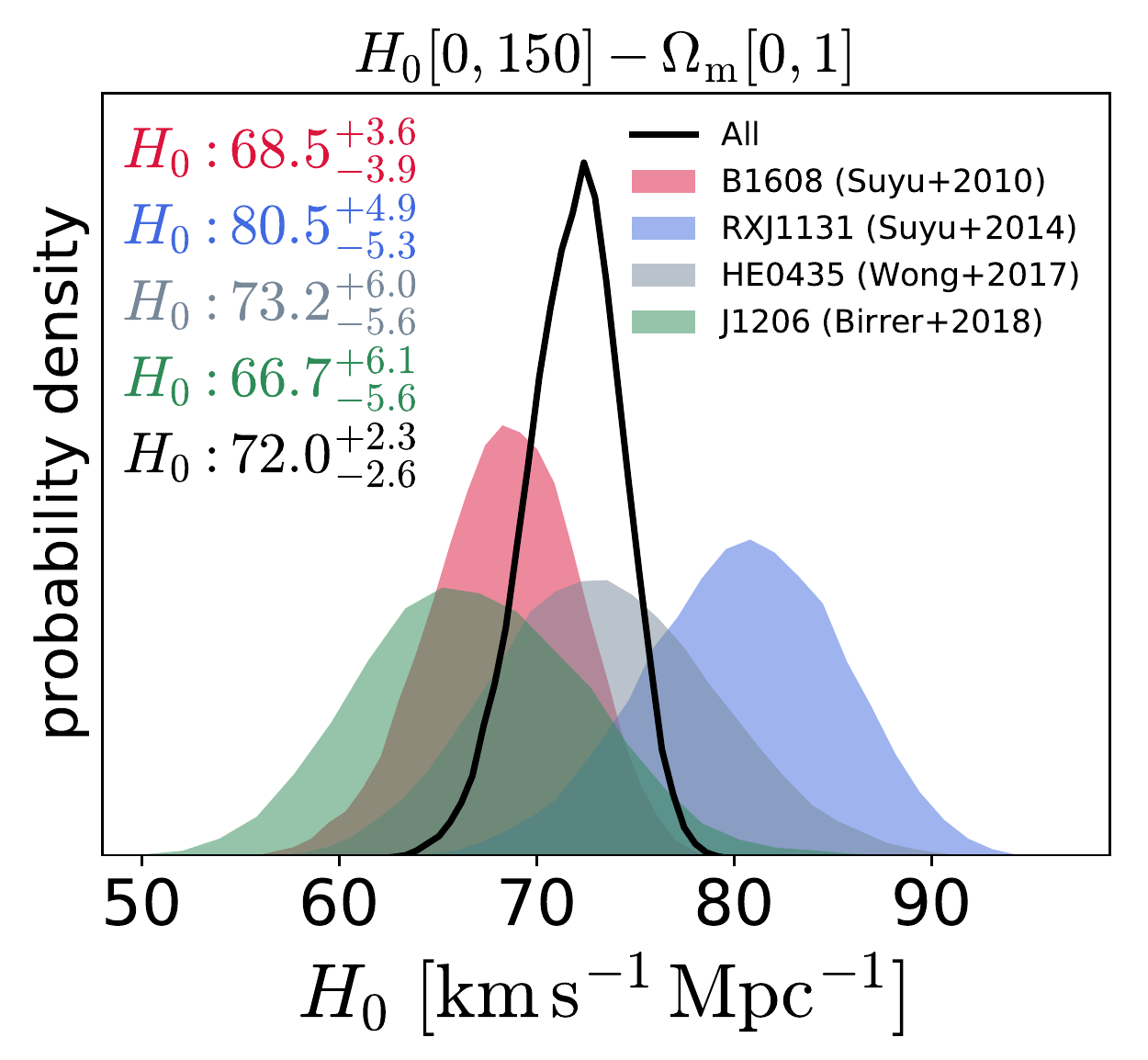}
  \caption{Combined result of the current H0LiCOW sample in terms of $H_0$ with a very conservative prior on $\Omega_{\rm m}$. Shown are the 1D marginalized posterior distribution form the measurements of the individual lenses with a uniform prior on $H_0$ in the range [0, 150] \Hunit and $\Omega_{\rm m}$ in the range [0, 1]. The combined posterior is shown in black.}
\label{fig:H0_combined_wide}
\end{figure}

We report a measurement of the Hubble constant of $H_0=$~\HFOU\ \Hunit for the four H0LiCOW lenses with a prior in $\Omega_{\rm m}$ in [0.05, 0.5] in a flat $\Lambda$CDM cosmology. For backward compatibility we illustrate the impact of the change in the prior on $\Omega_{\rm m}$ in Figure \ref{fig:H0_combined_wide}. In this case, the combined analysis with a uniform prior on $\Omega_{\rm m}$ in [0, 1] resulting in a measurement of $H_0=72.0^{+2.3}_{-2.6}$ \Hunit.

For completeness, we note that the lens RXJ1131-1231 of the H0LiCOW sample has been re-analyzed by \cite{Birrer:2016zy} using the same data (including a second \textit{HST} filter) and input from the H0LiCOW LOS analysis, and an early version of the \textsc{lenstronomy} software. They found a consistent result, but with significantly larger error bars on the Hubble constant than \cite{Suyu:2014aq}.

For conciseness, we do not repeat the cosmological analysis of \cite{Bonvin:2017} and we do not explore here the parameter inference of other cosmological models or in combination with other probes. This will be explored in our next H0LiCOW milestone paper (Wong et al., in preparation).

\section{Summary and conclusions} \label{sec:conclusion}

We presented a blind time-delay strong lensing (TDSL) cosmography measurement from the doubly imaged quasar \name. The measurement is based on a self-consistent analysis of the COSMOGRAIL time delay, deep \textit{HST} imaging data, stellar velocity dispersion of the main deflector measured from Keck spectroscopy, and extensive spectrophometric data to characterize the line of sight and immediate environment of the lens. In order to quantify the uncertainties arising from assumptions in the lens modeling, we construct 128 different models and combine their likelihoods and marginalize over the choices, based on an objective measure of goodness of fit that takes into account the varying degrees of freedom and number of data points. We take into
account the potential source of uncertainty arising from time-delay microlensing, which turns out to be almost negligible given the properties of \name.

Following the H0LiCOW protocol the analysis was kept blind until the very end, in order to prevent conscious and unconscious experimenter bias. Only after all the choices were frozen, the analysis was unblinded to reveal the absolute distance measurements and the inferred value of the Hubble Constant. We stress that our measurement of the Hubble Constant is completely independent of the local distance ladder method or any other cosmological probe. Our main results are:

\begin{itemize}
\item The measurement of the Hubble constant is comparable in precision and value with the previous measurements obtained by our collaboration using quadruply imaged quasars. Our measurement is based on the lens modeling code \textsc{lenstronomy} \citep{Birrer_lenstronomy, Birrer2015_basis_set}, which is completely independent of the code \textsc{GLEE} \citep{Suyu:2010jq, Suyu:2012rd} adopted in the analysis of previous H0LiCOW systems. 
\item Based on \name alone, we determine the Hubble constant to be $H_0$ = \HTWE\ \Hunit, in flat $\Lambda$CDM assuming uniform priors in $H_0$ [0, 150] \Hunit and in $\Omega_{\rm m}$ [0.05, 0.5].
\item By combining the \name likelihood with that of the three systems previously analyzed by our collaboration under the same prior we obtain $H_0$ = \HFOU\ \Hunit, a 3 per cent precision measurement.
\end{itemize}

In addition to the importance of a 3 per cent measurement of the Hubble constant, the analysis presented in this work has profound implications for the future of time delay cosmography. 

First, it demonstrates that, in cases when the host galaxy of the lensed quasar provides sufficient information on the lensing potential, doubles can also be used effectively for cosmography. Since doubles are five times more abundant than quads on the sky and generally easier to monitor, this proof of concept could lead to a significant increase in the number of lenses amenable for time delay cosmography. In turn, since the precision of the analysis is currently limited by sample size, extending the analysis to a new class of systems should lead to a boost in overall precision.

Second, the agreement with the previous results demonstrates that the systematic uncertainties related to lens modeling assumptions of the two codes are significantly smaller than our current random uncertainties.


Third, this is the first H0LiCOW analysis to infer $H_0$ simultaneously from
both, the time-delay distance, $D_{\Delta t}$, and the angular diameter
distance to the deflector, $D_{\rm d}$ \citep[by following][]{Birrer:2016zy}. This measurement
is independent of cosmological assumptions and provides more precision on $H_0$ as compared to an analysis focused only on $D_{\Delta t}$ \citep[][Jee et al. in prep]{Jee:2015, Birrer:2016zy}.


Finally, we remark that TDSL does not determine only the Hubble constant, but can be used to constrain a number of other cosmological parameters in combination with, e.g, CMB constraints, chiefly curvature and the equation of state parameter of dark energy \citep{Suyu:2014aq, Bonvin:2017}. For conciseness, we do not repeat the analysis of \cite{Bonvin:2017} with the new updated likelihood. Our likelihood will be made available upon acceptance of our manuscript.

A future milestone paper (Wong et al., in preparation) by the H0LiCOW collaboration will present the full exploration of cosmological constraints from \name and other lenses that are currently being analyzed by our team.

\section*{Acknowledgments}
We thank the referee, Peter Schneider, for the critical comments that improved the manuscript.
Support for this work was provided by NASA through grant number \textit{HST}-GO-15254 from the Space Telescope Science Institute, which is operated by AURA, Inc., under NASA contract NAS 5-26555. SB, TT and AJS acknowledge support by the Packard Foundation through a Packard Research fellowship to TT.
CER and CDF were funded through the NSF grant AST-1312329 and the \textit{HST} grant GO-12889.
FC, JC, GM and VB acknowledge support from the Swiss National Science Foundation.
VB, FC and JC thanks K. Rojas and M. Millon for their help in developing the micro-lensing time delay pipeline and PyCS analysis framework
G. C.-F. C. acknowledges support from the Ministry of Education in Taiwan via Government Scholarship to Study Abroad (GSSA).
SH acknowledges support by the DFG cluster of excellence \lq{}Origin and Structure of the Universe\rq{} (\href{http://www.universe-cluster.de}{\texttt{www.universe-cluster.de}}).
SHS thanks the Max Planck Society for support through the Max Planck Research Group.
K.C.W. is supported by an EACOA Fellowship awarded by the East Asia Core Observatories Association, which consists of the Academia Sinica Institute of Astronomy and Astrophysics, the National Astronomical Observatory of Japan, the National Astronomical Observatories of the Chinese Academy of Sciences, and the Korea Astronomy and Space Science Institute.

Lens modelling simulations were performed on the Hoffman2 cluster at University of California, Los Angeles.

This work has made use of \textsc{lenstronomy} \citep{Birrer_lenstronomy}, \textsc{Astropy} \citep{astropy}, \textsc{PyCS} \citep{Tewes:2013xr, Bonvin:2016}, \textsc{CosmoHammer} \citep{Akeret:2013nl}, \textsc{TOPCAT} \citep{Taylor:2005}, \textsc{corner} \citep{corner}, \textsc{Matplotlib} \citep{matplotlib} and standard \textsc{Python} libraries.

Based on observations made with the NASA/ESA \textit{Hubble Space Telescope}, obtained from the data archive at the Space Telescope Science Institute. STScI is operated by the Association of Universities for Research in Astronomy, Inc. under NASA contract NAS 5-26555.

Based on observations obtained at the Gemini Observatory, which is operated by the Association of Universities for Research in Astronomy, Inc., under a cooperative agreement with the NSF on behalf of the Gemini partnership: the National Science Foundation (United States), the National Research Council (Canada), CONICYT (Chile), Ministerio de Ciencia, Tecnolog\'{i}a e Innovaci\'{o}n Productiva (Argentina), and Minist\'{e}rio da Ci\^{e}ncia, Tecnologia e Inova\c{c}\~{a}o (Brazil).

Some of the data presented herein were obtained at the W. M. Keck Observatory, which is operated as a scientific partnership among the California Institute of Technology, the University of California and the National Aeronautics and Space Administration. The Observatory was made possible by the generous financial support of the W. M. Keck Foundation.

Based on observations obtained with WIRCam, a joint project of CFHT, Taiwan, Korea, Canada, France, at the Canada-France-Hawaii Telescope (CFHT) which is operated by the National Research Council (NRC) of Canada, the Institut National des Sciences de l'Univers of the Centre National de la Recherche Scientifique of France, and the University of Hawaii.

Based on observations at Kitt Peak National Observatory, National Optical Astronomy Observatory (NOAO Prop. ID 17A-0108; PI: C. Rusu), which is operated by the Association of Universities for Research in Astronomy (AURA) under a cooperative agreement with the National Science Foundation. 

COSMOGRAIL is made possible by the continuous work of all observers and technical staff obtaining the monitoring observations.

Funding for the SDSS IV has been provided by the Alfred P. Sloan Foundation, the U.S. Department of Energy Office of Science, and the Participating Institutions. SDSS acknowledges support and resources from the Center for High-Performance Computing at the University of Utah. The SDSS web site is www.sdss.org. SDSS is managed by the Astrophysical Research Consortium for the Participating Institutions of the SDSS Collaboration including the Brazilian Participation Group, the Carnegie Institution for Science, Carnegie Mellon University, the Chilean Participation Group, the French Participation Group, Harvard-Smithsonian Center for Astrophysics, Instituto de Astrofisica de Canarias, The Johns Hopkins University, Kavli Institute for the Physics and Mathematics of the Universe (IPMU) / University of Tokyo, the Korean Participation Group, Lawrence Berkeley National Laboratory, Leibniz Institut f\"{u}r Astrophysik Potsdam (AIP), Max-Planck-Institut f\"{u}r Astronomie (MPIA Heidelberg), Max-Planck-Institut f\"{u}r Astrophysik (MPA Garching), Max-Planck-Institut f\"{u}r Extraterrestrische Physik (MPE), National Astronomical Observatories of China, New Mexico State University, New York University, University of Notre Dame, Observatório Nacional / MCTI, The Ohio State University, Pennsylvania State University, Shanghai Astronomical Observatory, United Kingdom Participation Group, Universidad Nacional Autónoma de México, University of Arizona, University of Colorado Boulder, University of Oxford, University of Portsmouth, University of Utah, University of Virginia, University of Washington, University of Wisconsin, Vanderbilt University, and Yale University.

The authors recognize and acknowledge the very significant cultural role and reverence that the summit of Mauna Kea has always had within the indigenous Hawaiian community. We are most fortunate to have the opportunity to conduct observations from this superb mountain.

\vspace{5mm}




\bibliographystyle{mnras}
\bibliography{BibdeskLib}



\newpage
\appendix

\section{PSF estimation and point source modelling} \label{app:psf_iteration}
The quasar images dominate the observed flux over a large area and an accurate PSF estimate is necessary to obtain a reliable lens model inference based on the imaging data. The quasars in the lens system can not be taken into account in the PSF estimate without a sufficient subtraction of all other light components present. These other light components are model dependent and as such the PSF estimate when performed after model subtraction.

We estimate an initial PSF of the reduced image by a stack of two bright stars in the field. The individual stars are iteratively de-shifted to the center of a pixel (the inverse of an interpolated shift of the pixel grid). Additionally, we force the PSF to inherit a 90\textdegree-symmetry based on the symmetry inherited by the \textit{HST} optics. This allows us to rotate the stars 4 times and estimate the PSF with 8 stacks. We assign an additional error term (down-weighting of the image pixels mostly impacted by the PSF) based on the discrepancy between the 2 individual stars (8 stacks) that exceeds the S/N. We expect the stacking and interpolation to broaden the PSF estimate slightly with respect to the true underlining PSF.

\begin{figure*}
  \centering
  \includegraphics[angle=0, width=150mm]{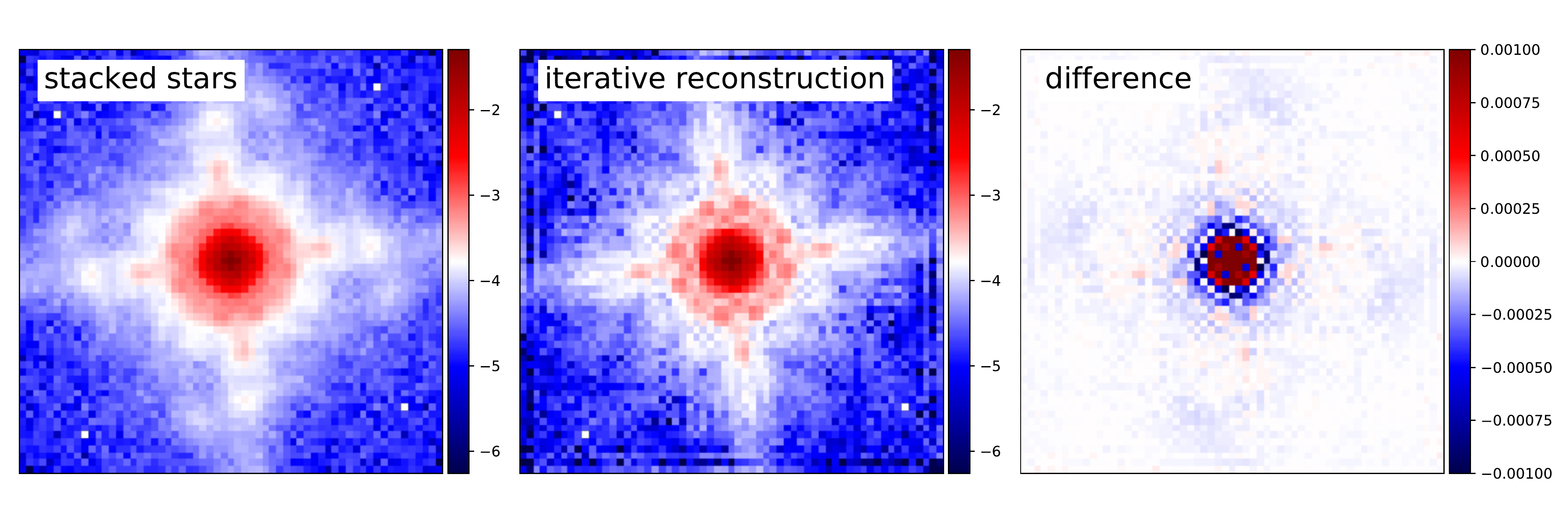}
  \caption{The result of the PSF estimation. \textbf{Left:} The PSF estimate from two bright stars in the field of the \textit{HST} exposures. \textbf{Middle:} The PSF estimated after an iterative approach taking into account the two quasar images. \textbf{Right:} The difference between the final estimate (middle image) and the initial guess (left image).}
\label{fig:psf_iteration}
\end{figure*}

In a next step, within the forward modelling of the imaging data, we sequentially subtract the best fit lens light and extended source surface brightness model and update the PSF with the additional two quasar point sources, similar to the procedure given by~\cite{Chen:2016} and~\cite{Birrer:2017a}. This procedure is repeated until a converged solution of the PSF for a given choice of lens model has been found. Figure~\ref{fig:psf_iteration} shows the result of those steps. The iterative PSF reconstruction provides a sharper PSF estimate with more prominent diffraction features. The forced 90\textdegree- symmetry in this process prevents the iterative reconstruction to over-constrain the PSF as the extended source surface brightness (Einstein ring) does not obey such a symmetry.

We repeat the iterative reconstruction three times within the PSO lens model optimization process to not be biased in the PSF model with respect to a specific choice of the model we assumed to perform this step.

\section{Estimating weighted galaxy count ratios from observational data} \label{sec:datawght}

To ensure that our method of computing galaxy weighted counts is unbiased, we must ensure that the LOS imaging data is of similar quality to that used to produce the CFHTLenS catalogues. Our GMOS and NIRI data were obtained with a pixel scale of $0.145\arcsec$ and $0.117\arcsec$, respectively. We used \texttt{Scamp} \citep{Bertin:2006} and \texttt{Swarp} \citep{Bertin:2002} to align them, correct for field distortion, and resample on a $0.187\arcsec$-scale pixel grid, matching CFHTLenS. We measure $5\sigma$ detection limits\footnote{$m_\mathrm{lim} = \mathrm{ZP} - 2.5 \log\left(5 \sqrt{N_\mathrm{pix}}\sigma_\mathrm{sky}\right)$, where ZP is the magnitude zero-point, $N_\mathrm{pix}$ is the number of pixels in a circle with radius 2.0\arcsec, and $\sigma_\mathrm{sky}$ is the sky-background noise variation. We derive the uncertainty as the standard deviation of the values in 10 empty regions across the frame.} of $24.54\pm0.04$ in the detection band $i$, matching the $24.54\pm0.19$ depth measured for CFHTLenS \citep{Erben:2013}, and $24.09\pm0.07$ ($g$), $24.88\pm0.04$ ($r$), $20.63\pm0.07$ ($Ks$, Vega-based).

We constructed the PSF in each band using unsaturated stars inside the FOV, combined together using the \texttt{IRAF PSF} task, and convolved them to a common seeing of $0.68\arcsec$, corresponding to the CFHTLenS $i$-band \citep{Erben:2013}. We performed object detections and photometric measurements using \texttt{Sextractor} \citep{sextractor}, with the same configuration used by CFHTLenS. As our resampling and convolutions can produce large noise correlation, which may significantly underestimate the photometric uncertainties measured with \texttt{Sextractor}, we use the technique described by \citet{Labbe:2003}, \citet{Gawiser:2006} and \citet{Quadri:2007} to correct for this effect.

We used the classification scheme of \citet{Hildebrandt:2012} to separate stars and galaxies, and further compared to the \textit{HST} image across the overlapping area. We used the 41 spectroscopic redshifts available over the $4\arcmin\times4\arcmin$ FOV to calibrate our photometric redshifts, which we computed with \texttt{BPZ} \citep{Benitez:2000}, and resulted in a photoz-specz scatter of 0.05, with an outlier fraction of $10$ per cent, comparable to the results for CFHTLenS \citep{Hildebrandt:2012}, even though we lack the $u$-band. Finally, in the uncertainties we report in Table \ref{tab:wghtcounts} we account for the following systematics: 1) sample variance using the four disjoined fields of CFHTLenS; 2) scatter from 10 samplings of the redshift and measured $i$-band magnitude of each galaxy in the aperture around the lensing system; 3) detections in the original-seeing $i$-band, not just the CFHTLenS value.

\section{Galaxy groups around the lensing system} \label{sec:groups}

\begin{figure}
\begin{center}
\includegraphics[angle=0,scale=0.65]{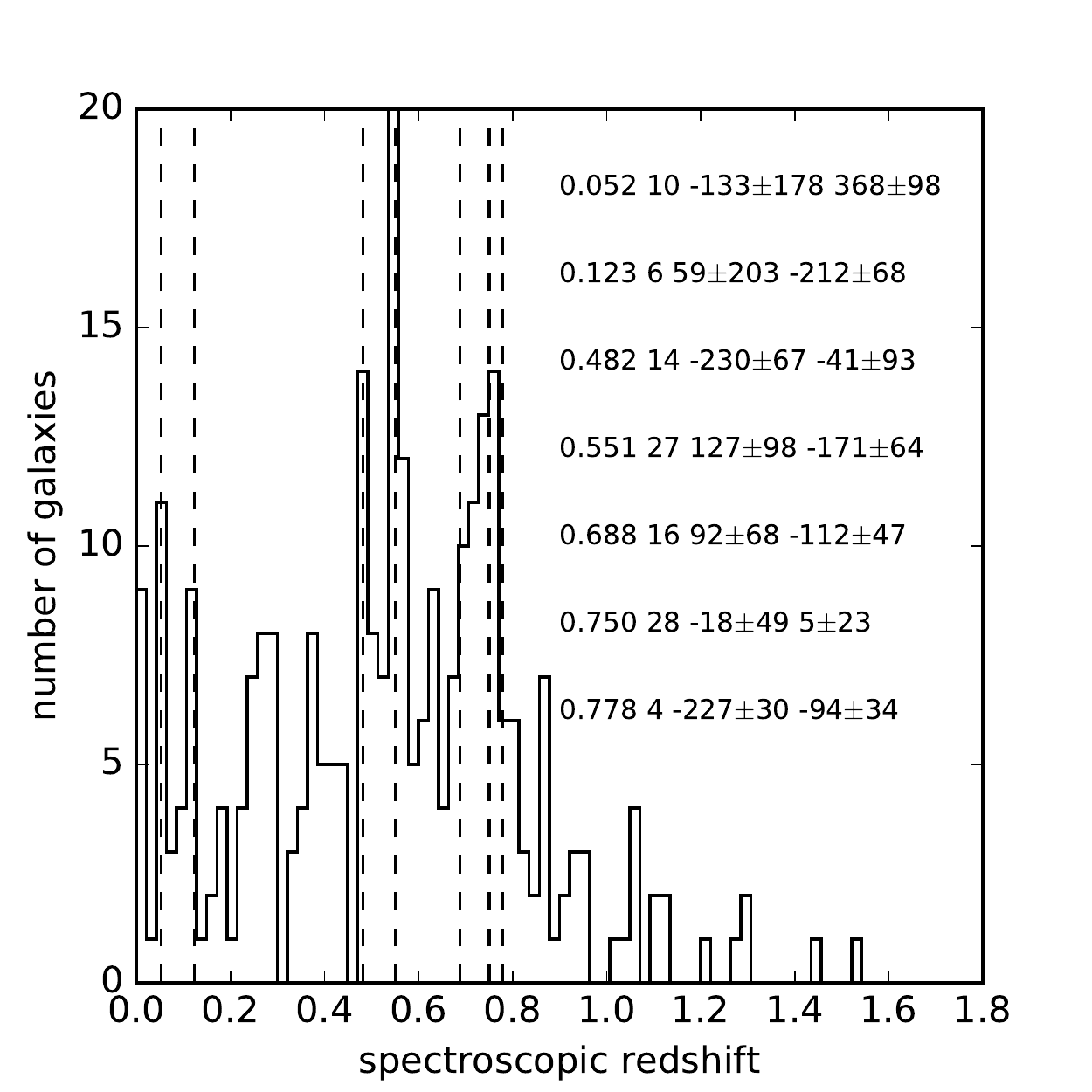}\\
\caption{Histogram of spectroscopic redshifts inside $14.5\arcmin$ around SDSS~J1206+4332, at $z<z_{\rm d}$, using $\Delta z=0.025$ bins. The redshifts of the groups identified following the methodology of \citet{Sluse:2017} are marked with vertical dashed lines. The numbers to the right specify the group redshift, number of group members, and the offset in arcseconds of the RA and DEC of the group centroid relative to the lens, respectively.}\label{fig:hist}
\end{center}
\end{figure}

Following the methodology of \citet{Sluse:2017}, we have conducted a search for galaxy groups in the available spectroscopic data. The result of this search is shown in Figure \ref{fig:hist}. We identify seven potential galaxy groups. One of the groups is consistent with the redshift of the lensing galaxy $z_{\rm d}$. This is one of the two richest groups, and its centroid is consistent with the position of the lensing galaxy (see Figure \ref{fig:fov}). Table \ref{tab:lens_group} presents the estimates of the group present at the lens redshift.

\begin{table}
\caption{The properties estimated for the group at the lens redshift.}
\begin{center}
\begin{threeparttable}
\begin{tabular}{l r r}
    
    \hline
    N members & = 15 \\
    redshift & = 0.74659 \\
    $\sigma_v$ & = 401 & $\pm$ 90 km/s \\
    Centroid (RA) & 181.61985833 & $\pm$ 30\arcsec \\
    Centroid (DEC) & 43.54065778 & $\pm$ 25\arcsec \\
    R$_{\text{vir}}$ & 1.4 & $\pm$ 0.3 Mpc \\
    R$_{200}$ & 1.2 & $\pm$ 0.3 Mpc \\
    Virial mass & 13.52 & $\pm$ 0.44 log$M_{\odot}$\\
    Distance to lens & 12.4\arcsec \\
     \hline
\end{tabular}
\begin{tablenotes}
\end{tablenotes}
\end{threeparttable}
\end{center}
\label{tab:lens_group}
\end{table}

\section{Summary of lens models with respect of the BIC value} \label{app:bic_models}
Table \ref{tab:bic_power_law} and \ref{tab:bic_composite} show the BIC ordered 128 models separated by the the two main deflector models considered in this work.

\begin{table*}
\caption{\texttt{SPEMD\_SERSIC} models ordered in increased BIC value. Shown are the 10 most promising models.}
\begin{center}
\begin{threeparttable}
\begin{tabular}{l r r r r r r r}
    \hline
    Main deflector & source model & perturber  & shear type  & mask  &  BIC  & $\Delta$BIC & posterior weight \\
    \hline
SPEMD\_SERSIC & SERSIC + n$_{max}$=8 & TRIPLET+2 & FOREGROUND & 3.0" & 5383 & 0 & 1 \\ 
SPEMD\_SERSIC & SERSIC + n$_{max}$=8 & TRIPLET+2 & SIMPLE & 3.2" & 5408 & 24 & 0.83575 \\ 
SPEMD\_SERSIC & SERSIC + n$_{max}$=8 & TRIPLET+2 & FOREGROUND & 3.2" & 5421 & 38 & 0.7815 \\ 
SPEMD\_SERSIC & SERSIC + n$_{max}$=8 & TRIPLET+2 & SIMPLE & 3.0" & 5496 & 112 & 0.389 \\ 
SPEMD\_SERSIC & SERSIC + n$_{max}$=8 & TRIPLET+2 & FOREGROUND & 3.0" & 5498 & 115 & 0.396 \\ 
SPEMD\_SERSIC & SERSIC + n$_{max}$=8 & TRIPLET+2 & SIMPLE & 3.0" & 5530 & 146 & 0.27625 \\ 
SPEMD\_SERSIC & SERSIC + n$_{max}$=8 & TRIPLET & SIMPLE & 3.0" & 5538 & 155 & 0.23525 \\ 
SPEMD\_SERSIC & SERSIC + n$_{max}$=8 & TRIPLET+2 & FOREGROUND & 3.2" & 5545 & 161 & 0.243 \\ 
SPEMD\_SERSIC & SERSIC + n$_{max}$=8 & TRIPLET & FOREGROUND & 3.0" & 5562 & 179 & 0.184 \\ 
SPEMD\_SERSIC & SERSIC + n$_{max}$=5 & TRIPLET+2 & FOREGROUND & 3.0" & 5581 & 197 & 0.1425 \\ 
SPEMD\_SERSIC & SERSIC + n$_{max}$=8 & TRIPLET & FOREGROUND & 3.2" & 5599 & 216 & 0.0905 \\ 
SPEMD\_SERSIC & SERSIC + n$_{max}$=8 & TRIPLET+2 & SIMPLE & 3.2" & 5623 & 239 & 0.07825 \\ 
SPEMD\_SERSIC & SERSIC + n$_{max}$=5 & TRIPLET+2 & SIMPLE & 3.0" & 5630 & 247 & 0.06475 \\ 
SPEMD\_SERSIC & SERSIC + n$_{max}$=8 & TRIPLET & FOREGROUND & 3.0" & 5653 & 269 & 0.04575 \\ 
SPEMD\_SERSIC & SERSIC + n$_{max}$=8 & TRIPLET & SIMPLE & 3.2" & 5654 & 270 & 0.0475 \\ 
SPEMD\_SERSIC & SERSIC + n$_{max}$=8 & TRIPLET & SIMPLE & 3.0" & 5663 & 279 & 0.0345 \\ 
SPEMD\_SERSIC & SERSIC + n$_{max}$=8 & TRIPLET & SIMPLE & 3.2" & 5666 & 282 & 0.03575 \\ 
SPEMD\_SERSIC & SERSIC + n$_{max}$=5 & TRIPLET+2 & SIMPLE & 3.2" & 5683 & 300 & 0.0295 \\ 
SPEMD\_SERSIC & SERSIC + n$_{max}$=8 & TRIPLET & FOREGROUND & 3.2" & 5698 & 315 & 0.0175 \\ 
SPEMD\_SERSIC & SERSIC + n$_{max}$=5 & TRIPLET & FOREGROUND & 3.0" & 5715 & 331 & 0.01525 \\ 
SPEMD\_SERSIC & SERSIC + n$_{max}$=5 & TRIPLET+2 & SIMPLE & 3.0" & 5729 & 346 & 0.0065 \\ 
SPEMD\_SERSIC & SERSIC + n$_{max}$=5 & TRIPLET+2 & FOREGROUND & 3.2" & 5737 & 353 & 0.007 \\ 
SPEMD\_SERSIC & SERSIC + n$_{max}$=5 & TRIPLET+2 & SIMPLE & 3.2" & 5833 & 449 & 0.00075 \\ 
SPEMD\_SERSIC & SERSIC + n$_{max}$=5 & TRIPLET & SIMPLE & 3.2" & 5903 & 519 & 0.00025 \\ 
SPEMD\_SERSIC & SERSIC + n$_{max}$=5 & TRIPLET & FOREGROUND & 3.2" & 5928 & 544 & 0.00025 \\ 
SPEMD\_SERSIC & SERSIC + n$_{max}$=5 & TRIPLET & SIMPLE & 3.0" & 5965 & 581 & 0.0 \\ 
SPEMD\_SERSIC & SERSIC + n$_{max}$=5 & TRIPLET & FOREGROUND & 3.0" & 5966 & 582 & 0.0 \\ 
SPEMD\_SERSIC & SERSIC + n$_{max}$=5 & TRIPLET+2 & FOREGROUND & 3.0" & 5989 & 606 & 0.0 \\ 
SPEMD\_SERSIC & SERSIC + n$_{max}$=5 & TRIPLET+2 & FOREGROUND & 3.2" & 5996 & 612 & 0.0 \\ 
SPEMD\_SERSIC & SERSIC + n$_{max}$=5 & TRIPLET & FOREGROUND & 3.2" & 6032 & 649 & 0.0 \\ 
SPEMD\_SERSIC & SERSIC + n$_{max}$=5 & TRIPLET & SIMPLE & 3.0" & 6066 & 682 & 0.0 \\ 
SPEMD\_SERSIC & SERSIC + n$_{max}$=5 & TRIPLET & SIMPLE & 3.2" & 6153 & 769 & 0.0 \\ 
SPEMD\_SERSIC & SERSIC + n$_{max}$=2 & TRIPLET+2 & SIMPLE & 3.0" & 6171 & 787 & 0.0 \\ 
SPEMD\_SERSIC & SERSIC + n$_{max}$=2 & TRIPLET+2 & FOREGROUND & 3.0" & 6314 & 930 & 0.0 \\ 
SPEMD\_SERSIC & SERSIC + n$_{max}$=2 & TRIPLET+2 & FOREGROUND & 3.2" & 6319 & 935 & 0.0 \\ 
SPEMD\_SERSIC & SERSIC + n$_{max}$=2 & TRIPLET & FOREGROUND & 3.0" & 6325 & 941 & 0.0 \\ 
SPEMD\_SERSIC & SERSIC + n$_{max}$=2 & TRIPLET+2 & FOREGROUND & 3.2" & 6374 & 991 & 0.0 \\ 
SPEMD\_SERSIC & SERSIC + n$_{max}$=2 & TRIPLET+2 & SIMPLE & 3.0" & 6388 & 1005 & 0.0 \\ 
SPEMD\_SERSIC & SERSIC + n$_{max}$=2 & TRIPLET & FOREGROUND & 3.0" & 6411 & 1028 & 0.0 \\ 
SPEMD\_SERSIC & SERSIC + n$_{max}$=2 & TRIPLET+2 & SIMPLE & 3.2" & 6561 & 1177 & 0.0 \\ 
SPEMD\_SERSIC & SERSIC + n$_{max}$=2 & TRIPLET+2 & SIMPLE & 3.2" & 6570 & 1186 & 0.0 \\ 
SPEMD\_SERSIC & SERSIC + n$_{max}$=2 & TRIPLET & SIMPLE & 3.2" & 6666 & 1283 & 0.0 \\ 
SPEMD\_SERSIC & SERSIC + n$_{max}$=2 & TRIPLET & FOREGROUND & 3.2" & 6711 & 1327 & 0.0 \\ 
SPEMD\_SERSIC & SERSIC + n$_{max}$=2 & TRIPLET+2 & FOREGROUND & 3.0" & 6711 & 1327 & 0.0 \\ 
SPEMD\_SERSIC & SERSIC & TRIPLET+2 & FOREGROUND & 3.2" & 6735 & 1352 & 0.0 \\ 
SPEMD\_SERSIC & SERSIC + n$_{max}$=2 & TRIPLET & FOREGROUND & 3.2" & 6759 & 1376 & 0.0 \\ 
SPEMD\_SERSIC & SERSIC & TRIPLET & FOREGROUND & 3.0" & 6771 & 1388 & 0.0 \\ 
SPEMD\_SERSIC & SERSIC & TRIPLET+2 & SIMPLE & 3.2" & 6774 & 1390 & 0.0 \\ 
SPEMD\_SERSIC & SERSIC + n$_{max}$=2 & TRIPLET & SIMPLE & 3.0" & 6779 & 1395 & 0.0 \\ 
SPEMD\_SERSIC & SERSIC & TRIPLET+2 & SIMPLE & 3.0" & 6788 & 1405 & 0.0 \\ 
SPEMD\_SERSIC & SERSIC + n$_{max}$=2 & TRIPLET & SIMPLE & 3.2" & 6831 & 1447 & 0.0 \\ 
SPEMD\_SERSIC & SERSIC & TRIPLET & FOREGROUND & 3.2" & 6880 & 1497 & 0.0 \\ 
SPEMD\_SERSIC & SERSIC & TRIPLET+2 & SIMPLE & 3.2" & 6904 & 1520 & 0.0 \\ 
SPEMD\_SERSIC & SERSIC & TRIPLET+2 & FOREGROUND & 3.2" & 6913 & 1530 & 0.0 \\ 
SPEMD\_SERSIC & SERSIC & TRIPLET+2 & FOREGROUND & 3.0" & 6932 & 1548 & 0.0 \\ 
SPEMD\_SERSIC & SERSIC & TRIPLET+2 & FOREGROUND & 3.0" & 6942 & 1558 & 0.0 \\ 
SPEMD\_SERSIC & SERSIC & TRIPLET & SIMPLE & 3.2" & 7013 & 1629 & 0.0 \\ 
SPEMD\_SERSIC & SERSIC + n$_{max}$=2 & TRIPLET & SIMPLE & 3.0" & 7018 & 1634 & 0.0 \\ 
SPEMD\_SERSIC & SERSIC & TRIPLET & FOREGROUND & 3.0" & 7036 & 1653 & 0.0 \\ 
SPEMD\_SERSIC & SERSIC & TRIPLET & FOREGROUND & 3.2" & 7073 & 1690 & 0.0 \\ 
SPEMD\_SERSIC & SERSIC & TRIPLET+2 & SIMPLE & 3.0" & 7098 & 1714 & 0.0 \\ 
SPEMD\_SERSIC & SERSIC & TRIPLET & SIMPLE & 3.0" & 7111 & 1727 & 0.0 \\ 
SPEMD\_SERSIC & SERSIC & TRIPLET & SIMPLE & 3.2" & 7153 & 1770 & 0.0 \\ 
SPEMD\_SERSIC & SERSIC & TRIPLET & SIMPLE & 3.0" & 7171 & 1788 & 0.0 \\
     \hline
\end{tabular}
\begin{tablenotes}
\end{tablenotes}
\end{threeparttable}
\end{center}
\label{tab:bic_power_law}
\end{table*}

\begin{table*}
\caption{\texttt{COMPOSITE} models ordered in increased BIC value. Shown are the 10 most promising models.}
\begin{center}
\begin{threeparttable}
\begin{tabular}{l r r r r r r r}
    \hline
    Main deflector & source model & perturber  & shear type  & mask  &  BIC  & $\Delta$BIC & posterior weight \\
    \hline
COMPOSITE & SERSIC + n$_{max}$=8 & TRIPLET+2 & SIMPLE & 3.0" & 4859 & 0 & 1 \\ 
COMPOSITE & SERSIC + n$_{max}$=8 & TRIPLET & FOREGROUND & 3.0" & 4921 & 61 & 0.6575 \\ 
COMPOSITE & SERSIC + n$_{max}$=8 & TRIPLET+2 & FOREGROUND & 3.0" & 4951 & 91 & 0.49125 \\ 
COMPOSITE & SERSIC + n$_{max}$=8 & TRIPLET+2 & FOREGROUND & 3.0" & 4958 & 98 & 0.48425 \\ 
COMPOSITE & SERSIC + n$_{max}$=8 & TRIPLET+2 & SIMPLE & 3.2" & 4964 & 104 & 0.4235 \\ 
COMPOSITE & SERSIC + n$_{max}$=8 & TRIPLET+2 & SIMPLE & 3.0" & 4979 & 119 & 0.38125 \\ 
COMPOSITE & SERSIC + n$_{max}$=8 & TRIPLET & SIMPLE & 3.0" & 5015 & 156 & 0.217 \\ 
COMPOSITE & SERSIC + n$_{max}$=8 & TRIPLET & FOREGROUND & 3.2" & 5044 & 184 & 0.1755 \\ 
COMPOSITE & SERSIC + n$_{max}$=8 & TRIPLET+2 & SIMPLE & 3.2" & 5047 & 187 & 0.1595 \\ 
COMPOSITE & SERSIC + n$_{max}$=8 & TRIPLET & FOREGROUND & 3.2" & 5112 & 252 & 0.063 \\ 
COMPOSITE & SERSIC + n$_{max}$=8 & TRIPLET & SIMPLE & 3.2" & 5114 & 254 & 0.05625 \\ 
COMPOSITE & SERSIC + n$_{max}$=8 & TRIPLET+2 & FOREGROUND & 3.2" & 5134 & 274 & 0.0455 \\ 
COMPOSITE & SERSIC + n$_{max}$=8 & TRIPLET & FOREGROUND & 3.0" & 5148 & 288 & 0.027 \\ 
COMPOSITE & SERSIC + n$_{max}$=8 & TRIPLET+2 & FOREGROUND & 3.2" & 5164 & 304 & 0.02275 \\ 
COMPOSITE & SERSIC + n$_{max}$=5 & TRIPLET & FOREGROUND & 3.0" & 5185 & 326 & 0.014 \\ 
COMPOSITE & SERSIC + n$_{max}$=5 & TRIPLET+2 & FOREGROUND & 3.2" & 5209 & 349 & 0.007 \\ 
COMPOSITE & SERSIC + n$_{max}$=5 & TRIPLET+2 & FOREGROUND & 3.0" & 5211 & 351 & 0.008 \\ 
COMPOSITE & SERSIC + n$_{max}$=5 & TRIPLET+2 & SIMPLE & 3.2" & 5254 & 394 & 0.00525 \\ 
COMPOSITE & SERSIC + n$_{max}$=5 & TRIPLET+2 & FOREGROUND & 3.2" & 5291 & 431 & 0.00025 \\ 
COMPOSITE & SERSIC + n$_{max}$=5 & TRIPLET+2 & FOREGROUND & 3.0" & 5342 & 482 & 0.0 \\ 
COMPOSITE & SERSIC + n$_{max}$=8 & TRIPLET & SIMPLE & 3.2" & 5394 & 534 & 0.0 \\ 
COMPOSITE & SERSIC + n$_{max}$=5 & TRIPLET+2 & SIMPLE & 3.0" & 5397 & 537 & 0.0 \\ 
COMPOSITE & SERSIC + n$_{max}$=5 & TRIPLET+2 & SIMPLE & 3.2" & 5400 & 540 & 0.0 \\ 
COMPOSITE & SERSIC + n$_{max}$=5 & TRIPLET & SIMPLE & 3.0" & 5425 & 565 & 0.0 \\ 
COMPOSITE & SERSIC + n$_{max}$=5 & TRIPLET & FOREGROUND & 3.2" & 5476 & 616 & 0.0 \\ 
COMPOSITE & SERSIC + n$_{max}$=8 & TRIPLET & SIMPLE & 3.0" & 5491 & 631 & 0.0 \\ 
COMPOSITE & SERSIC + n$_{max}$=5 & TRIPLET & SIMPLE & 3.2" & 5493 & 633 & 0.0 \\ 
COMPOSITE & SERSIC + n$_{max}$=5 & TRIPLET & SIMPLE & 3.2" & 5504 & 644 & 0.0 \\ 
COMPOSITE & SERSIC + n$_{max}$=5 & TRIPLET+2 & SIMPLE & 3.0" & 5566 & 706 & 0.0 \\ 
COMPOSITE & SERSIC + n$_{max}$=5 & TRIPLET & FOREGROUND & 3.0" & 5578 & 718 & 0.0 \\ 
COMPOSITE & SERSIC + n$_{max}$=5 & TRIPLET & FOREGROUND & 3.2" & 5623 & 763 & 0.0 \\ 
COMPOSITE & SERSIC + n$_{max}$=2 & TRIPLET+2 & FOREGROUND & 3.0" & 5656 & 796 & 0.0 \\ 
COMPOSITE & SERSIC + n$_{max}$=2 & TRIPLET+2 & FOREGROUND & 3.2" & 5704 & 844 & 0.0 \\ 
COMPOSITE & SERSIC + n$_{max}$=2 & TRIPLET+2 & FOREGROUND & 3.0" & 5716 & 857 & 0.0 \\ 
COMPOSITE & SERSIC + n$_{max}$=2 & TRIPLET+2 & FOREGROUND & 3.2" & 5748 & 888 & 0.0 \\ 
COMPOSITE & SERSIC + n$_{max}$=2 & TRIPLET & SIMPLE & 3.0" & 5752 & 892 & 0.0 \\ 
COMPOSITE & SERSIC + n$_{max}$=5 & TRIPLET & SIMPLE & 3.0" & 5774 & 914 & 0.0 \\ 
COMPOSITE & SERSIC + n$_{max}$=2 & TRIPLET & SIMPLE & 3.0" & 5775 & 915 & 0.0 \\ 
COMPOSITE & SERSIC + n$_{max}$=2 & TRIPLET & SIMPLE & 3.2" & 5805 & 945 & 0.0 \\ 
COMPOSITE & SERSIC + n$_{max}$=2 & TRIPLET & FOREGROUND & 3.2" & 5815 & 956 & 0.0 \\ 
COMPOSITE & SERSIC + n$_{max}$=2 & TRIPLET+2 & SIMPLE & 3.2" & 5839 & 979 & 0.0 \\ 
COMPOSITE & SERSIC + n$_{max}$=2 & TRIPLET+2 & SIMPLE & 3.0" & 5903 & 1043 & 0.0 \\ 
COMPOSITE & SERSIC + n$_{max}$=2 & TRIPLET+2 & SIMPLE & 3.2" & 5933 & 1074 & 0.0 \\ 
COMPOSITE & SERSIC + n$_{max}$=2 & TRIPLET+2 & SIMPLE & 3.0" & 5936 & 1076 & 0.0 \\ 
COMPOSITE & SERSIC + n$_{max}$=2 & TRIPLET & SIMPLE & 3.2" & 5942 & 1082 & 0.0 \\ 
COMPOSITE & SERSIC & TRIPLET+2 & FOREGROUND & 3.0" & 6019 & 1159 & 0.0 \\ 
COMPOSITE & SERSIC & TRIPLET & SIMPLE & 3.2" & 6077 & 1217 & 0.0 \\ 
COMPOSITE & SERSIC & TRIPLET+2 & FOREGROUND & 3.2" & 6092 & 1232 & 0.0 \\ 
COMPOSITE & SERSIC & TRIPLET+2 & FOREGROUND & 3.0" & 6105 & 1245 & 0.0 \\ 
COMPOSITE & SERSIC & TRIPLET+2 & SIMPLE & 3.0" & 6112 & 1253 & 0.0 \\ 
COMPOSITE & SERSIC & TRIPLET & SIMPLE & 3.0" & 6114 & 1254 & 0.0 \\ 
COMPOSITE & SERSIC + n$_{max}$=2 & TRIPLET & FOREGROUND & 3.0" & 6147 & 1287 & 0.0 \\ 
COMPOSITE & SERSIC & TRIPLET+2 & SIMPLE & 3.0" & 6212 & 1352 & 0.0 \\ 
COMPOSITE & SERSIC & TRIPLET & FOREGROUND & 3.0" & 6213 & 1353 & 0.0 \\ 
COMPOSITE & SERSIC & TRIPLET+2 & SIMPLE & 3.2" & 6265 & 1405 & 0.0 \\ 
COMPOSITE & SERSIC + n$_{max}$=2 & TRIPLET & FOREGROUND & 3.2" & 6300 & 1440 & 0.0 \\ 
COMPOSITE & SERSIC & TRIPLET+2 & SIMPLE & 3.2" & 6323 & 1463 & 0.0 \\ 
COMPOSITE & SERSIC & TRIPLET & SIMPLE & 3.2" & 6390 & 1530 & 0.0 \\ 
COMPOSITE & SERSIC & TRIPLET & FOREGROUND & 3.0" & 6400 & 1541 & 0.0 \\ 
COMPOSITE & SERSIC & TRIPLET & SIMPLE & 3.0" & 6411 & 1551 & 0.0 \\ 
COMPOSITE & SERSIC & TRIPLET & FOREGROUND & 3.2" & 6417 & 1557 & 0.0 \\ 
COMPOSITE & SERSIC & TRIPLET+2 & FOREGROUND & 3.2" & 6545 & 1685 & 0.0 \\ 
COMPOSITE & SERSIC + n$_{max}$=2 & TRIPLET & FOREGROUND & 3.0" & 6691 & 1831 & 0.0 \\ 
COMPOSITE & SERSIC & TRIPLET & FOREGROUND & 3.2" & 6899 & 2039 & 0.0 \\
    \hline
\end{tabular}
\begin{tablenotes}
\end{tablenotes}
\end{threeparttable}
\end{center}
\label{tab:bic_composite}
\end{table*}



\bsp    
\label{lastpage}
\end{document}